\documentclass{vldb}
\usepackage{graphicx}
\usepackage{graphicx}
\usepackage{balance}
\usepackage{times}
\usepackage{subfigure}
\usepackage{hyperref}
\usepackage{comment}

\usepackage{algorithm}
\usepackage[noend]{algorithmic}

\newtheorem{definition}{Definition}
\newtheorem{theorem}{Theorem}
\newtheorem{lemma}{Lemma}

\usepackage{xcolor}

\definecolor{green}{rgb}{0.0, 0.55, 0.13}

\vldbTitle{Scalable Mining of Maximal Quasi-Cliques: An Algorithm-System Codesign Approach}
\vldbAuthors{Guimu Guo, Da Yan, M.\ Tamer \"{O}zsu, Zhe Jiang, }
\vldbDOI{https://doi.org/10.14778/xxxxxxx.xxxxxxx}
\vldbVolume{12}
\vldbNumber{xxx}
\vldbYear{2019}

\begin{document}

\title{Scalable Mining of Maximal Quasi-Cliques: An Algorithm-System Codesign Approach}

\author{
{Guimu	Guo$^*$,\ \ \ \ \ \ Da Yan$^*$,\ \ \ \ \ \ M.\ Tamer \"{O}zsu$^\dag$},\ \ \ \ \ \ Zhe Jiang$^\ddag$,\ \ \ \ \ \ Jalal Majed Khalil$^*$
\vspace{1mm}\\
\fontsize{10}{10}\selectfont\itshape\rmfamily Guimu	Guo and Da Yan are parallel first authors
\vspace{2mm}\\
\fontsize{10}{10}\selectfont\itshape\rmfamily $^*$Department of Computer Science,  The University of Alabama at Birmingham\ \ 
\fontsize{9}{9}\selectfont\ttfamily\upshape \{guimuguo, yanda, jalalk\}@uab.edu\\
\fontsize{10}{10}\selectfont\itshape\rmfamily $^\dag$David R.\ Cheriton School of Computer Science, University of Waterloo\ \ \ \ \ \ \ \ 
\fontsize{9}{9}\selectfont\ttfamily\upshape tamer.ozsu@uwaterloo.ca\ \ \ \ \ \ \ \ \\
\fontsize{10}{10}\selectfont\itshape\rmfamily $^\ddag$Department of Computer Science, University of Alabama\ \ \ \ \ \ \ \ \ \ \ \ \ \ \ \ \ 
\fontsize{9}{9}\selectfont\ttfamily\upshape zjiang@cs.ua.edu\ \ \ \ \ \ \ \ \ 
}

\maketitle

\begin{abstract}
Given a user-specified minimum degree threshold $\gamma$, a $\gamma$-quasi-clique is a subgraph $g=(V_g,E_g)$ where each vertex $v\in V_g$ connects to at least $\gamma$ fraction of the other vertices (i.e., $\lceil \gamma\cdot(|V_g|-1)\rceil$ vertices) in $g$. Quasi-clique is one of the most natural definitions for dense structures useful in finding communities in social networks and discovering significant biomolecule structures and pathways.
However, mining maximal quasi-cliques is notoriously expensive.

In this paper, we design parallel algorithms for mining maximal quasi-cliques on G-thinker,  a recent distributed framework targeting divide-and-conquer graph mining algorithms that decomposes the mining into compute-intensive tasks to fully utilize CPU cores. However, we found that directly using G-thinker results in the straggler problem due to (i)~the drastic load imbalance among different tasks and (ii)~the difficulty of predicting the task running time and the time growth with task-subgraph size. 
We address these challenges by redesigning G-thinker's execution engine to prioritize long-running tasks for mining, and by utilizing a novel 
timeout 
%time-delayed divide-and-conquer 
strategy to effectively decompose the mining workloads of long-running tasks to improve load balancing. While this system redesign applies to many other expensive dense subgraph mining problems, this paper verifies the idea by adapting the state-of-the-art quasi-clique algorithm, Quick, to our redesigned G-thinker. We improve Quick by integrating new pruning rules, and fixing some missed boundary cases that could lead to missed results.
Extensive experiments verify that our new solution scales well with the number of CPU cores, achieving 201$\times$ runtime speedup when mining a graph with 3.77M vertices and 16.5M edges in a 16-node cluster.
\end{abstract}

\section{Introduction}\label{sec:intro}
Given a degree threshold $\gamma$ and an undirected graph $G$, a $\gamma$-quasi-clique is a subgraph of $G$, denoted by $g=(V_g,E_g)$, where each vertex connects to at least $\lceil \gamma\cdot(|V_g|-1)\rceil$ other vertices in $g$. Quasi-clique is a natural generalization of clique that is useful in mining various networks, such as finding protein complexes or biologically relevant functional groups~\cite{bhattacharyya2009mining,MatsudaIH99,bader2003automated,bu2003topological,hu2005mining,ucar2006improving}, and social communities~\cite{li2014uncovering,hopcroft2004tracking} that can correspond to cybercriminals~\cite{weiss2015tracking}, botnets~\cite{ecrime,weiss2015tracking} and spam/phishing email sources~\cite{sac,sheng2009empirical}.

Mining maximal quasi-cliques is notoriously expensive~\cite{bigdata18} and the state-of-the-art algorithms~\cite{quick,Pei05,cocain} were only tested on small graphs. For example, Quick~\cite{quick}, the best among existing algorithms, was only tested on graphs with thousands of vertices~\cite{quick}. This has hampered its use in real applications involving big graphs.

In this paper, we design parallel algorithms for mining maximal quasi-cliques that scale to big graphs. Our algorithms follow the idea of divide and conquer which partitions the problem of mining a big graph into tasks that mine smaller subgraphs for concurrent execution, which has been made possible recently by the G-thinker~\cite{gthinker} framework for distributed graph mining that avoids the IO bottleneck for data movement that exists in other existing data-intensive systems. In fact, it is found that using conventional IO-bound data-intensive systems could result in a throughput comparable or even less than a single-threaded program~\cite{cost2,ChuC12}, making it a must to use a compute-intensive framework like G-thinker.

However, we found that porting such a divide-and-conquer algorithm (hereafter called divisible algorithm for simplicity) directly to the current G-thinker implementation still leads to the straggler problem. This is because the state-of-the-art divisible algorithms for mining dense subgraphs such as quasi-cliques and $k$-plexes~\cite{kplex1} are much more difficult than the applications that G-thinker already implemented, such as maximum clique finding and triangle counting~\cite{gthinker}. Specifically, \cite{bigdata18} showed that even the problem of detecting whether a given quasi-clique in a graph is maximal is NP-hard, while~\cite{kplex1} showed that ``(i)~maximal $k$-plexes are even more numerous than maximal cliques'', and that ``(ii)~the most efficient algorithms in the literature for computing maximal $k$-plexes can only be used on small-size graphs''. Unlike those simpler problems considered in~\cite{gthinker} where the runtimes of individual tasks are relatively short compared with the entire mining workloads, quasi-clique mining generates tasks of drastically different running time which was not sufficiently handled by the G-thinker engine.

We remark that while the existing execution engine of G-thinker is insufficient, its graph-divisible computing paradigm is a perfect fit for dense subgraph mining problems, and all we need to do is to redesign G-thinker's engine to address the straggler problem. After all, before G-thinker, such parallelization was not easy: \cite{bigdata18} makes it a future work ``Can the algorithms for quasi-cliques be parallelized effectively?'', while~\cite{kplex1} indicated that ``We are not aware of parallel techniques for implementing the all\_plexes() sub-routine, and we leave this for future work''. Addressing the load balancing issue of G-thinker would not only benefit parallel quasi-clique mining, but also the parallelization of many other graph-divisible algorithms for mining dense subgraphs~\cite{kplex0,kplex1,kplex2,biclique,kecc,maxclique,ocs}.

We adopt an algorithm-system codesign approach to parallelize quasi-clique mining, and the main contributions are as follows:
\vspace{-2mm}
\begin{itemize}
\setlength\itemsep{-0.2em}
\item We redesigned G-thinker's execution engine to prioritize the execution of big tasks that tend to be stragglers. Specifically, we add a global task queue to keep big tasks which is shared by all mining threads in a machine for prioritized fetching; task stealing is used to balance big tasks among machines.
\item We improved Quick by integrating new pruning rules that are highly effective, and fixing some missed boundary cases in Quick that could lead to missed results. The new algorithm, called Quick+, is then parallelized using G-thinker API.
\item We achieve effective and early decomposition of big tasks by a novel timeout 
% time-delayed task decomposition 
strategy, without the need to predict task running time which is very difficult.
\end{itemize}
\vspace{-1mm}

The efficiency of our parallel solution has been extensively verified over various real graph datasets.  For example, in our 16-node cluster, we are able to obtain 201$\times$ speedup when mining 0.89-quasi-cliques on the {\em Patent} graph with 3.77M vertices and 16.5M edges in a 16-node cluster: the total serial mining time of 25,369 seconds are computed by our parallel solution in 126 seconds.

The rest of this paper is organized as follows. Section~\ref{sec:related} reviews those related work closely related to quasi-clique mining and graph computing time prediction. Section~\ref{sec:preliminaries} formally defines our notations, the general divisible algorithmic framework for dense subgraph mining which is also adopted by Quick and our Quick+, and which is amenable to parallelization in G-thinker. Section~\ref{sec:challenges} then demonstrates that the tasks of Quick+ can have drastically different running time, and describes the straggler problem that we faced.
Section~\ref{sec:gthinker} then reviews the original execution engine of G-thinker and describes our redesign to prioritize big tasks for execution. Section~\ref{sec:quick} then outlines our Quick+ algorithm and Section~\ref{sec:algo} presents its adaptation on G-thinker as well as another version of it using 
%time-delayed task decomposition. 
timeout-based task decomposition. 
Finally, Section~\ref{sec:results} reports our experiments and Section~\ref{sec:conclude} concludes this paper.

\section{Related Work}\label{sec:related}
A few seminal works devised branch-and-bound subgraph searching algorithms for mining quasi-cliques, such as Crochet~\cite{Pei05,Pei09} and Cocain~\cite{cocain} which finally led to the Quick algorithm~\cite{quick} that integrated all previous search space pruning techniques and added new %ones, especially a lower bound base pruning that is shown to speed up mining by 192.48$\times$. 
effective ones.
However, we find that % the pruning rules 
some pruning techniques
are not utilized or fully utilized by Quick. Even worse, Quick may miss results. We will elaborate on these weaknesses in Section~\ref{sec:quick}.

%Yang et al.~\cite{yangyi} studied the problem of mining a set of diversified temporal subgraph patterns from a temporal graph, where each subgraph is associated with the time interval that the pattern spans. The dense subgraph definition uses $\gamma$-quasi-cliques, and the algorithm is essentially adapted from Quick to include the temporal aspects.

Sanei-Mehri et al.~\cite{bigdata18} noticed that if $\gamma'$-quasi-cliques ($\gamma'>\gamma$) are mined first using Quick which are faster to find, then it is more efficient to expand these ``kernels'' to generate $\gamma$-quasi-cliques than to mine them from the original graph. Their kernel expansion is conducted only on those largest $\gamma'$-quasi-cliques extracted by postprocessing, in order to find big $\gamma$-quasi-cliques as opposed to all of them to keep time tractable. However, this work does not fundamentally address the scalability issue: (1)~it only studies the problem of enumerating $k$ big maximal quasi-cliques containing kernels rather than all valid ones, and these subgraphs can be clustered in one region (e.g., they overlap on a smaller clique) while missing results on other parts of the data graph, compromising result diversity; (2)~the method still needs to first find some $\gamma'$-quasi-cliques to grow from and this first step is still using Quick; and (3)~the method is not guaranteed to return exactly the set of top-$k$ maximal quasi-cliques. We remark that the kernel-based acceleration technique is orthogonal to our parallel algorithm and can be easily incorporated; however, as Section~\ref{sec:results} shall show, the performance of this solution is only faster than our exact solution when $k$ is very small.

Other than~\cite{bigdata18}, quasi-cliques have seldom been considered in a big graph setting. Quick~\cite{quick} was only tested on two small graphs: %a yeast interaction network with 4932 vertices (proteins) and 17201 edges (interactions), and an {\em E.\ coli} interaction network with 1846 vertices and 5929 edges. 
one with 4,932 vertices and 17,201 edges, and the other with 1,846 vertices and 5,929 edges.
In fact, earlier works~\cite{Pei05,Pei09,cocain} formulate quasi-clique mining as frequent pattern mining problems where the goal is to find quasi-clique patterns that appear in a significant portion of small graph transactions in a graph database. Some works consider big graphs but not the problem of finding all valid quasi-cliques, but rather those that contain a particular vertex or a set of query vertices~\cite{query_driven,target_vertex,density_community} to aggressively narrow down the search space by sacrificing result diversity, with some additional pruning rules beyond Quick, some in the query-vertex context.

There is another definition of quasi-clique based on edge density~\cite{density_def0,density_def1,density_community} rather than vertex degree, but it is essentially a different kind of dense subgraph definition. As \cite{density_community} indicates, the edge-density based quasi-cliques are less dense than our degree-based quasi-cliques, and thus we focus on degree-based quasi-cliques in this paper as in~\cite{density_community}.
%Quasi-clique can also be defined based on the total number of edges, i.e., the edge density of a subgraph should pass a user-defined threshold~\cite{density_def0,density_def1,density_community}. 
Brunato et al.~\cite{brunato2007effectively} further consider both vertex degree and edge density. There are also many other definitions of dense subgraphs~\cite{kplex0,kplex1,kplex2,biclique,kecc,maxclique,ocs}, and they all follow a similar divisible algorithmic framework as Quick (c.f.\ Section~\ref{sec:preliminaries}).

% G-thinker~\cite{gthinker} is a recent graph mining system that is able to achieve high CPU utilization and scale to big graphs. For example, G-thinker is able to find the maximum clique (with 129 vertices) of the big Friendster social network containing 65.6 M vertices and 1,806 M edges using only 252 seconds in total and 3.1 GB memory per machine in a small cluster~\cite{gthinker}. While the current engine design is sufficient for many graph mining applications, quasi-clique mining tasks have so variable running times that a system redesign becomes necessary to scale.

A recent work proposed to use machine learning to predict the running time of graph computation for workload partitioning~\cite{wenfei}, but the graph algorithms considered there are iterative algorithms that do not have unpredictable pruning rules and thus the running time can be easily estimated. This is not the case in quasi-clique mining (c.f.\ Section~\ref{sec:challenges}),  and dense subgraph mining in general which adopts divide-and-conquer (and often recursive) algorithms, calling for a new solution for effective task workload partitioning.

\section{Preliminaries}\label{sec:preliminaries}
%This section prepares the notations for subsequent algorithmic description, formally defines our mining problem, and provides the intuition of the algorithmic framework for concurrency-amenable mining, i.e., the set-enumeration tree for search space partitioning. We then review the programming model of G-thinker and explain how the algorithmic framework naturally maps to G-thinker's API.

\noindent {\bf Graph Notations.} We consider an undirected graph $G = (V, E)$ where $V$ (resp.\ $E$) is the set of vertices (resp.\ edges). The vertex set of a graph $G$ can also be explicitly denoted as $V(G)$. We use $G(S)$ to denote the subgraph of $G$ induced by a vertex set $S\subseteq V$, and use $|S|$ to denote the number of vertices in $S$. We also abuse the notation and use $v$ to mean the singleton set $\{v\}$.
We denote the set of neighbors of a vertex $v$ in $G$ by $N(v)$, and denote the degree of $v$ in $G$ by $d(v)=|N(v)|$. Given a vertex subset $V'\subseteq V$, we define $N_{V'}(v)=\{u\,|\,(u, v)\in E, u\in V'\}$, i.e., $N_{V'}(v)$ is the set of $v$'s neighbors inside $V'$, and we also define $d_{V'}(v)=|N_{V'}(v)|$.

\begin{figure}[t]
\centering
\includegraphics[width=0.6\columnwidth]{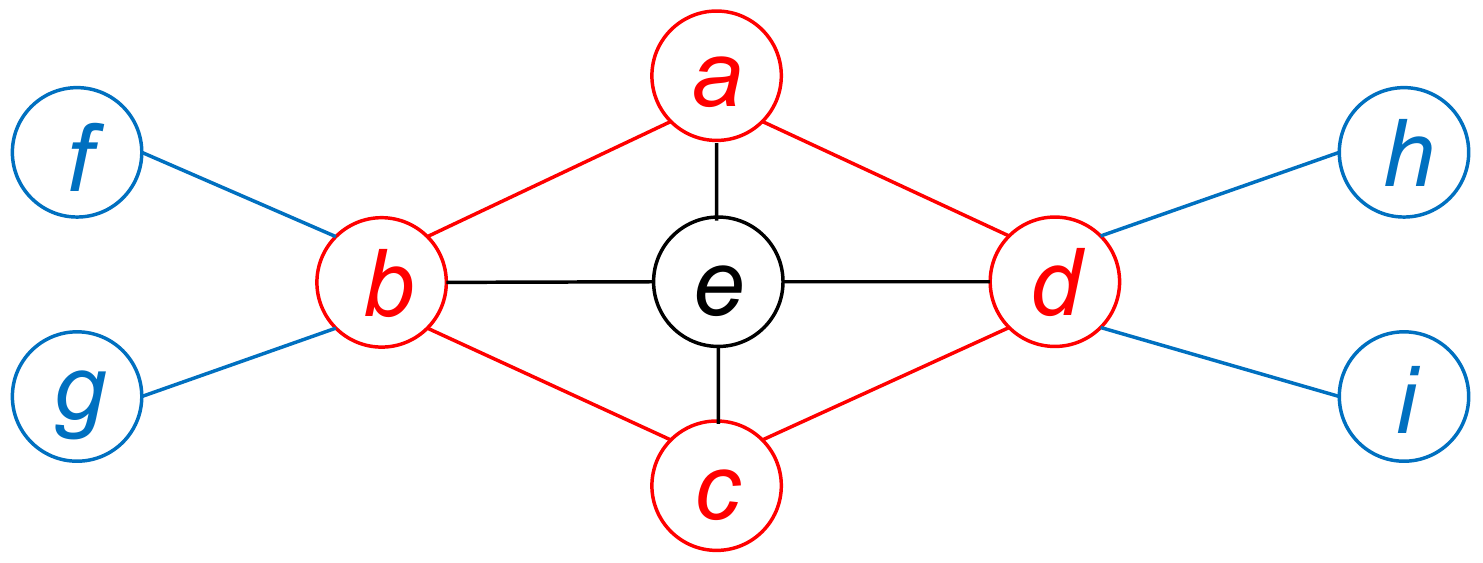}
\vspace{-2mm}
\caption{An Illustrative Graph}\label{qc}
\vspace{-4mm}
\end{figure}

To illustrate the notations, consider the graph $G$ shown in Figure~\ref{qc}. Let us use $v_a$ to denote Vertex~\textcircled{a} (the same for other vertices),  thus we have $N(v_d)=\{v_a, v_c, v_e, v_h, v_i\}$ and $d(v_d)=5$. Also, let $S=\{v_a, v_b, v_c, v_d, v_e\}$, then $G(S)$ is the subgraph of $G$ consisting of the vertices and edges in red and black.

Given two vertices $u, v\in V$, we define $\delta(u, v)$ as the number of edges on the shortest path between $u$ and $v$. We call $G$ as connected if $\delta(u, v)<\infty$ for any $u, v\in V$. We further define $N_{k}(v)=\{u\,|\,\delta(u,v)=k\}$ and define $N_k^+(v)=\{u\,|\,\delta(u,v)\leq k\}$. In a nutshell, $N_k^+(v)$ are the set of vertices reachable from $v$ within $k$ hops, and $N_{k}(v)$ are the set of vertices reachable from $v$ in $k$ hops but not in $(k-1)$ hops. Then, we have $N_0(v)=v$ and $N_1(v)=N(v)$, and
$N_k^+(v) = N_0(v)+N_1(v)+\ldots+N_k(v).$
For 2-hop neighbors, we define $B(v)=N_2(v)$ and $\mathbb{B}(v)=N_2^+(v)$.

To illustrate using Figure~\ref{qc}, we have $N(v_e)=\{v_a, v_b, v_c, v_d\}$, $B(v_e)=\{v_f, v_g, v_h, v_i\}$, and $\mathbb{B}(v_e)$ consisting of all vertices.

\vspace{1mm}
\noindent {\bf Problem Definition.} We next formally define our problem. %Given a user-specified minimum degree threshold $\gamma$, a $\gamma$-quasi-clique is a subgraph $G$ where each vertex connects to at least $\gamma$ fraction of the other vertices in $G$. Formally,

\vspace{-2mm}

\begin{definition}[$\gamma$-quasi-clique]
{\em % to cancel italics
A graph $G = (V, E)$ is a $\gamma$-quasi-clique ($0\leq\gamma\leq1$) if $G$ is connected, and for every vertex $v\in V$, its degree $d(v)\geq \lceil \gamma\cdot(|V|-1)\rceil$.
}
\end{definition}

\vspace{-2mm}

If a graph is a $\gamma$-quasi-clique, then its subgraphs usually become uninteresting, so we only mine maximal $\gamma$-quasi-clique as follows:

\vspace{-2mm}

\begin{definition}[Maximal $\gamma$-quasi-clique]
{\em % to cancel italics
Given graph $G = (V, E)$ and a vertex set $S\subseteq V$, $G(S)$ is a maximal $\gamma$-quasi-clique of $G$ if $G(S)$ is a $\gamma$-quasi-clique, and there does not exist a superset $S'\supset S$ such that $G(S')$ is a $\gamma$-quasi-clique.
}
\end{definition}

\vspace{-2mm}

To illustrate using Figure~\ref{qc}, consider $S_1=\{v_a, v_b, v_c, v_d\}$ (i.e., vertices in red) and $S_2=S_1\cup v_e$. If we set  $\gamma=0.6$, then both $S_1$ and $S_2$ are $\gamma$-quasi-cliques: every vertex in $S_1$ has at least 2 neighbors in $G(S_1)$ among the other 3 vertices (and $2/3>0.6$), while every vertex in $S_2$ has at least 3 neighbors in $G(S_2)$ among the other 4 vertices (and $3/4>0.6$). Also, since $S_1\subset S_2$, $G(S_1)$ is not a maximal $\gamma$-quasi-clique.

In the literature of dense subgraph mining, researchers usually only strive to find big dense subgraphs, such as the largest dense subgraph~\cite{biclique,kplex1,maxclique,query_driven}, the top-$k$ largest ones~\cite{bigdata18}, and those larger than a predefined size threshold~\cite{kplex1,kplex2,quick}. There are two reasons. (i)~Small dense subgraphs are common and thus statistically insignificant and not interesting. For example, a single vertex itself is a quasi-clique for any $\gamma$, and so is an edge with its two end-vertices. (ii)~The number of dense subgraphs grows exponentially with the graph size and is thus intractable unless we focus on finding large ones. In fact, it has been shown that even the problem of detecting if a given quasi-clique is maximal is NP-hard~\cite{bigdata18}, and it is well recognized that those clique relaxation definitions (aka.\ pseudo-clique) are much more expensive than clique mining~ \cite{bigdata18,kplex0,kplex1,kplex2} which is already NP-hard per se. In fact, there are algorithms that simply guess the maximum pseudo-clique size in order to utilize effective size-based pruning, and adjust the guess if the search fails~\cite{kplex1,maxclique}. Following~\cite{quick}, 
we use a minimum size threshold $\tau_{size}$ to return only large quasi-cliques.

% exponential time may be needed just to print the output. kplex0
% maximal k-plexes are even more numerous than maximal cliques, kplex1
% the number of pseudo-cliques grows exponentially, at an even faster pace than that of cliques;

\vspace{-2mm}

\begin{definition}[Problem Statement]\label{def}
{\em % to cancel italics
Given a graph $G = (V, E)$, a minimum degree threshold $\gamma\in[0, 1]$ and a minimum size threshold $\tau_{size}$, we aim to find all the vertex sets $S$ such that $G(S)$ is a maximal $\gamma$-quasi-cliques of $G$, and that $|S|\geq\tau_{size}$.
}
\end{definition}

\vspace{-2mm}

For ease of presentation, when $G(S)$ is a valid quasi-clique, we simply say that $S$ is a valid quasi-clique.

\begin{figure}[t]
\centering
\includegraphics[width=0.9\columnwidth]{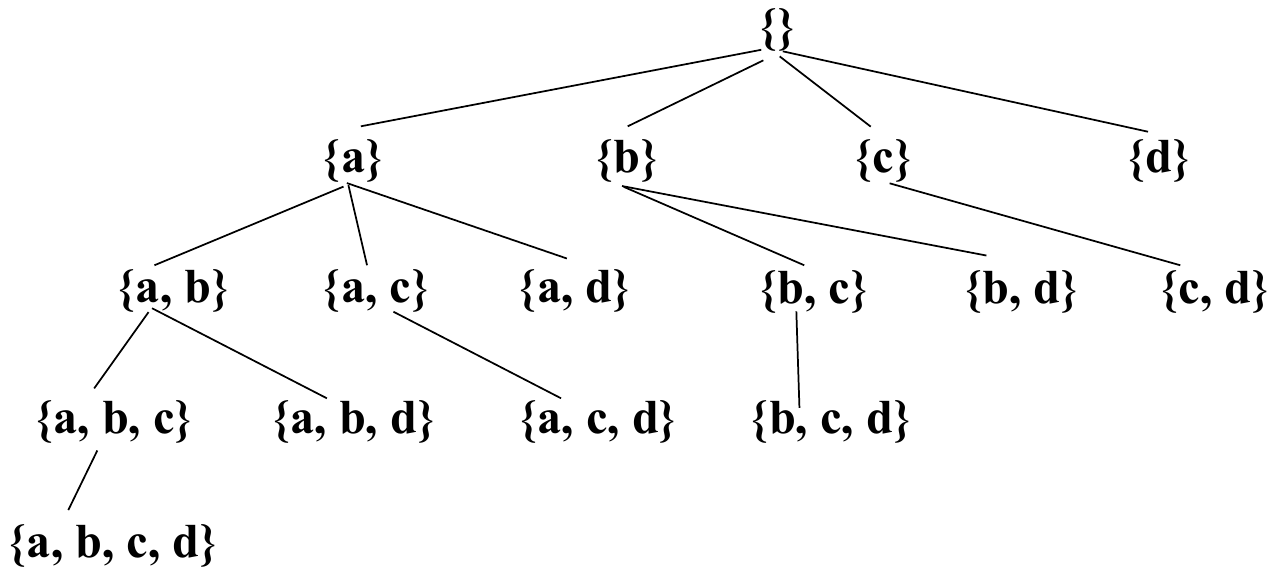}
\vspace{-4mm}
\caption{Set-Enumeration Tree}\label{set_enum}
\vspace{-4mm}
\end{figure}

\vspace{1mm}
\noindent {\bf Framework for Recursive Mining.} In general pseudo-clique mining problems (including ours), the giant search space of a graph $G=(V, E)$, i.e., $V$'s power set, can be organized as a set-enumera-tion tree~\cite{quick}. %=====note the split enumera-tion
Figure~\ref{set_enum} shows the set-enumeration tree $T$ for a graph $G$ with four vertices $\{a, b, c, d\}$ where $a<b<c<d$ (ordered by ID). Each tree node represents a vertex set $S$, and only vertices larger than the largest vertex in $S$ are used to extend $S$. For example, in Figure~\ref{set_enum}, node $\{a,c\}$ can be extended with $d$ but not $b$ as $b<c$; in fact, $\{a,b,c\}$ is obtained by extending $\{a,b\}$ with $c$.

Let us denote $T_S$ as the subtree of the set-enumeration tree $T$ rooted at a node with set $S$. Then, $T_S$ represents a search space for all possible %$\gamma$-quasi-cliques 
pseduo-cliques
that contain all vertices in $S$. In other words, let $Q$ be a %$\gamma$-quasi-clique 
pseduo-clique
found by $T_S$, then $Q\supseteq S$.

We represent the task of mining $T_S$ as a pair $\langle S, ext(S)\rangle$, where $S$ is the set of vertices assumed to be already included, and $ext(S)\subseteq(V-S)$ keeps those vertices that can extend $S$ further into a $\gamma$-quasi-clique. As we shall see, many vertices cannot form a $\gamma$-quasi-clique together with $S$ and can thus be safely pruned from $ext(S)$; therefore, $ext(S)$ is usually much smaller than $(V-S)$.

Note that the mining of $T_S$ can be recursively decomposed into the mining of the subtrees rooted at the children of node $S$ in $T_S$, denoted by $S'\supset S$. Note that since $ext(S')\subset ext(S)$, the subgraph induced by nodes of a child task $\langle S', ext(S')\rangle$ is smaller.

This set-enumeration approach typically requires postprocessing to remove non-maximal 
%quasi-cliques 
pseudo-cliques 
from the set of valid 
pseudo-cliques 
%quasi-cliques 
found~\cite{quick}. 
For example, when processing task that mines $T_{\{b\}}$, vertex~$a$ is not considered and thus the task has no way to determine that $\{b, c, d\}$ is not maximal, even if $\{b, c, d\}$ is invalidated by $\{a, b, c, d\}$ which happens to be a valid 
pseudo-clique, 
%quasi-clique, 
since $\{a, b, c, d\}$  is processed by the task mining $T_{\{a\}}$. But this postprocessing is efficient especially when the number of valid 
pseudo-cliques 
%quasi-cliques 
is not big (as we only find large pseduo-cliques).
% small which is often the case as users give selective parameters (i.e., relatively large $\gamma$ and $\tau_{size}$) to mine significant quasi-cliques~\cite{bigdata18}.

%%%%%%%%%%%%%%%%%%%% NEW SECTION BEGIN %%%%%%%%%%%%%%%%%%%%

\section{Challenges in Load Balancing}\label{sec:challenges}

% - summary, emphasize that subgraphs are already pruned
% - 2 tables on top-10 tasks, showing straggler and wide time span
% - ML features (top-10 deg, core->advanced)
% - expoential + pruning >> wenfei
% - per feature check -> complete: see ArXiv

We explain the straggler problem using two large graphs {\em YouTube} and {\em Patent} shown in Table~\ref{data} of Section~\ref{sec:results}. We show that (1)~the running time of tasks spans a wide range, (2)~even tasks with subgraphs of similar size- and degree-related features can have drastically different running time, and hence (3)~expensive tasks cannot be effectively predicted using regression models in machine learning.

To conduct these experiments, we run quasi-clique mining using G-thinker where each task is spawned from a vertex $v$ and mines the entire set-enumeration subtree $T_{\{v\}}$ (i.e., $S=\{v\}$) in serial without generating any subtasks. As we shall see from pruning rules (P1) and (P2) in Section~\ref{sec:quick}, vertices with low degrees can be pruned using a $k$-core algorithm, and vertices in $ext(S)$ have to be within $f(\gamma)$ hops from $v$. Our reported experiments has applied these pruning rules so that (i)~low-degree vertices are directly pruned without generating tasks, (ii)~the subgraphs have been pruned not to include low-degree vertices and vertices beyond $f(\gamma)$ hops.

Also, we only report the actual time of mining $T_{\{v\}}$ for each task, not including any system-level overheads for task scheduling and vertex data requesting, though the latter cost is never a bottleneck: when some tasks are being scheduled or waiting for vertex data needed, other ready-tasks are being mined so almost all mining threads are busy on the actual mining workloads when there are enough tasks to process (i.e., not near the end of a job)~\cite{gthinker}.

\begin{table}[t]
\centering
\caption{Features of the 10 Most Expensive Tasks on {\em YouTube}}\label{youtube_table}
\includegraphics[width=\columnwidth]{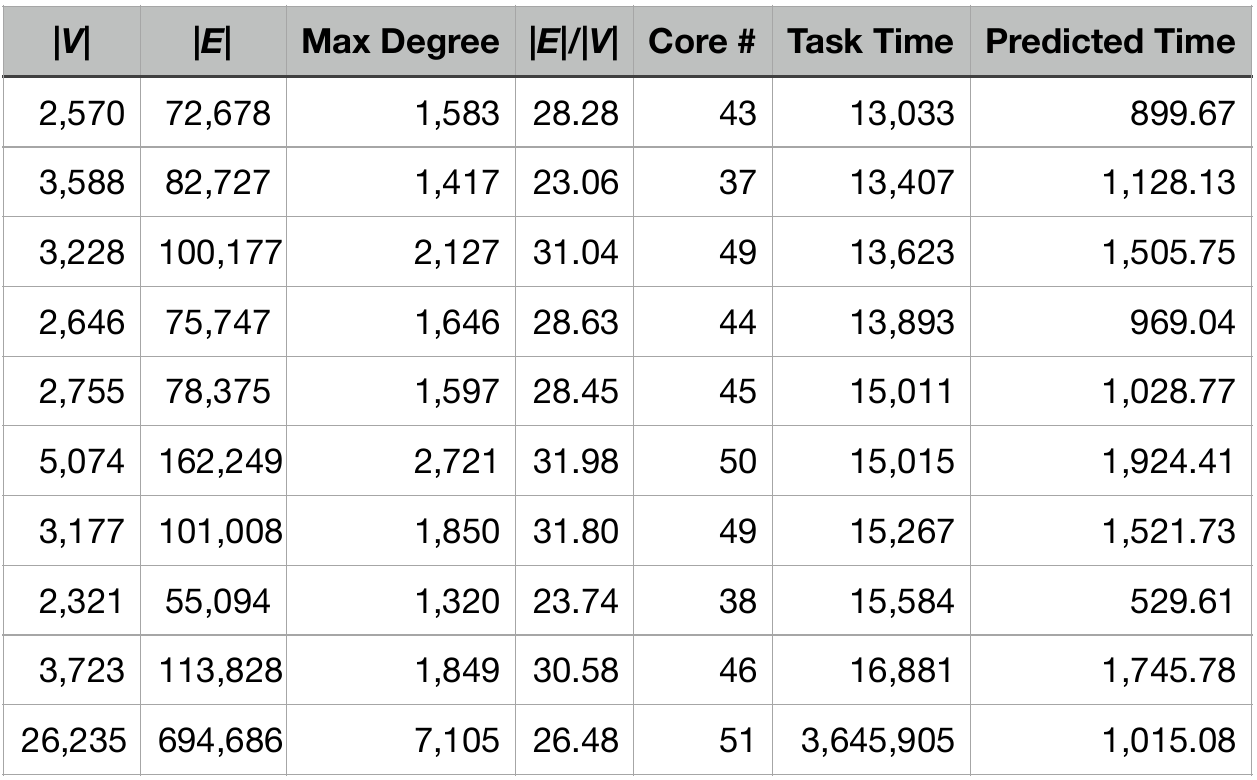}
\vspace{-6mm}
\end{table}

\begin{table}[t]
\centering
\caption{Features of the 10 Most Expensive Tasks on {\em Patent}}\label{patent_table}
\includegraphics[width=\columnwidth]{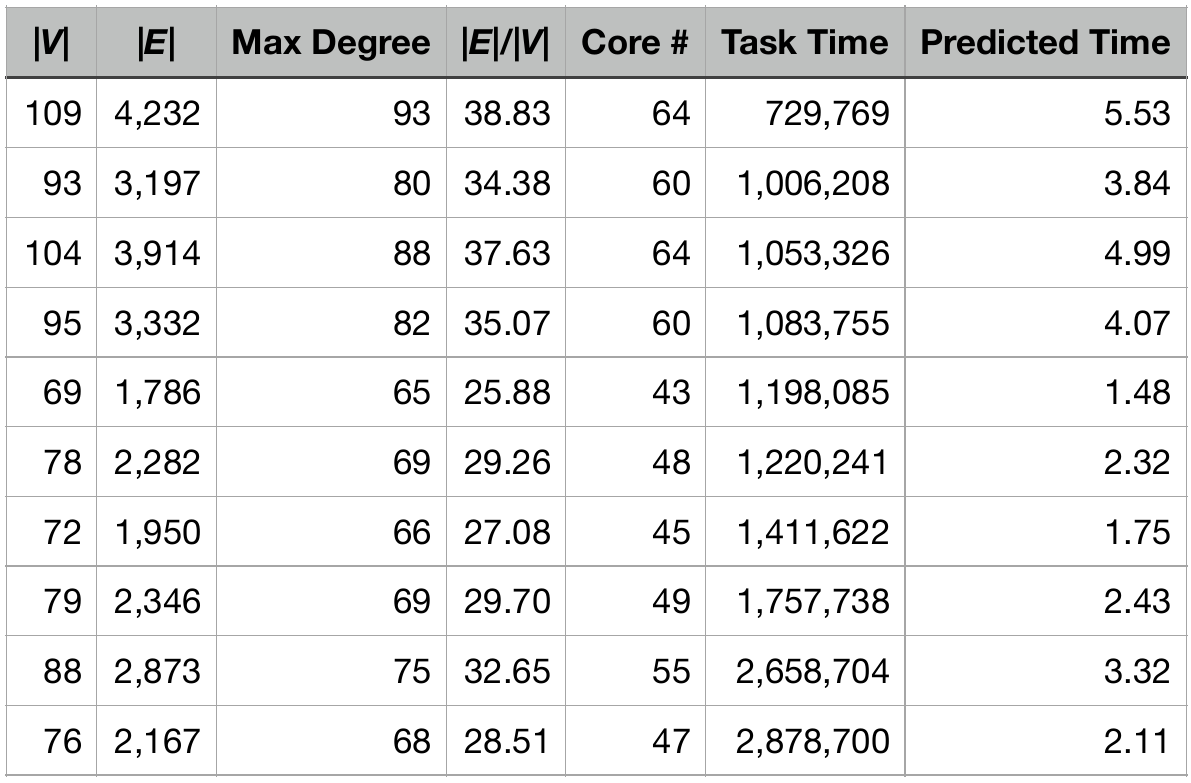}
\vspace{-6mm}
\end{table}

Table~\ref{youtube_table} (resp.\ Table~\ref{patent_table}) shows the task-subgraph features of the top-10 longest-running tasks on {\em YouTube} with $\gamma=0.9$ (resp.\ {\em Patent} with $\gamma=0.89$) including the number of vertices and edges, the maximum and average vertex degree, the $k$-core number (aka.\ degeneracy) of the subgraph, the actual serial mining time on the subgraph, along with the predicted time using support vector regression. The tasks are listed in ascending order of the running time (c.f.\ Column ``Task Time''), and the time unit is millisecond (ms).

In Table~\ref{youtube_table}, the last task takes more than 1 hour (3645.9 seconds) to complete, while the entire G-thinker job only takes 61 minutes and 33.2 seconds, clearly indicating that this task is a straggler. In fact, even if we sum the mining time of all tasks, the total is just 5.5 times that of this straggler task, meaning that the speedup ratio is locked at 5.5$\times$ if we do not further decompose an expensive task.

In Table~\ref{patent_table}, the last 9 tasks all take more than 1000 seconds, so unlike {\em YouTube} with one particularly expensive tasks, {\em Patent} has a few of them, so the computing thread that gets assigned most of those tasks will become a straggler. In fact, the job takes only 55 minutes and 25.4 seconds, but the last task alone takes 2878.7 seconds, clearly a straggler. In fact, on both graphs, there are tasks taking less than 1 ms, so the task time spans 8 orders of magnitude!

Note that in the tables, we already have size- and degree-based features of a task-subgraph, as well as the more advanced feature of the $k$-core number of the subgraph that reflects the graph density. We have extensively tested the various machine learning models for task-time regression using the above input features along with the top-10 highest vertex degrees and top-10 vertex core indices (computed by core decomposition), but none of the models can effectively predict the time-consuming tasks. In both Tables~\ref{youtube_table} and~\ref{patent_table}, the last column shows the predicted time using a support vector regression model trained using all the task statistics, and we can see that the predicted times are way off the ground truth.

We remark that this difficulty is because the set-enumeration search is exponential in nature, and the timing when pruning rules are applicable changes dynamically during the mining depending on the vertex connections, and cannot be effectively predicted other than conducting the actual divisible mining. This is different from the existing work of~\cite{wenfei} that considers low-order polynomial-time graph computation problems that do not use pruning rules and the polynomial coefficients can be easily learned from the job profile.

\begin{figure*}[p]
\centering
\includegraphics[width=\columnwidth]{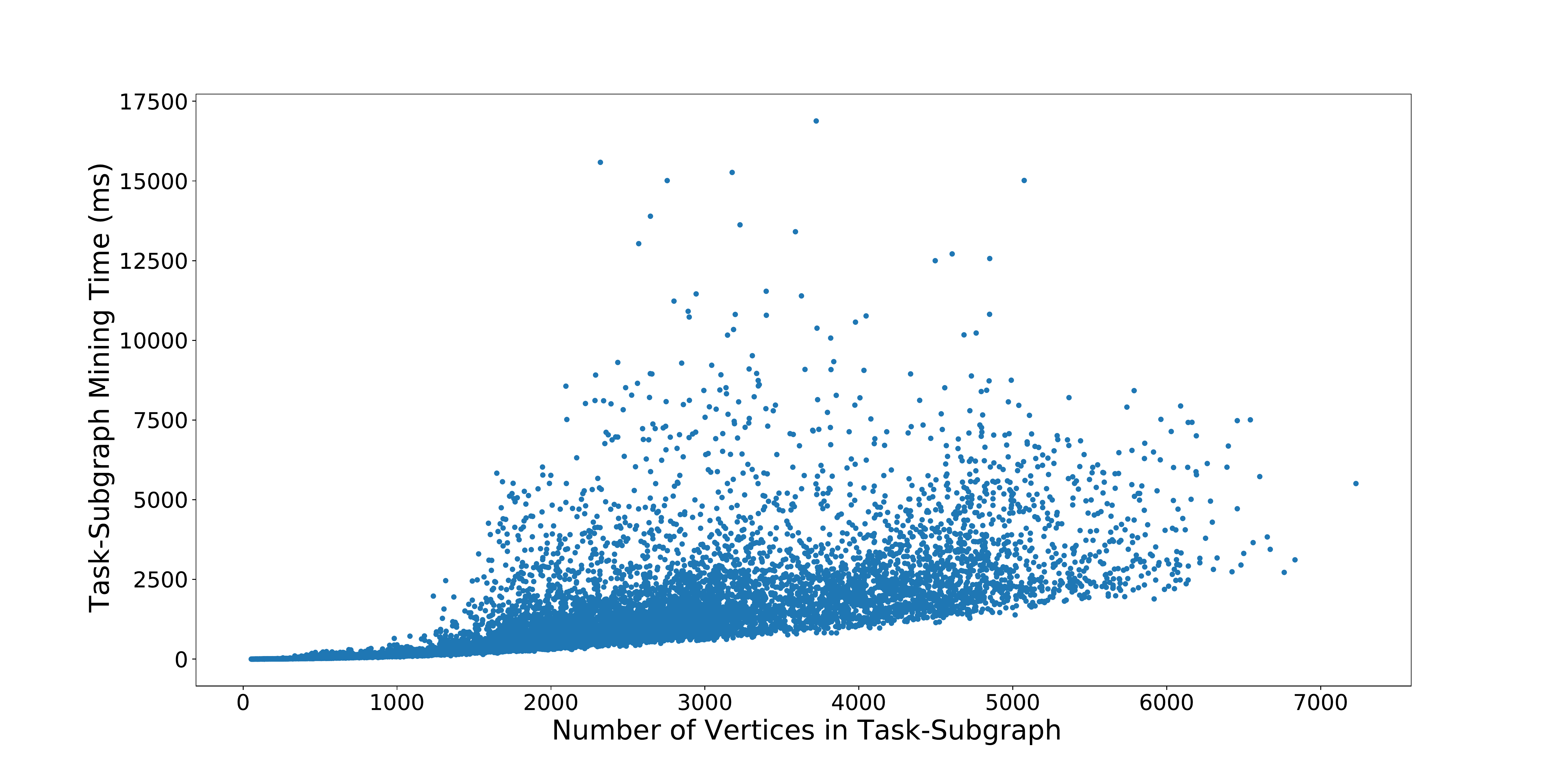}
\includegraphics[width=\columnwidth]{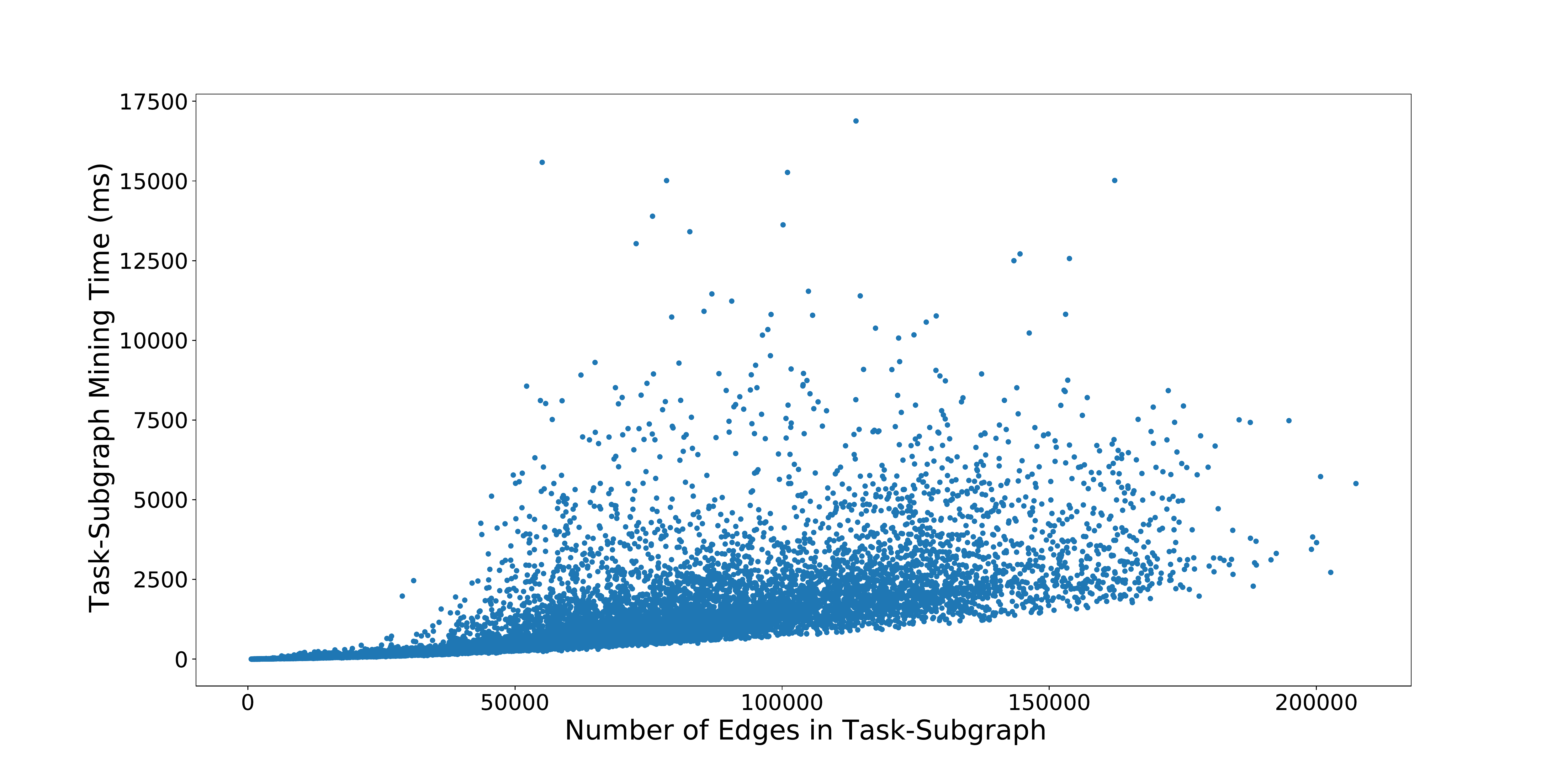}
\includegraphics[width=\columnwidth]{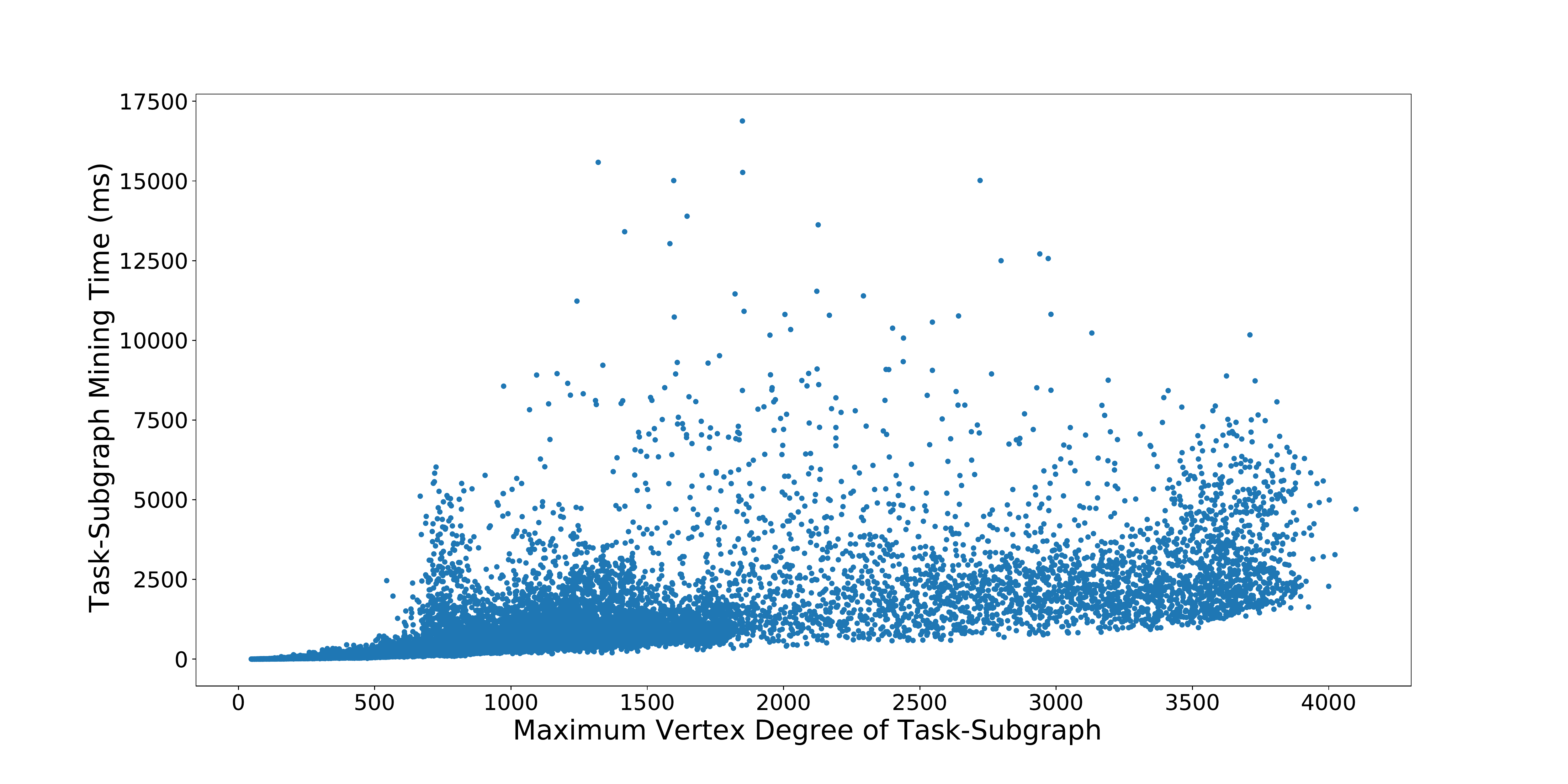}
\includegraphics[width=\columnwidth]{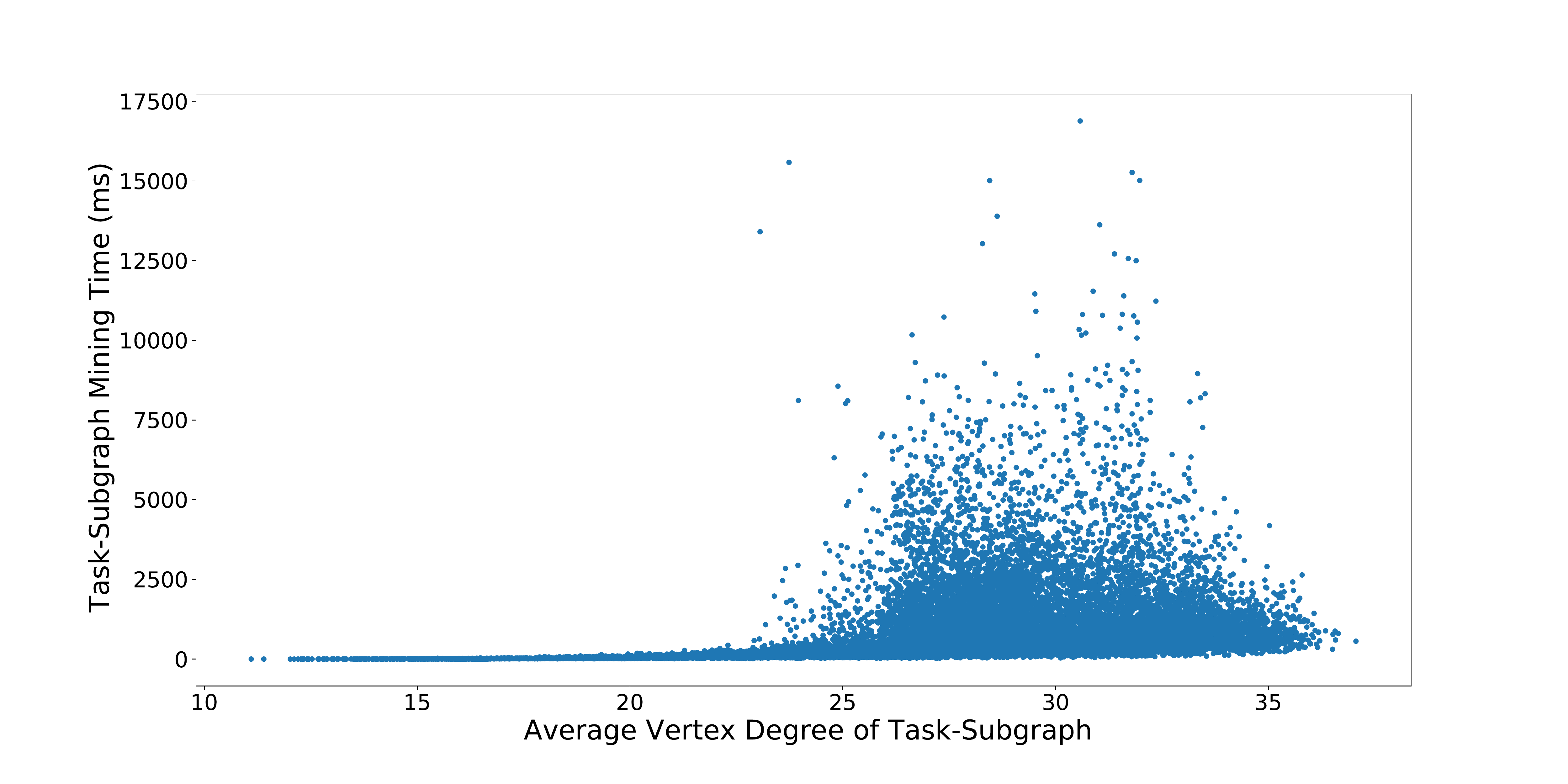}
\includegraphics[width=\columnwidth]{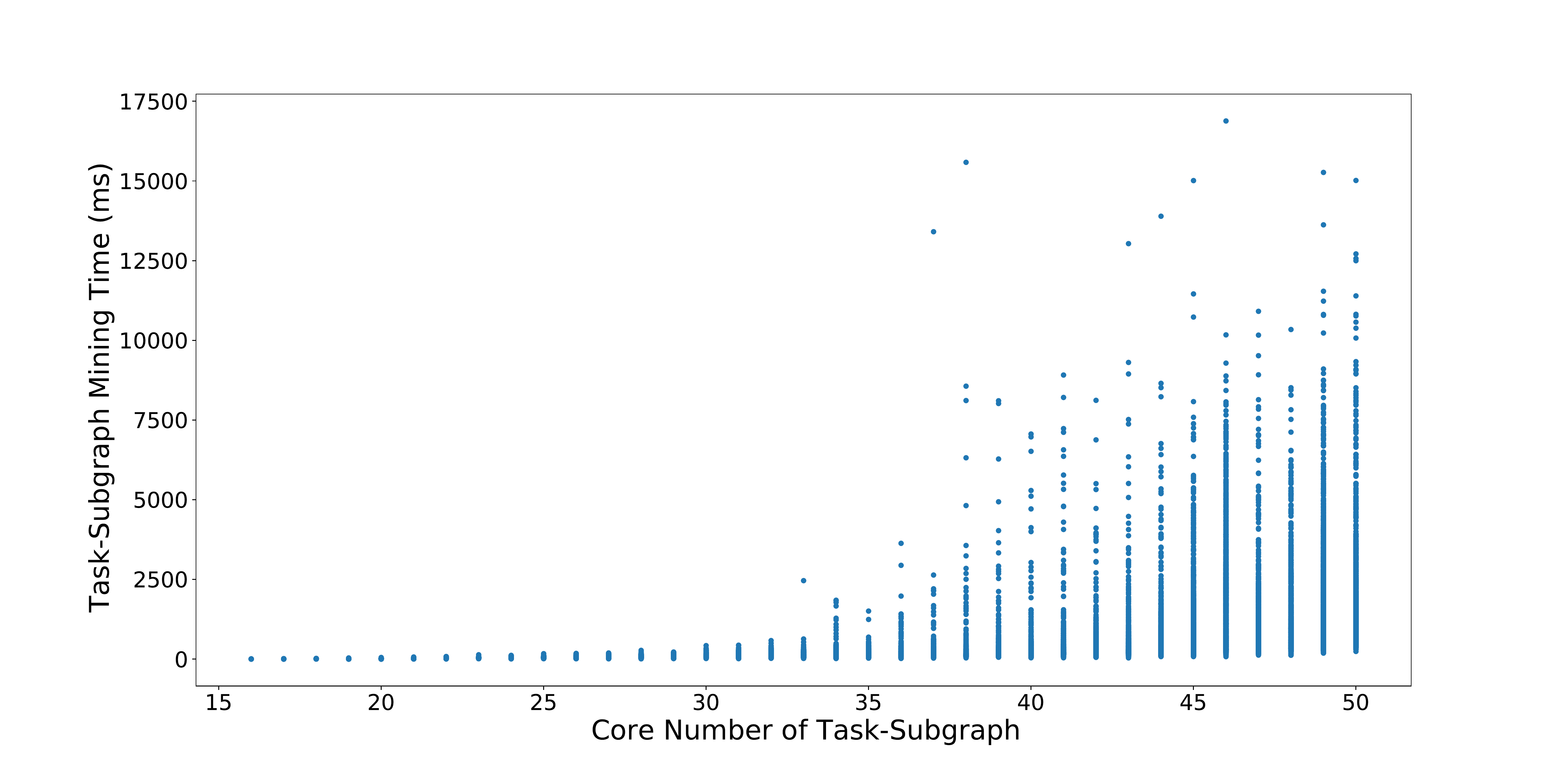}
\caption{Subgraph Features v.s.\ Task Time on {\em YouTube}}\label{V}
\end{figure*}

To visualize how each subgraph feature impacts the task running time, we plot the impacts of $|V|$, $|E|$, maximum degree, average degree, and core \# in the five subplots in Figure~\ref{V} for the {\em YouTube} graph, where we excluded the sole straggler task that takes 3645.9 seconds which would otherwise flatten other points to near 0 on the y-axis. We can see that for about the same feature values, the time can vary a lot along the vertical direction, and this happens unless the subgraph is very small (e.g., less than 1000 vertices or average degree less than 20). No wonder that the expensive tasks cannot be predicted from these features.

\begin{figure*}[p]
\centering
\includegraphics[width=\columnwidth]{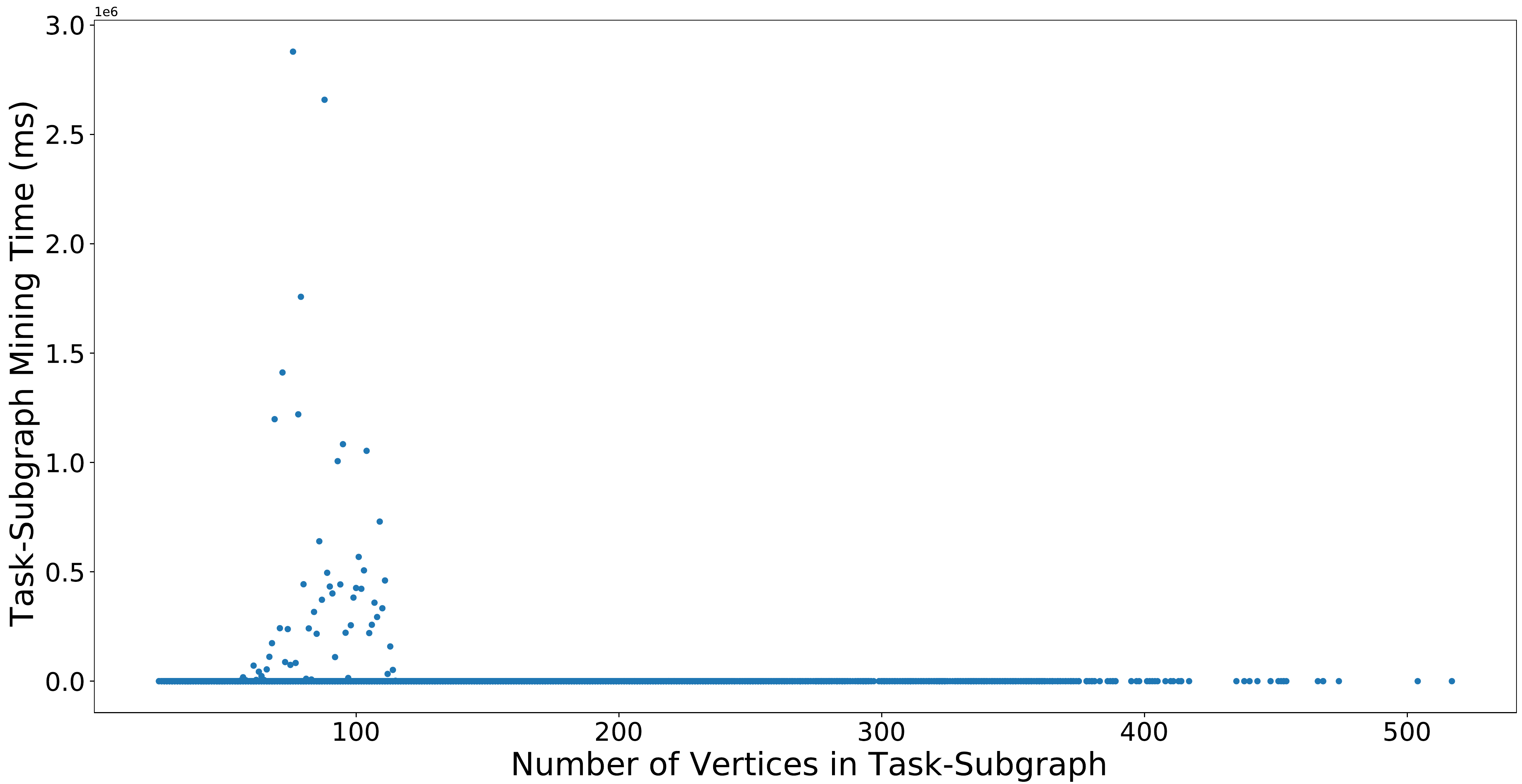}
\includegraphics[width=\columnwidth]{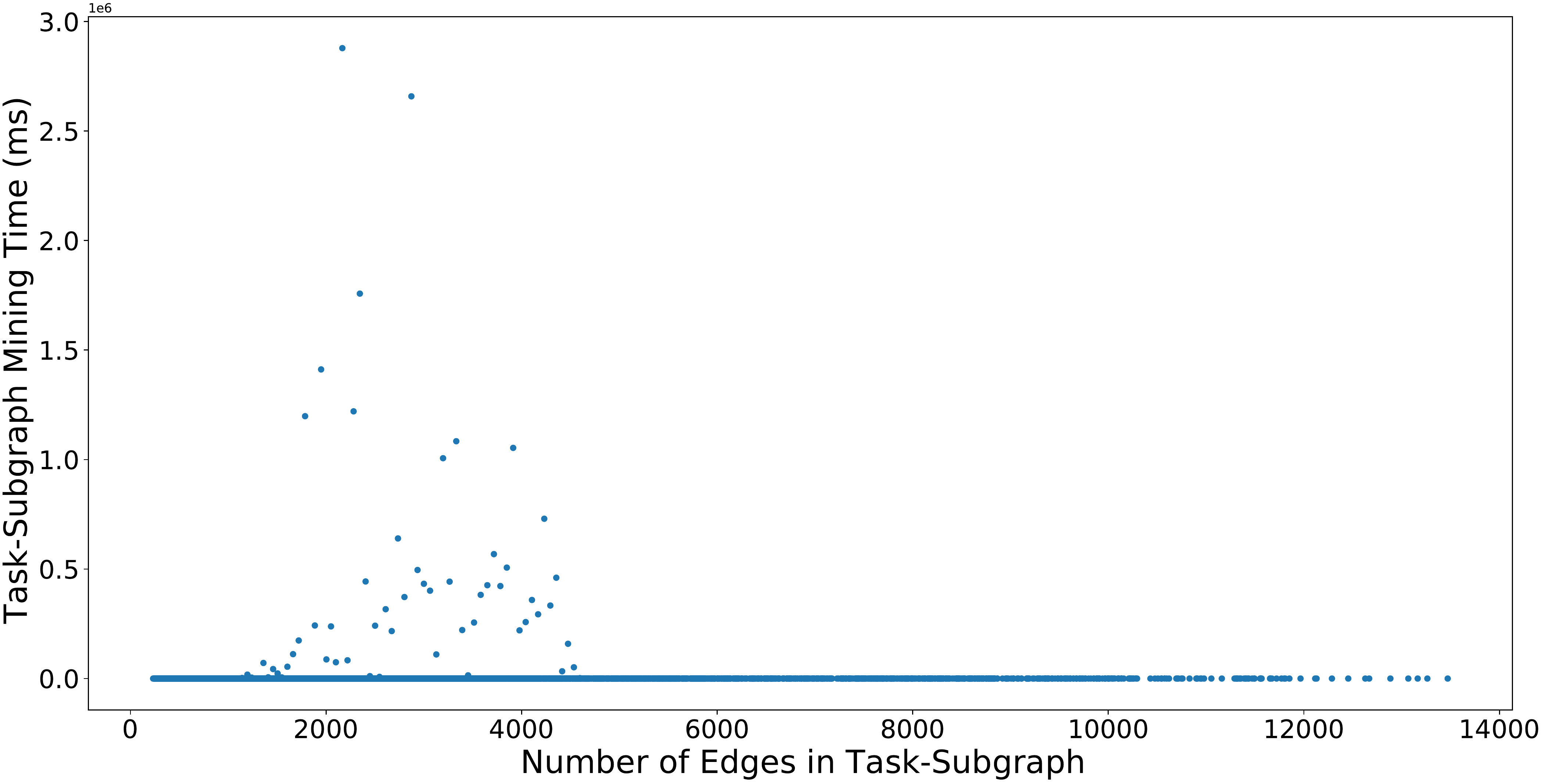}
\includegraphics[width=\columnwidth]{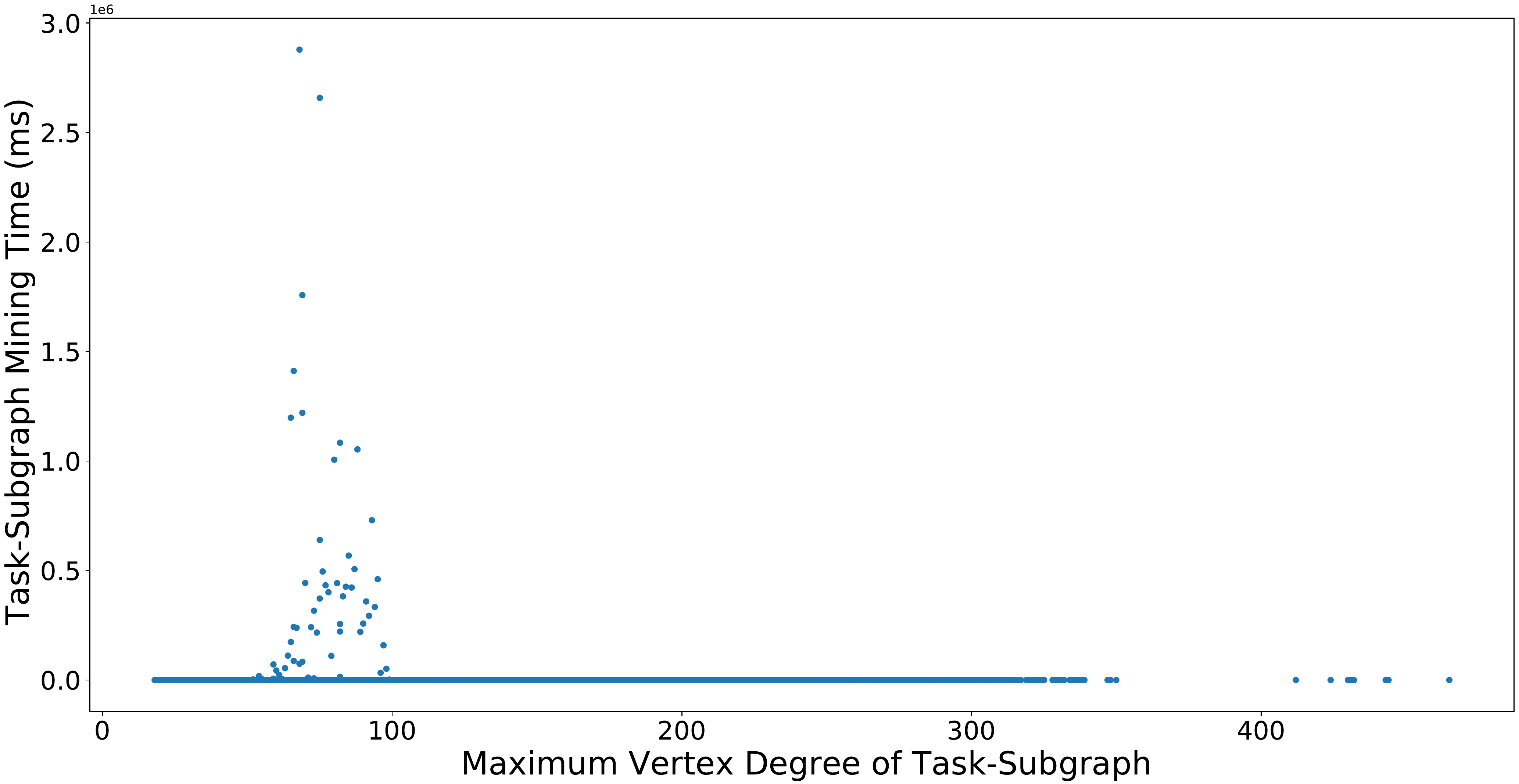}
\includegraphics[width=\columnwidth]{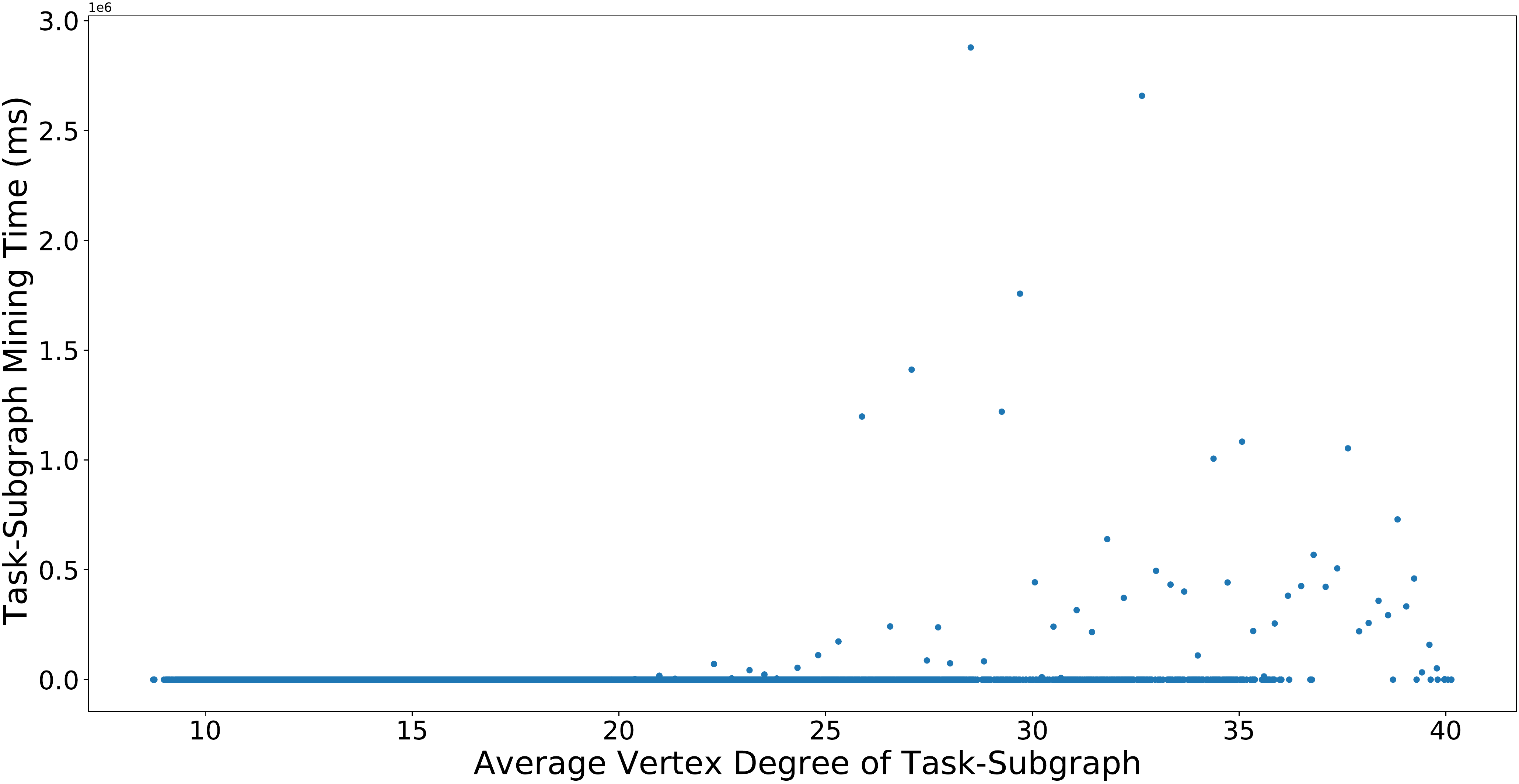}
\includegraphics[width=\columnwidth]{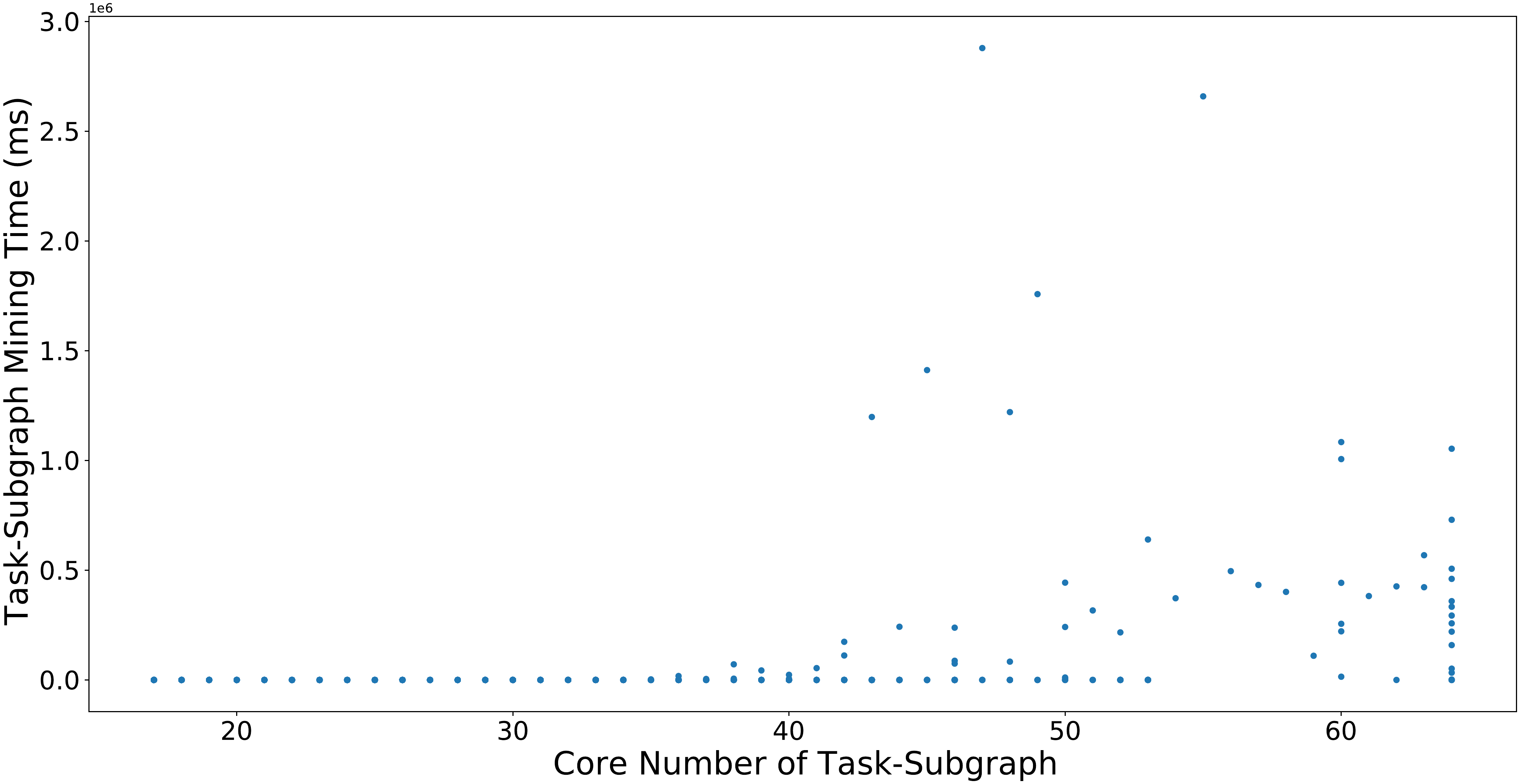}
\caption{Subgraph Features v.s.\ Task Time on {\em Patent}}\label{V2}
\end{figure*}

\begin{figure*}[p]
\centering
\includegraphics[width=\columnwidth]{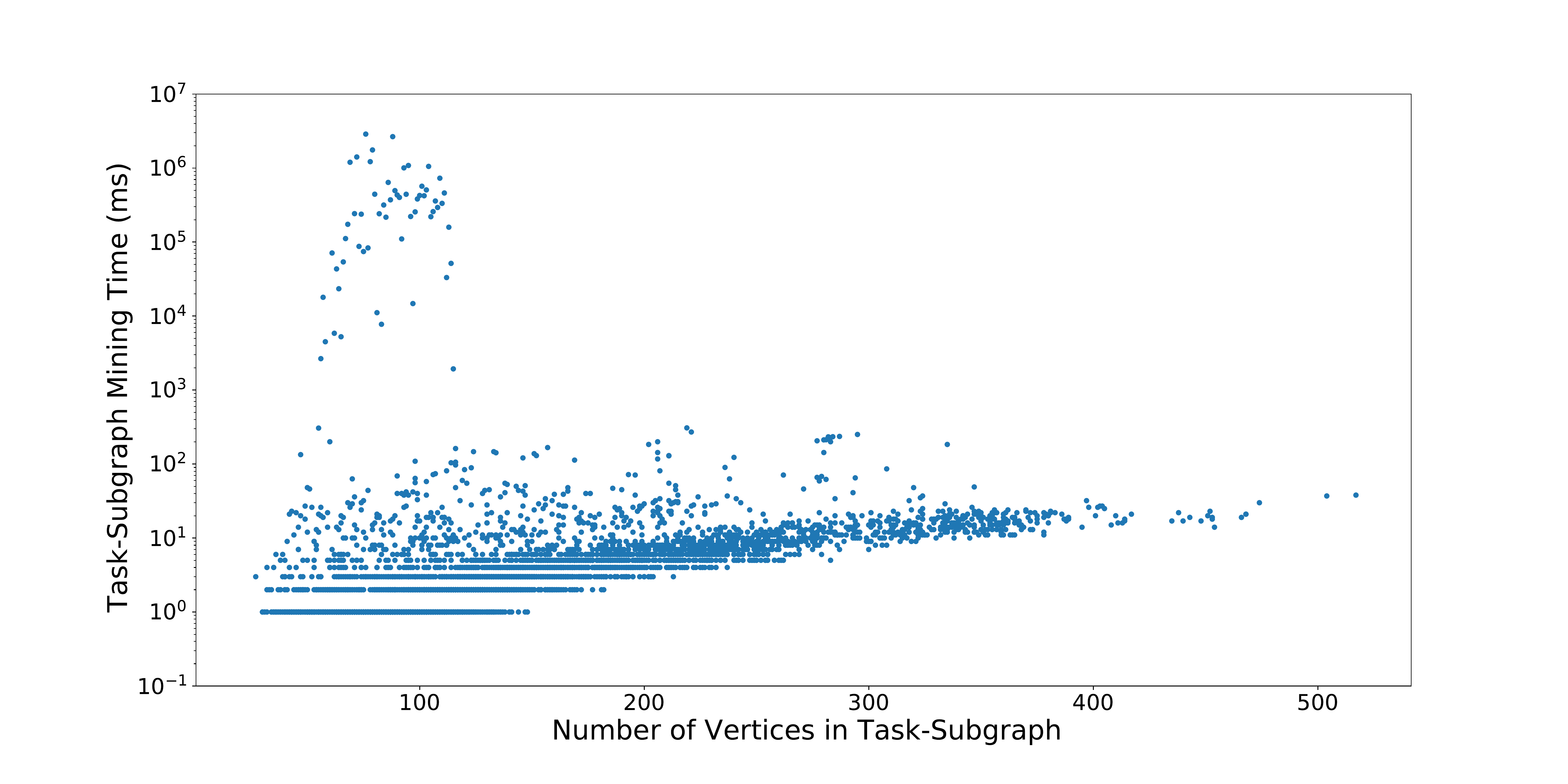}
\includegraphics[width=\columnwidth]{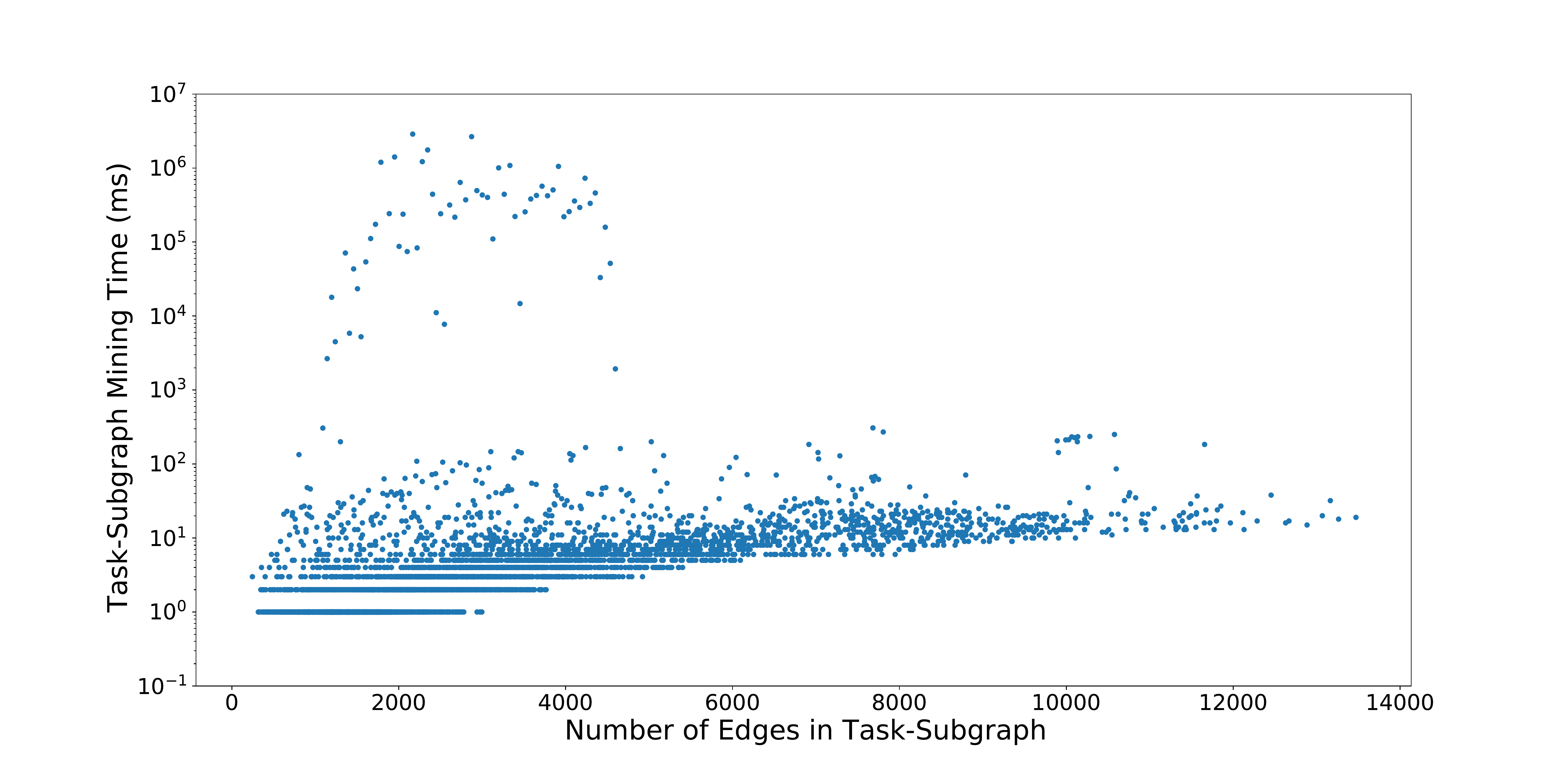}
\includegraphics[width=\columnwidth]{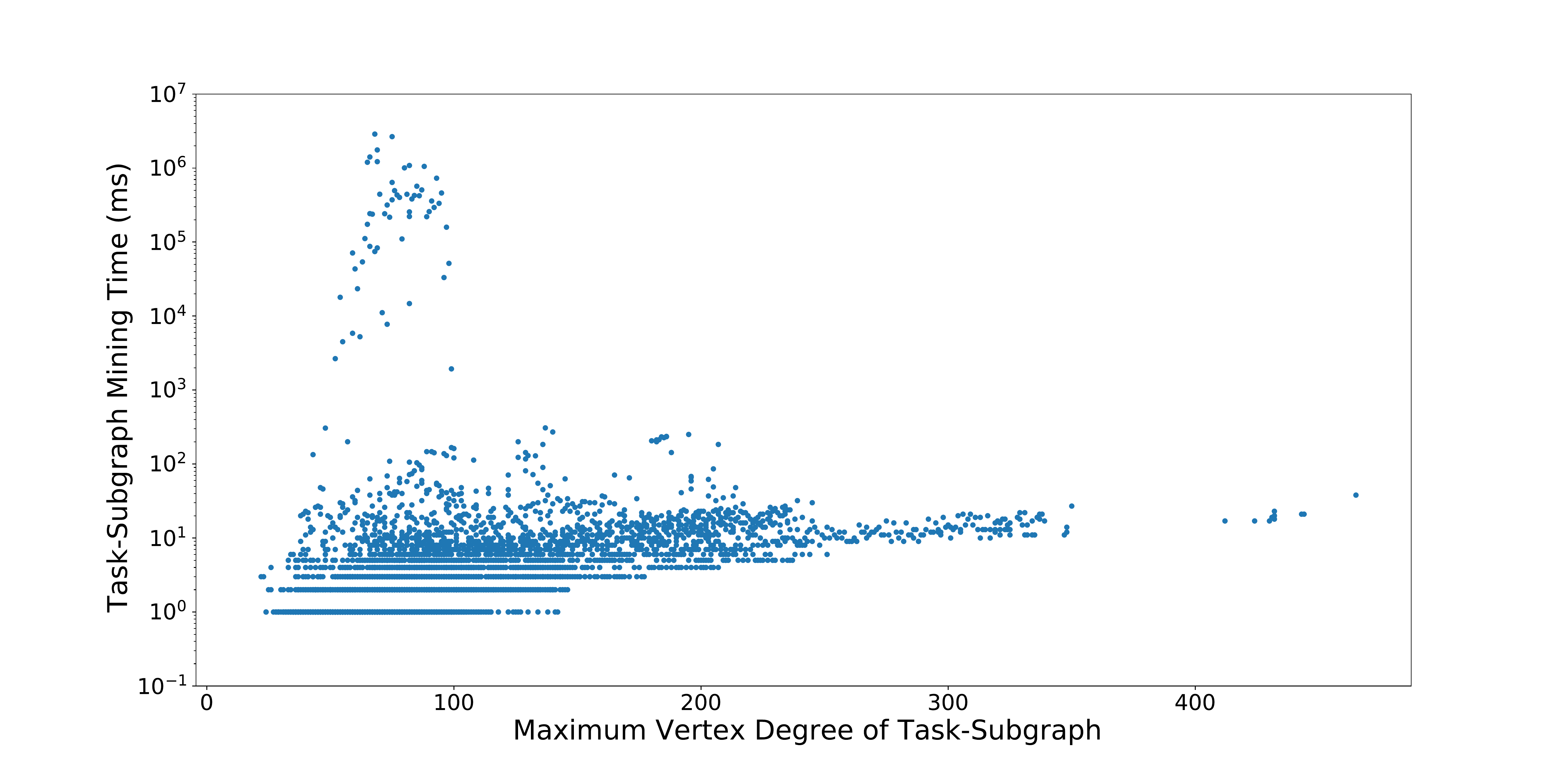}
\includegraphics[width=\columnwidth]{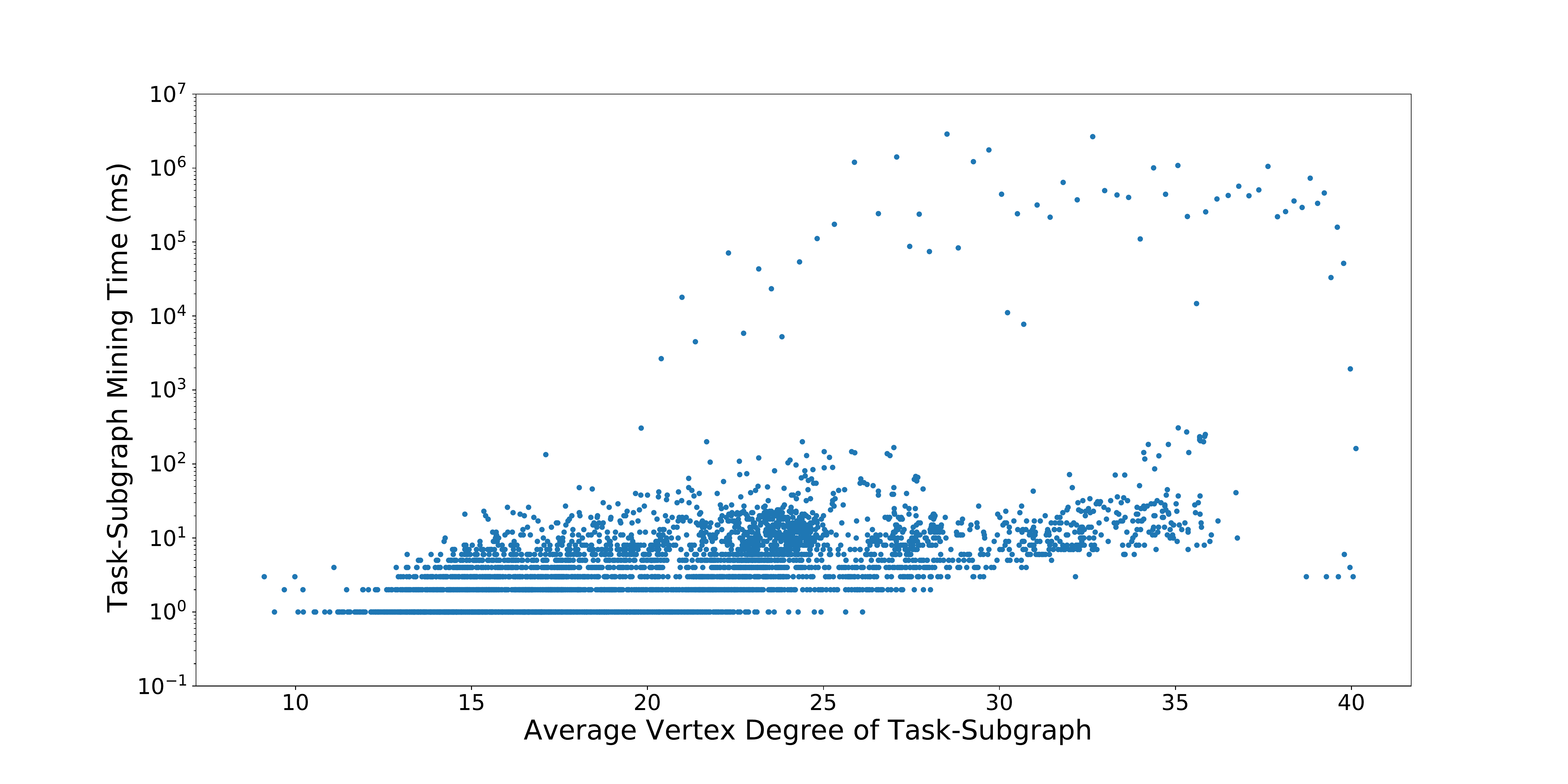}
\includegraphics[width=\columnwidth]{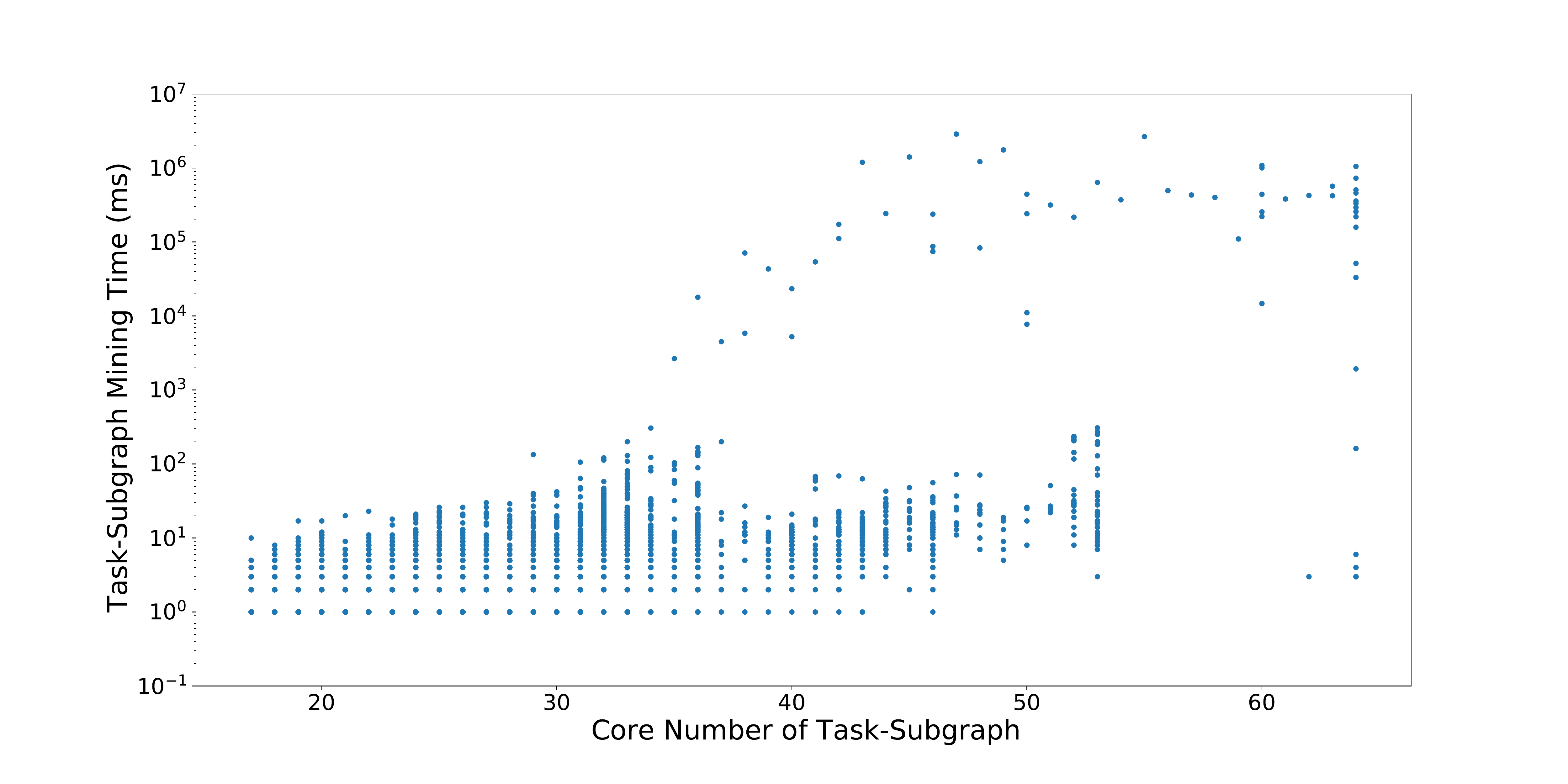}
\caption{Subgraph Features v.s.\ Logarithmic Task Time on {\em Patent}}\label{V3}
\end{figure*}

For {\em Patent}, we plot the impacts of $|V|$, $|E|$, maximum degree, average degree, and core \# in the five subplots in Figure~\ref{V2}. Similar to Figure~\ref{V}, we can see that for about the same feature values, the time can vary a lot along the vertical direction. The difference is that the task time varies even more where some tasks are so much more time-consuming that most other tasks have their time flatten to be close to 0 along the y-axis. To mitigate this issue, we also plot the diagrams by making the time in log scale. The plots are shown in Figure~\ref{V3}, where we can observe that the time still varies a lot along the vertical direction for similar feature values.

All the relevant analyses are shared as jupyter notebook files at \url{https://github.com/yanlab19870714/gthinkerQC_taskTimeDistribution}.

\vspace{1mm}
\noindent {\bf Solution Overview.} We address the above challenges from both the algorithmic and the system perspectives. {\bf Algorithmically}, straggler tasks need to be divided into subtasks with controllable running time even though the actual running time needed by a task is difficult to predict; this will be addressed in Section~\ref{sec:algo}. However, even with effective task decomposition algorithms, the {\bf system} still needs to have a mechanism to schedule straggler tasks early so that its workloads can be partitioned and concurrently processed as early as possible; we address this in Section~\ref{sec:gthinker} below.

%%%%%%%%%%%%%%%%%%%% NEW SECTION END %%%%%%%%%%%%%%%%%%%%

\begin{figure*}[t]
\centering
\includegraphics[width=2.1\columnwidth]{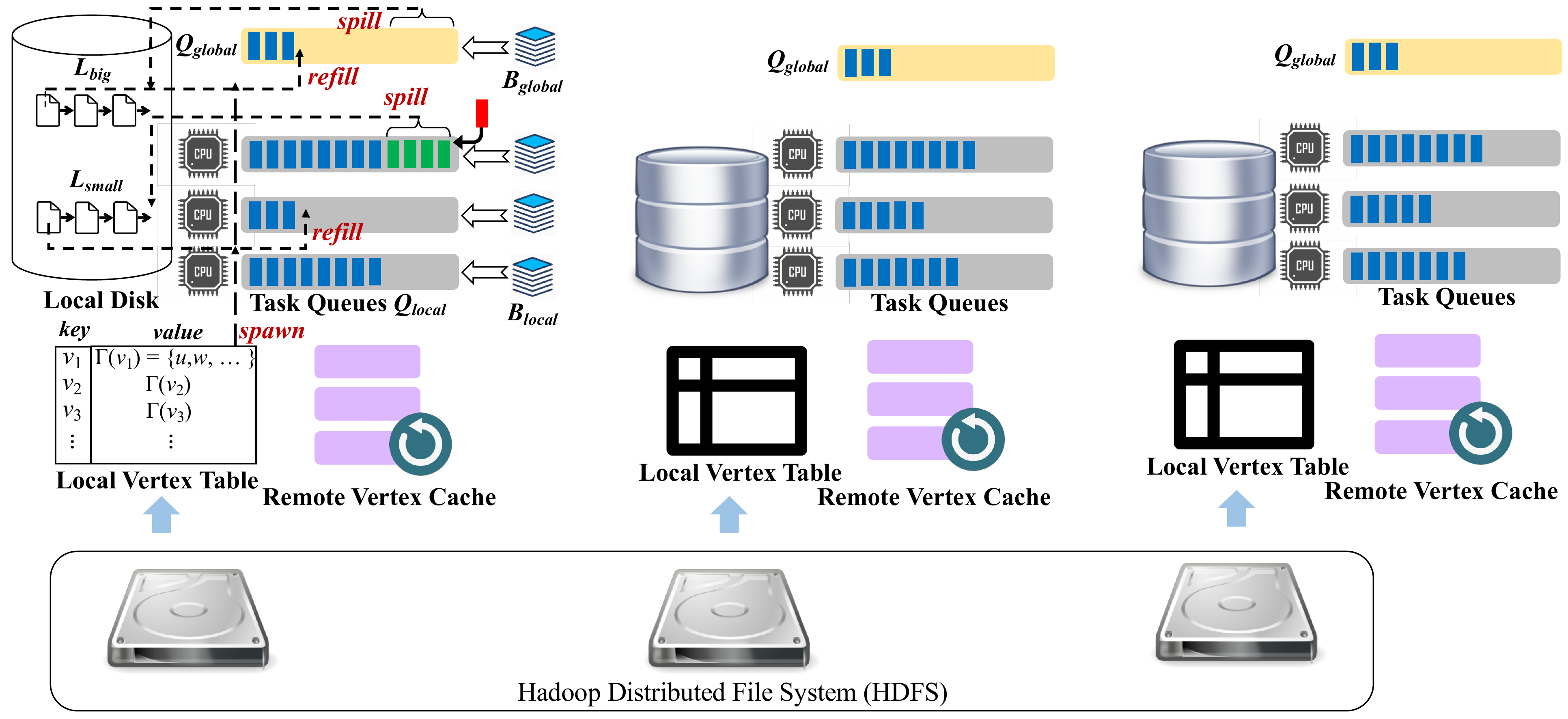}
\caption{G-thinker Architecture Overview}\label{overview}
\end{figure*}

\section{G-thinker and Its Redesign}\label{sec:gthinker}

%!!!! 强调对quasi-clique通用

\noindent {\bf G-thinker API.} The distributed system G-thinker~\cite{gthinker} computes in the unit of tasks. A task $t$ maintains a subgraph $g$ that it constructs and then mines. Each initial task is spawned from an individual vertex $v$ and requests for the adjacency lists of its surrounding vertices (whose IDs are in $v$'s adjacency list). When the one-hop neighbors of $v$ are received, $t$ can continue to grow its subgraph $g$ by requesting the second-hop neighbors. When $g$ is fully constructed, $t$ can then mine it or decompose it to generate smaller tasks.

To avoid double-counting, a vertex $v$ only requests those vertices with ID $>v$. In Figure~\ref{set_enum}, each level-1 singleton node $\{v\}$ corresponds to a G-thinker task spawned from $v$, and it only examines those vertices with ID $> v$, so that a quasi-clique whose smallest vertex is $v$ is  found exactly in the set-enumeration subtree $T_{\{v\}}$ (recall Figure~\ref{set_enum}) by the task spawned from $v$.

To write a G-thinker algorithm, a user only implements 2 user-defined functions (UDFs): (1)~{\em spawn}($v$) indicating how to spawn a task from each individual vertex of the input graph; (2)~{\em compute}($t$, {\em frontier}) indicating how a task $t$ processes an iteration where {\em frontier} keeps the adjacency lists of the requested vertices in the previous iteration. In a UDF, users may request for the adjacency list of a vertex $u$ to expand the subgraph $g$ of a task $t$, or even to decompose $g$ by creating multiple new tasks with smaller subgraphs, which corresponds to branching a node into its children in Figure~\ref{set_enum}.

UDF {\em compute}($t$, {\em frontier}) is called in iterations for growing task $t$'s subgraph in a breadth-first manner. If some requested vertices~are not locally available, $t$ will be suspended so that its mining thread can continue to process other tasks; $t$ will be scheduled to call {\em compute}(.) again once all its requested data become locally available.

UDF {\em compute}($t$, {\em frontier}) returns {\em true} if the task $t$ needs to call {\em compute}(.) for more iterations for further processing; it returns {\em false} if $t$ is finished so that G-thinker will delete $t$ to release space.

In this paper, we maintain G-thinker's programming interface as described above while redesigning its parallel execution engine so that big tasks can be scheduled early to partition its computations.

\vspace{1mm}
\noindent{\bf The Original System Architecture.} Figure~\ref{overview} shows the architecture (components) of G-thinker on a cluster of machines (yellow global task queues are the new additions by our redesign).

We assume that a graph is stored as a set of vertices, where each vertex $v$ is stored with its adjacency list $N(v)$ that keeps its neighbors. G-thinker loads an input graph from HDFS. As Figure~\ref{overview} shows, each machine only loads a fraction of vertices along with their adjacency lists into its memory, kept in a local vertex table. Vertices are assigned to machines by hashing their vertex IDs, and the aggregate memory of all machines is used to keep a big graph. The local vertex tables of all machines constitute a distributed key-value store where any task can request for $N(v)$ using $v$'s ID.

G-thinker spawns initial tasks from each individual vertex $v$ in the local vertex table.
As Figure~\ref{overview} shows, each machine also maintains a remote vertex cache to keep the requested vertices (and their adjacency lists) that are not in the local vertex table, for access by tasks via the input argument {\em frontier} to UDF {\em compute}($t$, {\em frontier}). This allows multiple tasks to share requested vertices to minimize redundancy. In {\em compute}($t$, {\em frontier}), task $t$ is supposed to save the needed vertices and edges in {\em frontier} into its subgraph, as G-thinker releases $t$'s hold of those vertices in {\em frontier} right after {\em compute}($t$, {\em frontier}) returns, and they may be evicted from the vertex cache.

Our distributed vertex store and cache are designed to allow a graph to be processed even if it cannot fit entirely in the memory of a machine. But if the machine memory is large enough, a pulled vertex will never be evicted, so every vertex will be pulled for at most once, the cost of which is no larger than if we pre-load the entire graph into the memory of every machine.

If {\em compute}($t$, {\em frontier}) returns {\em true}, $t$ is added to a task queue to be scheduled to call {\em compute}(.) for more iterations; while if it returns {\em false}, $t$ is finished and thus deleted to release space.

In the original G-thinker, each mining thread keeps a task queue $Q_{local}$ of its own to stay busy and to avoid contention. Since tasks are associated with subgraphs that may overlap, it is infeasible to keep all tasks in memory. G-thinker only keeps a pool of active tasks in memory at any time by controlling the pace of task spawning. If a task is waiting for its requested vertices, it is suspended so that the mining thread can continue to process the next task in its queue; the suspended task will be added to a task buffer $B_{local}$ by the data serving module once all its requested vertices become locally available, to be fetched by the mining thread for calling {\em compute}(.), and adding it to $Q_{local}$ if {\em compute}(.) returns {\em true}.

Note that a task queue can become full if a task generates many subtasks into its queue, or if many tasks that are waiting for data become ready all at once. To keep the number of in-memory tasks bounded, if a task queue is full but a new task is to be inserted, we spill a batch of $C$ tasks at the end of the queue as a file to local disk to make room. As the upper-left corner of Figure~\ref{overview} shows, each machine maintains a list $\mathcal{L}_{small}$ of task files spilled from the task queues of mining threads. To minimize the task volume on disks, when a thread finds that its task queue is about to become empty, it will first refill tasks into the queue from a task file (if it exists), before choosing to spawn more tasks from vertices in local vertex table. Note that tasks are spilled to disks and loaded back in batches to be IO-efficient. For load balancing, machines about to become idle will steal tasks from busy ones by prefetching a batch of tasks and adding them as a file to $\mathcal{L}_{small}$. These tasks will be loaded by a mining thread for processing when its task queue needs a refill.

Note that while we materialize subgraphs for tasks, the above design ensures that only a pool of tasks are in memory and spilled tasks are temporarily kept on local disks. This is important to keep memory usage bounded as the number of tasks can grow exponentially with graph size. Moreover, since the computations over subgraphs are the bottleneck rather than IO for subgraph creation/moving, G-thinker is designed to be distributed mainly to use the CPU cores on all machines in a cluster, rather than to use the aggregate IO bandwidth as in a conventional data-intensive system. The IO and locking operations are well overlapped with and thus hidden by task computations~\cite{gthinker}.

\vspace{1mm}
\noindent{\bf System Redesign.} Recall that a task in pseduo-clique mining can be very time consuming. If we only let each mining thread to buffer pending tasks in its own local queue, big tasks in the queue cannot be moved around to idle threads in time until they reach the queue head, and they can be stuck by other time-consuming big tasks located earlier in the queue, causing the straggler problem. We now describe how we redesign the execution engine to allow big tasks to be scheduled as soon as possible, always before small tasks.

We maintain separate task containers for big tasks and small ones, and always prioritize the containers for big tasks for processing.
Note that for the new engine to function, we also need our new task decomposition algorithms in Section~\ref{sec:algo} to ensure that a big task will not be computed for a long time before being decomposed, so that later big tasks can be timely scheduled for processing.

Specifically, we use the local task queues of the respective mining threads and the associated task containers (i.e., file list $\mathcal{L}_{small}$ and ready-task buffer $B_{local}$) to keep small tasks only. We similarly maintain a global task queue $Q_{global}$ to keep big tasks shared by all computing threads, along with its associated task containers as shown in Figure~\ref{overview}, including file list $\mathcal{L}_{big}$ to buffer big tasks spilled from $Q_{global}$, and task buffer $B_{global}$ to hold those big tasks that have their requested data ready for computation.

We define a user-specified threshold $\tau_{split}$ so that if a task $t=\langle S, ext(S)\rangle$ has a subgraph with potentially more than $\tau_{split}$ vertices to check, it is appended to $Q_{global}$; otherwise, it is appended to $Q_{local}$ of the current thread. Here, it is difficult to decide the subgraph size of $t$ as it is changing. So when $t$ is still requesting vertices to construct its subgraph, we consider $t$ as a big task iff the number of vertices to pull in the current iteration of {\em compute}(.) is at least $\tau_{split}$, which prioritizes its execution to construct the potentially big subgraph early; while when $t$ is mining its constructed subgraph, we consider $t$ as a big task iff $|ext(S)|>\tau_{split}$, since there are $|ext(S)|$ vertices to check to expand $S$.

\begin{algorithm}[t]
\caption{Old Execution Procedure of a Computing Thread}\label{algo:comper}
\begin{algorithmic}[1]
\WHILE{job end tag is not set by the main thread}
\IF{memory capacity permits}
\STATE {\bf if} $Q_{local}$ does not have enough tasks {\bf then} refill $Q_{local}$
\STATE pop a task $t$ from $Q_{local}$ and provide requested vertices
\STATE if all vertices are ready, repeat {\em compute}($t$, {\em frontier})
\STATE if $t$ is not finished, suspend $t$ to wait for data
\ENDIF
\STATE obtain a task $t'$ from $B_{local}$
\STATE repeat {\em compute}($t'$, {\em frontier}) till some vertex is not available
\STATE if $t'$ is not finished, append $t'$ to $Q_{local}$
\ENDWHILE
\end{algorithmic}
\end{algorithm}
\setlength{\textfloatsep}{5pt}

In the original G-thinker, each thread loops two operations:
\begin{itemize}
\item Algorithm~\ref{algo:comper} Lines 4-6 ``pop'': to fetch a task $t$ from $Q_{local}$ and to feed its requested vertices; if any remote vertex is not in the vertex cache, $t$ will be suspended to wait for data;
\item Algorithm~\ref{algo:comper} Lines 7-9 ``push'': to fetch a task from the thread's local ready-buffer $B_{local}$ for computation, which is then appended to $Q_{local}$ if further processing is needed.
\end{itemize}

``Pop'' is only done if there is enough space left in the vertex cache and task containers, otherwise only ``push'' is conducted to process partially computed tasks so that their requested vertices can be released to make room, which is necessary to keep tasks flowing.

Task refill is conducted right before ``pop'' if the number of tasks in $Q_{local}<$ task batch size $C$, with the priority order of getting a task batch from $\mathcal{L}_{small}$, then from $B_{local}$, and then spawning from vertices in the local vertex table that have not spawned tasks yet. This order is to digest old/spilled tasks before spawning new tasks.

In our redesigned G-thinker engine, we prioritize big tasks for execution and the procedure in Algorithm~\ref{algo:comper} has three major changes.

The first change is with ``push'': a mining thread keeps flowing those tasks that have their requested data ready to compute, by (i)~first fetching a big task from $B_{global}$ for computing. The task may need to be appended back to $Q_{global}$, or may be decomposed into smaller tasks to be appended either to $Q_{global}$ or the thread's $Q_{local}$. (ii)~If $B_{global}$ is, however, found to be empty, the thread will instead fetch a small task from its $B_{local}$ for computing.

The second change is with ``pop'': a computing thread always fetches a task from $Q_{global}$ first. If (I)~$Q_{global}$ is locked by another thread (i.e., a try-lock failure), or if (II)~$Q_{global}$ is found to be empty, the thread will then pop a task from its local queue $Q_{local}$.

In Case~(I) if $Q_{global}$ is successfully locked, if its number of tasks is below a batch size $C$, the thread will try to refill a batch of tasks from $\mathcal{L}_{big}$. We do not check $B_{global}$ for refill since it is shared by all mining threads which will incur frequent locking overheads. Note that ``push'' already keeps flowing big tasks with data ready.

In Case~(II) when there is no big task to pop, a mining thread will check its $Q_{local}$ to pop, before which if the number of tasks therein is below a batch, task refill happens where lies our third change.

Specifically, the thread will refill tasks from $\mathcal{L}_{small}$, and then from its $B_{local}$ in this prioritized order to minimize the number of partially processed tasks buffered on local disk tracked by $\mathcal{L}_{small}$.

If both $\mathcal{L}_{small}$ and $B_{local}$ are still empty, the computing thread will then spawn a batch of new tasks from vertices in the local vertex table for refill.  However, we stop as soon as a spawned task is big, which is then added to $Q_{global}$ (previous tasks are added to $Q_{local}$). This avoids generating many big tasks out of one refill.

Finally, since the main performance bottleneck is caused by big tasks, task stealing is conducted only on big tasks to balance them among machines. The number of pending big tasks (in $Q_{global}$ plus $\mathcal{L}_{big}$) in each machine is periodically collected by a master (every 1 second), which computes the average and generates stealing plans to make the number of big tasks on every machine close to this average. If a machine needs to take (resp.\ give) less than a batch of $C$ tasks, these tasks are taken from (resp.\ appended to) the global task queue $Q_{global}$; otherwise, we allow at most one task file (containing $C$ tasks) to be transmitted to avoid frequent task thrashing that overloads the network bandwidth. Note that in one load balancing cycle (i.e., 1 second) at most $C$ tasks are moved at each machine.

\section{Proposed Recursive Algorithm}\label{sec:quick}
It has been pointed out that ``the key to an efficient set-enumeration search is the pruning strategies that are applied to remove entire branches from consideration''~\cite{Bayardo98}. Without pruning, the search space is exponential and thus intractable. Different pseudo-clique mining algorithms propose different sophisticated pruning rules, and in the context of quasi-clique, Quick~\cite{quick} uses the most complete set of pruning rules. To further improve the efficiency, this section presents our Quick+ algorithm that integrates Quick with new pruning rules. We also fix some missed boundary cases that could lead to missed results in the original Quick algorithm.

\subsection{Pruning Rules}\label{ssec:prune}

Recall the set-enumeration tree in Figure~\ref{set_enum}, where each node represents a mining task, denoted by $t_S=\langle S, ext(S)\rangle$. Task $t_S$ mines the set-enumeration subtree $T_S$: it assumes that vertices in $S$ are already included in a result quasi-clique to find, and continues to expand $G(S)$ with vertices of $ext(S)\subseteq(V-S)$ into a valid quasi-clique. Task $t_S$ that mines $T_S$ can be recursively decomposed into the mining of the subtrees $\{T_{S'}\}$ where $S'\supset S$ are child nodes of node $S$. Our recursive serial algorithm basically examines the set-enumeration search tree in depth-first order, while the parallel algorithm in the next section will utilize the concurrency among child nodes $\{S'\}$ of node $S$ in the set-enumeration tree.

To reduce search space, we consider two categories of pruning rules that can effectively prune either candidate vertices in $ext(S)$ from expansion, or simply the entire subtree $T_S$. Formally, we have

\begin{itemize}
\item {\bf Type~I: Pruning $\mathbf{ext(S)}$.} In such a rule, if a vertex $u\in ext(S)$ satisfies certain conditions, $u$ can be pruned from $ext(S)$ since there must not exist a vertex set $S'$ such that $(S\cup u)\subseteq S'\subseteq(S\cup ext(S))$ and $G(S')$ is a $\gamma$-quasi-clique. 
\item {\bf Type~II: Pruning $\mathbf{S}$.} In such a rule, if a vertex $v\in S$ satisfies certain conditions, there must not exist a vertex set $S'$ such that $S\subseteq S'\subseteq(S\cup ext(S))$ and $G(S')$ is a $\gamma$-quasi-clique, and thus there is no need to extend $S$ further.
\end{itemize}

Type-II pruning invalidates the entire $T_S$. A variant invalidates $G(S')$, $S\subset S'\subseteq(S\cup ext(S))$ from being a valid quasi-clique, but node $S$ is not pruned (i.e., $G(S)$ may be a valid quasi-clique).

We identify 7 groups of pruning rules that are utilized by our algorithm, where each rule either belongs to Type~I, or Type~II, or sometimes both. Below we summarize these groups as (P1)--(P7), respectively.

\vspace{2mm}

\noindent{\bf (P1) Graph-Diameter Based Pruning.} Theorem~1 of~\cite{Pei05} defines the upper bound of the diameter of a $\gamma$-quasi-clique as a function $f(\gamma)$. Often, we only consider the case where $\gamma\geq0.5$, in which case the diameter is bounded by 2. To see this, consider any two vertices $u,v\in V$ in a quasi-clique $G$ that are not direct neighbors: since both $u$ and $v$ can be adjacent to at least $\lceil0.5\cdot(|V|-1)\rceil$ other vertices, they must share a neighbor (and thus are within 2 hops) or otherwise, there exist $2\cdot\lceil0.5\cdot(|V|-1)\rceil=\lceil|V|-1\rceil$ vertices in $V$ other than $u$ and $v$, leading to a contradiction since there will be more than $|V|$ vertices in $G$ when adding $u$ and $v$.

Without loss of generality, we use 2 as the diameter upper bound in our algorithm description, but it is straightforward to generalize it to the case $\gamma<0.5$ by considering vertices $f(\gamma)$ hops away. Since a vertex $u\in ext(S)$ must be within 2 hops from any $v\in S$, i.e., $u\in\mathbb{B}(v)$, we obtain the following theorem:

\begin{theorem}[Diameter Pruning]\label{prune:diameter}
{\em % to cancel italics
Given a mining task $\langle S, \\%===== note the split
ext(S)\rangle$, we have $ext(S)\ \ \subseteq\ \ \bigcap_{v\in S} \mathbb{B}(v).$
}
\end{theorem}

This is a Type-I pruning since if $u\not\in\bigcap_{v\in S} \mathbb{B}(v)$, $u$ can be pruned from $ext(S)$.

\vspace{2mm}
\noindent{\bf (P2) Size-Threshold Based Pruning.} A valid $\gamma$-quasi-clique $Q\subseteq V$ should contain at least $\tau_{size}$ vertices (i.e., $|Q|\geq\tau_{size}$), and therefore for any $v\in Q$, its degree $d(v)\geq\lceil\gamma\cdot(|Q|-1)\rceil\geq\lceil\gamma\cdot(\tau_{size}-1)\rceil$. We thus have:

\begin{theorem}[Size Threshold Pruning]\label{prune:size}
{\em % to cancel italics
If a vertex $u$ has $d(u)<\lceil\gamma\cdot(\tau_{size}-1)\rceil$, then $u$ cannot appear in any quasi-clique $Q$ with $|Q|\geq\tau_{size}$.
}
\end{theorem}

In other words, we can prune any such vertex $u$ from $G$. It is a Type-I pruning as $u\not\in ext(S)$, and also a Type-II pruning as $u\not\in S$. Note that a higher $\tau_{size}$ significantly reduces the search~space. Let us define  $k=\lceil\gamma\cdot(\tau_{size}-1)\rceil$, this rule essentially shrinks~$G$~into its $k$-core, which is defined as the maximal subgraph of $G$ where every vertex has degree $\geq k$. The $k$-core of a graph $G=(V, E)$ can be computed in $O(|E|)$ time using a peeling algorithm~\cite{kcore_peel}, which repeatedly deletes vertices with degree $<k$ until there is no such vertex. We thus always shrink a graph $G$ into its $k$-core before running our mining algorithm, which effectively reduces the search space.

\vspace{2mm}
\noindent{\bf (P3) Degree-Based Pruning.} There are two degree-based pruning rules, which belong to Type~I and Type~II, respectively. Recall that $d_{V'}(v)=|N_{V'}(v)|$, and thus $d_S(v)$ denotes the number of $v$'s neighbors inside $S$, and $d_{ext(S)}(v)$ denotes the number of $v$'s neighbors inside $ext(S)$. These two degrees are frequently used in our pruning rules to be presented subsequently.

\begin{theorem}[Type~I Degree Pruning]\label{prune:deg1}
{\em % to cancel italics
Given a vertex $u\in ext(S)$, if Condition~(i): $d_S(u) + d_{ext(S)}(u) < \lceil\gamma\cdot(|S| +d_{ext(S)}(u))\rceil$ holds, then $u$ can be pruned from $ext(S)$.
}
\end{theorem}

% [PROOF BEGIN]

This theorem is a result of the following lemma proved by~\cite{ZengWZK07}:

\begin{lemma}\label{prune:lemma2}
{\em % to cancel italics
If $a+n<\lceil\gamma\cdot(b+n)\rceil$ where $a,b,n\geq0$, then $\forall i\in[0, n]$, we have $a+i<\lceil\gamma\cdot(b+i)\rceil$.
}
\end{lemma}

Theorem~\ref{prune:deg1} follows since for any valid quasi-clique $Q=S\cup V'$ where $u\in V'$ and $V'\subseteq ext(S)$, according to Condition~(i) and Lemma~\ref{prune:lemma2} we have $d_S(u) + d_{V'}(u) < \lceil\gamma\cdot(|S| +d_{V'}(u))\rceil \leq \lceil\gamma\cdot(|Q|-1)\rceil$ (since $d_{V'}(u)\leq |V'|-1$ and $Q=S\cup V'$), which contradicts with the fact that $Q$ is a $\gamma$-quasi-clique.

% [PROOF END]

\begin{theorem}[Type~II Degree Pruning]\label{prune:deg2}
{\em % to cancel italics
Given vertex $v\in S$, if (i)~$d_S(v) < \lceil\gamma\cdot|S|\rceil$ and $d_{ext(S)}(v)=0$, or (ii)~if $d_S(v) + d_{ext(S)}(v)<\lceil\gamma(|S|-1+d_{ext(S)}(v))\rceil$, then for any $S'$ such that $S\subset S'\subseteq(S\cup ext(S))$, $G(S')$ cannot be a $\gamma$-quasi-clique.\\
\indent If Condition~(ii) applies for any $v\in S$, then for any $S'$ such that $S\subseteq S'\subseteq(S\cup ext(S))$, $G(S')$ cannot be a $\gamma$-quasi-clique.
}
\end{theorem}

% [PROOF BEGIN]
Theorem~\ref{prune:deg2} Condition~(ii) also follows Lemma~\ref{prune:lemma2}: $d_S(v) + d_{V'}(v)\\%===== note the split
 < \lceil\gamma\cdot(|S| -1+d_{V'}(v))\rceil \leq \lceil\gamma\cdot(|Q|-1)\rceil$ (since $d_{V'}(v)\leq|V'|$ and $Q=S\cup V'$). Note that as long as we find one such $v\in S$, there is no need to extend $S$ further. If $d_{ext(S)}(v)=0$ in Condition~(ii), then we obtain $d_S(v)<\lceil\gamma(|S|-1)\rceil$ which is contained in Condition~(i). Note that Condition~(ii) applies to the case $S'=S$ since $i$ can be 0 in Lemma~\ref{prune:lemma2}.
 
Condition~(i) allows more effective pruning and is correct since for any valid quasi-clique $Q\supset S$ extended from $S$, we have $d_Q(v)\leq d_S(v)+d_{ext(S)}(v)=d_S(v)<\lceil\gamma(|Q|-1)\rceil$ (since $d_S(v) < \lceil\gamma\cdot|S|\rceil$ and $|S|\leq|Q|-1$), which contradicts with the fact that $Q$ is a $\gamma$-quasi-clique. Note that the pruning of Condition~(i) does not include the case where $S'=S$.
% [PROOF END]
 
% [No Proof Version to Replace]:
% In Theorem~\ref{prune:deg2} Condition~(ii), as long as we find one such $v\in S$, there is no need to extend $S$ further; also if $d_{ext(S)}(v)=0$, then we obtain $d_S(v)<\lceil\gamma(|S|-1)\rceil$ which is contained in Condition~(i). Note that Condition~(ii) also applies to the case $S'=S$ while Condition~(i) does not, but Condition~(i) allows more effective pruning.

\begin{figure}[t]
\centering
\includegraphics[width=0.7\columnwidth]{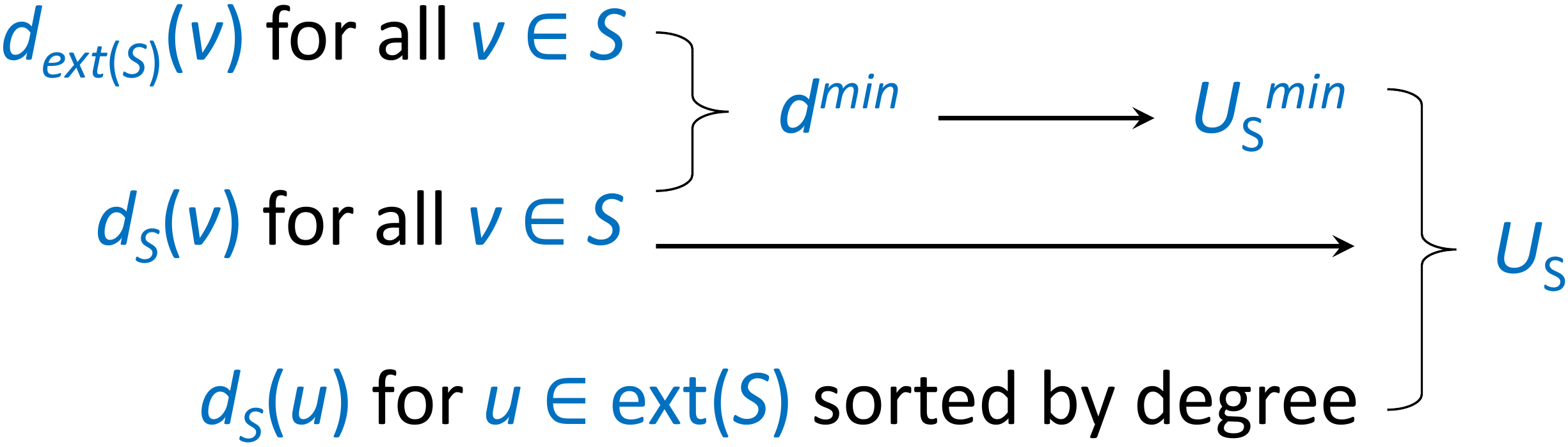}
\caption{Upper Bound Derivation}\label{ub}
\end{figure}
\setlength{\textfloatsep}{2mm}

\vspace{1mm}
\noindent{\bf (P4) Upper Bound Based Pruning.} We next define an upper bound on the number of vertices in $ext(S)$ that can be added to $S$ concurrently to form a $\gamma$-quasi-clique, denoted by $U_S$. The definition of $U_S$ is based on $d_S(v)$ and $d_{ext(S)}(v)$ of all vertices $v\in S$ and on $d_S(u)$ of vertices $u\in ext(S)$ as summarized by Figure~\ref{ub}, which we describe next.

We first define $d^{min}$ as the minimum degree of any vertex in $S$:

\vspace{-2mm}

\begin{equation}\label{eq:dmin}
d^{min}=\min_{v\in S}\{d_S(v)+d_{ext(S)}(v)\}.
\end{equation}

% [PROOF BEGIN]
Now consider any $S'$ such that $S\subseteq S'\subseteq(S\cup ext(S))$. For any $v\in S$, we have $d_S(v)+d_{ext(S)}(v)\geq d_{S'}(v)\geq\lceil\gamma(|S'|-1)\rceil$, and therefore, $d^{min}\geq\lceil\gamma(|S'|-1)\rceil$. As a result, $\lfloor d^{min}/\gamma\rfloor \geq \lfloor \lceil\gamma(|S'|-1)\rceil/\gamma\rfloor \geq \lfloor\gamma(|S'|-1)/\gamma\rfloor = |S'|-1$, which gives the following upper bound on $|S'|$:

\vspace{-2mm}

\begin{equation}\label{eq:usmin0}
|S'|\leq\lfloor d^{min}/\gamma\rfloor+1.
\end{equation}

Since $|S|$ vertices are already included, we obtain an upper bound $U_S^{min}$ on the number of vertices from $ext(S)$ that can further extend $S$ to form a valid quasi-clique:
% [PROOF END]

% [No Proof Version to Replace]:
% We can derive an upper bound $U_S^{min}$ on the number of vertices from $ext(S)$ that can further extend $S$ to form a valid quasi-clique:

\vspace{-2mm}

\begin{equation}\label{eq:usmin}
U_S^{min}=\lfloor d^{min}/\gamma\rfloor+1-|S|.
\end{equation}

% [PROOF BEGIN]
We next tighten this upper bound using vertices in $ext(S)=\{u_1,u_2,\ldots,u_n\}$, assuming that the vertices are listed in non-increa-%===== note the split
sing order of $d_S(u)$. Then, we have:

\vspace{-1mm}
\begin{lemma}\label{lemma:us}
{\em % to cancel italics
Given an integer $k$ such that $1\leq k\leq n$, if $\sum_{v\in S}\\%===== note the split
d_S(v) + \sum_{i:1\leq i\leq k}d_S(u_i) < |S|\cdot\lceil\gamma(|S|+k-1)\rceil$, then for any vertex set $Z\subseteq ext(S)$ with $|Z|=k$, $S\cup Z$ is not a $\gamma$-quasi-clique.
}
\end{lemma}
\vspace{-1mm}

Note that if $S'$ is a $\gamma$-quasi-clique, then $d_{S'}(v)\geq\lceil\gamma(|S'|-1)\rceil$ for any $v\in S'$, and therefore, for any $S\subseteq S'$, we have $\sum_{v\in S} d_{S'}(v)\geq |S|\cdot\lceil\gamma(|S'|-1)\rceil$. Thus, to prove Lemma~\ref{lemma:us}, we only need to show that $\sum_{v\in S} d_{S\cup Z}(v)<|S|\cdot\lceil\gamma(|S|+|Z|-1)\rceil$, which is because:
\begin{eqnarray}
\sum_{v\in S} d_{S\cup Z}(v) & = & \sum_{v\in S} d_S(v) + \sum_{v\in S} d_Z(v)\nonumber\\
 & = &  \sum_{v\in S} d_S(v) + \sum_{u\in Z} d_S(u)\nonumber\\
 & \leq & \sum_{v\in S} d_S(v) + \sum_{i: 1\leq i\leq|Z|} d_S(u_i)\nonumber\\
 & < & |S|\cdot\lceil\gamma(|S|+|Z|-1)\rceil.\nonumber
\end{eqnarray}

Based on Lemma~\ref{lemma:us}, we define a tightened upper bound $U_S$ as follows:
% [PROOF END]

% [No Proof Version to Replace]:
% We can then tighten this upper bound using vertices in $ext(S)=\{u_1,u_2,\ldots,u_n\}$, assuming that the vertices are listed in non-increas-%===== note the split
%ing order of degree. We can derive the following upper bound $U_S$:
\vspace{-1mm} % [Comment it if Proof is saved]

\vspace{-5mm}

\begin{eqnarray}\label{eq:us}
U_S & = & \max\left\{t\ \bigg|\ \left(1\leq t\leq U_S^{min}\right)\ \ \bigwedge\ \ \left(\sum_{v\in S}d_S(v)+\right.\right.\nonumber\\
& & \left.\left.\sum_{i:1\leq i\leq t}d_S(u_i)\ \geq\ |S|\cdot\lceil\gamma(|S|+t-1)\rceil\right)\right\}.
\end{eqnarray}

If such a $t$ cannot be found, then $S$ cannot be extended to generate a valid quasi-clique, which is a Type~II pruning. Otherwise, we further consider two pruning rules based on $U_S$.

\vspace{-2mm} 

\begin{theorem}[Type~I Upper Bound Pruning]\label{prune:ub1}
{\em % to cancel italics
Given a vertex $u\in ext(S)$, if $d_S(u)+U_S-1 < \lceil\gamma\cdot(|S|+U_S-1)\rceil$, then $u$ can be pruned from $ext(S)$.
}
\end{theorem}

\vspace{-2mm} 

% [PROOF BEGIN]

Consider any valid quasi-clique $Q=S\cup V'$ where $u\in V'$ and $V'\subseteq ext(S)$. If the condition in Theorem~\ref{prune:ub1} holds, i.e., $d_S(u)+U_S-1 < \lceil\gamma\cdot(|S|+U_S-1)\rceil$, then based on Lemma~\ref{prune:lemma2} and the fact that $|V'|\leq U_S$, we have:
\begin{equation}\label{eq:ub1}
d_S(u) + |V'| - 1\ \ <\ \ \lceil\gamma\cdot(|S|+|V'|-1)\rceil\ \ =\ \ \lceil\gamma\cdot(|Q|-1)\rceil,
\end{equation}
and therefore, $d_Q(u)=d_S(u)+d_{V'}(u)\leq d_S(u)+|V'|-1<\lceil\gamma\cdot(|Q|-1)\rceil$, which contradicts with the fact that $Q$ is a $\gamma$-quasi-clique.

% [PROOF END]

\vspace{-2mm} 

\begin{theorem}[Type~II Upper Bound Pruning]\label{prune:ub2}
{\em % to cancel italics
Given a \\%===== note the split
vertex $v\in S$, if $d_S(v)+U_S< \lceil\gamma\cdot(|S|+U_S-1)\rceil$, then for any $S'$ such that $S\subseteq S'\subseteq(S\cup ext(S))$, $G(S')$ cannot be a $\gamma$-quasi-clique.
}
\end{theorem}

\vspace{-2mm} 

% [PROOF BEGIN]

Theorem~\ref{prune:ub2} follows Lemma~\ref{prune:lemma2} and the fact that $d_{V'}(v)\leq|V'|$, as can be proved similarly to Eq~(\ref{eq:ub1}). Note that as long as we find one such $v\in S$, there is no need to extend $S$ further. Since $i$ can be 0 in Lemma~\ref{prune:lemma2}, the pruning of Theorem~\ref{prune:ub2} includes the case where $S'=S$, which is different from Theorem~\ref{prune:deg2}.

% [PROOF END]

% [No Proof Version to Replace]:
% Note that Theorem~\ref{prune:ub2} includes the case where $S'=S$, which is different from Theorem~\ref{prune:deg2}.

\begin{figure}[t]
\centering
\includegraphics[width=0.7\columnwidth]{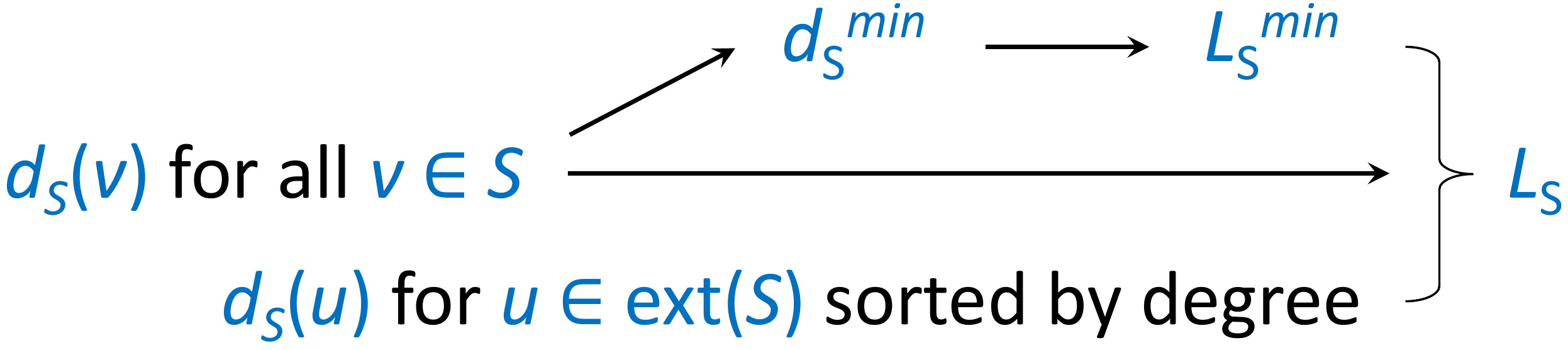}
\caption{Lower Bound Derivation}\label{lb}
\end{figure}
\setlength{\textfloatsep}{2mm}

\vspace{2mm}
\noindent{\bf (P5) Lower Bound Based Pruning.} Given a vertex set $S$, if some vertex $v\in S$ has $d_S(v)<\lceil\gamma\cdot(|S|-1)\rceil$, then at least a certain number of vertices need to be added to $S$ to increase the degree of $v$ in order to form a $\gamma$-quasi-clique. We denote this lower bound as $L_{min}$, which is defined based on $d_S(v)$ of all vertices $v\in S$ and on $d_S(u)$ of vertices $u\in ext(S)$ as summarized by Figure~\ref{lb}, which we describe next.

We first define $d_S^{min}$ as the minimum degree of any vertex in $S$:

\vspace{-2mm}

\begin{equation}\label{eq:dsmin}
d_S^{min}=\min_{v\in S}\ d_S(v).
\end{equation}

Then, a straightforward lower bound is given by:

\vspace{-2mm}

\begin{equation}\label{eq:lsmin}
L_S^{min}=\min\{t\ |\ d_S^{min}+t\geq\lceil\gamma\cdot(|S|+t-1)\rceil\}.
\end{equation}

To find such $L_S^{min}$, we check $t=0,1,\cdots,|ext(S)|$, %toGuimu: t should begin from 0
and if none of them satisfies the inequality, $S$ and its extensions cannot produce a valid quasi-clique, which is a Type~II pruning.

% [PROOF BEGIN]

Otherwise, we further tighten the lower bound into $L_{S}$ below using Lemma~\ref{lemma:us}, assuming that vertices in $ext(S)=\{u_1,u_2,\ldots,u_n\}$ are listed in non-increasing order of $d_S(u)$:

% [PROOF END]

% [No Proof Version to Replace]:
% Otherwise, we can further tighten the lower bound into $L_{S}$ below assuming that vertices in $ext(S)=\{u_1,u_2,\ldots,u_n\}$ are listed in non-increasing order of degree:

\vspace{-5mm}

\begin{eqnarray}\label{eq:ls}
L_S = \min\left\{t\ \bigg|\ \left(L_S^{min}\leq t\leq n\right)\ \ \bigwedge\ \ \left(\sum_{v\in S}d_S(v)+\right.\right.\nonumber\\
\left.\left.\sum_{i:1\leq i\leq t}d_S(u_i)\ \geq\ |S|\cdot\lceil\gamma(|S|+t-1)\rceil\right)\right\}
\end{eqnarray}

% [PROOF BEGIN]

If such a $t$ cannot be found, then $S$ cannot be extended to generate a valid quasi-clique, which is a Type~II pruning. Otherwise, we further consider two pruning rules based on $L_S$ whose proofs are straightforward.

% [PROOF END]

% [No Proof Version to Replace]:
% If such a $t$ cannot be found, then $S$ cannot be extended to generate a valid quasi-clique, which is a Type~II pruning. Otherwise, we further consider two pruning rules based on $L_S$ as follows:

\vspace{-1mm}

\begin{theorem}[Type~I Lower Bound Pruning]\label{prune:lb1}
{\em % to cancel italics
Given a\\ % ===== note the split
vertex $u\in ext(S)$, if $d_S(u)+d_{ext(S)}(u) < \lceil\gamma\cdot(|S|+L_S-1)\rceil$, then $u$ can be pruned from $ext(S)$.
}
\end{theorem}

\begin{theorem}[Type~II Lower Bound Pruning]\label{prune:lb2}
{\em % to cancel italics
Given a\\ % ===== note the split
vertex $v\in S$, if $d_S(v)+d_{ext(S)}(v)< \lceil\gamma\cdot(|S|+L_S-1)\rceil$, then for any $S'$ such that $S\subseteq S'\subseteq(S\cup ext(S))$, $G(S')$ cannot be a $\gamma$-quasi-clique.
}
\end{theorem}

\noindent{\bf (P6) Critical-Vertex Based Pruning.} We next define the concept of {\em critical vertex} using the lower bound $L_S$ defined before.

\vspace{-1mm}

\begin{definition}[Critical Vertex]\label{def:cv}
{\em % to cancel italics
Let $S$ be a vertex set. If there exists a vertex $v\in S$ such that $d_S(v)+d_{ext(S)}(v)=\lceil\gamma\cdot(|S|+L_S-1)\rceil$, then $v$ is called a critical vertex of $S$.
}
\end{definition}

\vspace{-1mm}

Then, we have the following theorem:

\vspace{-1mm}

\begin{theorem}[Critical Vertex Pruning]\label{prune:cv}
{\em % to cancel italics
If $v\in S$ is a critical vertex, then for any vertex set $S'$ such that $S\subset S'\subseteq (S\cup ext(S))$, if $G(S')$ is a $\gamma$-quasi-clique, then $S'$ must contain every neighbor of $v$ in $ext(S)$, i.e., $N_{ext(S)}(v)\subseteq S'$.
}
\end{theorem}

\vspace{-1mm}

This is because if $u\in N_{ext(S)}(v)$ is not in $S'$, then $d_{S'}(v)<d_S(v)+d_{ext(S)}(v)=\lceil\gamma\cdot(|S|+L_S-1)\rceil\leq\lceil\gamma\cdot(|S'|-1)\rceil$, which contradicts with the fact that $S'$ is a $\gamma$-quasi-clique. Therefore, when extending $S$, if we find $v\in S$ is a critical vertex, we can directly add all vertices in $N_{ext(S)}(v)$ to $S$ for further mining.

\vspace{2mm}
\noindent{\bf (P7) Cover-Vertex Based Pruning.} Given a vertex $u\in ext(S)$, we will define a vertex set $C_S(u)\subseteq ext(S)$ such that for any $\gamma$-quasi-clique $Q$ generated by extending $S$ with vertices in $C_S(u)$, $Q\cup u$ is also a $\gamma$-quasi-clique. In other words, $Q$ is not maximal and can thus be pruned. We say that $C_S(u)$ is the set of vertices in $ext(S)$ that are covered by $u$, and that $u$ is the cover vertex.

To utilize $C_S(u)$ for pruning, we put vertices of $C_S(u)$ after all the other vertices in $ext(S)$ when checking the next level in the set-enumeration tree (see Figure~\ref{set_enum}), and only check until vertices of $ext(S)-C_S(u)$ are examined (i.e., the extension of $S$ using $V'\subseteq C_S(u)$ is pruned). To maximize the pruning effectiveness, we find $u\in ext(S)$ to maximize $|C_S(u)|$.

% [PROOF BEGIN]

We compute $C_S(u)$ as the intersection of (1)~$ext(S)$, (2)~$N(u)$, and (3)~$N(v)$ of any $v\in S$ that is not a neighbor of $u$:

% [PROOF END]

% [No Proof Version to Replace]:
% We can prove that $C_S(u)$ can be the intersection of (1)~$ext(S)$, (2)~$\Gamma(u)$, and (3)~$\Gamma(v)$ of any $v\in S$ that is not a neighbor of $u$:

\vspace{-2mm}

\begin{equation}\label{eq:cover}
C_S(u)\ \ =\ \ N_{ext(S)}(u)\ \ \cap\ \ \bigcap_{v\in S\ \wedge\ v\not\in N(u)}N(v)
\end{equation}

We compute $C_S(u)$ only if $d_S(u)\geq\lceil \gamma\cdot|S|\rceil$ and for any $v\in S$ that are not adjacent to $u$, it holds that $d_S(v)\geq\lceil\gamma\cdot|S|\rceil$; otherwise, we deem this pruning inapplicable as they are pruned by Theorems~\ref{prune:deg1} and~\ref{prune:deg2}.

% [PROOF BEGIN]

For any $\gamma$-quasi-clique $Q$ that extends $S$ with vertices in $C_S(u)$, we now explain why $Q\cup u$ is also a $\gamma$-quasi-clique by showing that for any vertex $v\in Q\cup u$, it holds that $d_{Q\cup u}(v)\geq\lceil\gamma\cdot(|Q\cup u|-1)\rceil=\lceil \gamma\cdot|Q|\rceil$. There are 4 cases for $v$: (1)~$v=u$: then since $u$ is adjacent to all the vertices in $C_S(u)$ and we require $d_S(u)\geq\lceil \gamma\cdot|S|\rceil$,  we have $d_{Q\cup u}(u)=d_S(u)+|Q|-|S|\geq\lceil \gamma\cdot|S|\rceil+|Q|-|S|\geq\lceil \gamma\cdot|Q|\rceil+|Q|-|Q|\geq\lceil \gamma\cdot|Q|\rceil$; (2)~$v\in S$ and $v\not\in N(u)$: then since $v$ is adjacent to all the vertices in $C_S(u)$ and we require $d_S(v)\geq\lceil \gamma\cdot|S|\rceil$, we have $d_{Q\cup u}(v)=d_S(v)+|Q|-|S|\geq\lceil \gamma\cdot|S|\rceil+|Q|-|S|\geq\lceil \gamma\cdot|Q|\rceil+|Q|-|Q|\geq\lceil \gamma\cdot|Q|\rceil$; (3)~$v\in S$ and $v\in N(u)$: then we have $d_{Q\cup u}(v)=d_Q(v)+1\geq\lceil\gamma\cdot(|Q|-1)\rceil+1\geq\lceil \gamma\cdot|Q|\rceil$; (4)~$v\in(Q-S)$: then we have $d_{Q\cup u}(v)=d_Q(v)+1\geq\lceil\gamma\cdot(|Q|-1)\rceil+1\geq\lceil \gamma\cdot|Q|\rceil$. In summary, $Q\cup u$ is a $\gamma$-quasi-clique and $Q$ is not maximal.

As a degenerate special case, initially when $S=\emptyset$, Eq~(\ref{eq:cover}) is essentially $C_S(u)=N_{ext(S)}(u)=N(u)$, i.e., we only need to find $u$ as the vertex with the maximum degree. Note that for any $\gamma$-quasi-clique $Q$ constructed out of vertices in $C_S(u)=N(u)$, adding $u$ to $Q$ still produces a $\gamma$-quasi-clique. We find $u$ as the vertex the maximum degree after $k$-core pruning by (P2) above, since otherwise, we may find a high-degree vertex without much pruning power (e.g., the center of a sparse star graph).

\subsection{The Recursive Algorithm}\label{ssec:quick}
We have summarized 7 categories of pruning rules (P1)--(P7). Next, we present our recursive algorithm for mining maximal quasi-cliques in topics (T1)--(T6) below, which effectively utilizes the pruning rules.

\vspace{2mm}
\noindent{\bf (T1) Size Threshold Pruning as a Preprocessing.} First consider the size-threshold based pruning established by Theorem~\ref{prune:size}, which says that any vertex with degree less than $k=\lceil\gamma\cdot(\tau_{size}-1)\rceil$ cannot be in a valid quasi-clique. %Quick somehow does not use this pruning rule, leading to a very poor scalability in our preliminary test. In fact, 
This rule essentially shrinks an input graph $G$ into its $k$-core, which is defined as the maximal subgraph of $G$ where every vertex has degree $\geq k$. The $k$-core of $G$ can be computed in $O(|E|)$ time using a peeling algorithm~\cite{kcore_peel}, which repeatedly deletes vertices with degree $<k$ until there is no such vertex. We thus always shrink a graph $G$ into its $k$-core before running the mining algorithm to be described next, and since the $k$-core of $G$ is much smaller than $G$ itself, this pruning effectively reduces the search space.

\vspace{2mm}
\noindent{\bf (T2) Degree Computation.} Since we are growing $G(S)$ into a valid quasi-clique by including more vertices in $ext(S)$, when we say we maintain $S$, we actually maintain $G(S)$: every vertex $v\in S$ is associated with an adjacency list in $G(S)$. Whenever we add a new vertex $u\in ext(S)$ to $G(S)$, for each $v\in N(u)\cap S$, we add $u$ (resp.\ $v$) to $v$'s (resp.\ $u$'s) adjacency list in $G(S)$.

Recall that our pruning rules use 4 kinds of vertex degrees:
\begin{itemize}
\setlength\itemsep{0pt}
\item SS-degrees: $d_S(v)$ for all $v\in S$;
\item SE-degrees: $d_S(u)$ for all $u\in ext(S)$;
\item ES-degrees: $d_{ext(S)}(v)$ for all $v\in S$;
\item EE-degrees: $d_{ext(S)}(u)$ for all $u\in ext(S)$.
\end{itemize}

As Figure~\ref{ub} shows, computing $U_S$ requires the first 3 kinds of degrees; and as Figure~\ref{lb} shows, computing $L_S$ requires the first 2 kinds of degrees. The EE-degrees are only used by Type~I pruning rules of Theorems~\ref{prune:deg1} and~\ref{prune:lb1}.

SS-degrees can be obtained from the adjacency list sizes of $G(S)$. {\em SE-degrees and ES-degrees can be calculated together}: for each  $u\in ext(S)$, and for each $v\in N(u)\cap S$, $(u, v)$ is an edge crossing $S$ and $ext(S)$ and thus we increment both $d_S(u)$ and $d_{ext(S)}(v)$. Finally, EE-degrees can be computed from adjacency lists of vertices in $ext(S)$, and since it is only needed by Type~I pruning rather than computing $U_S$ and $L_S$, {\em we can delay its computation to right before checking Type~I pruning rules}.

\vspace{2mm}

\noindent{\bf (T3) Type~II Pruning Rules.} We have described 3 major Type~II pruning rules in Theorems~\ref{prune:deg2}, \ref{prune:ub2} and~\ref{prune:lb2}, which share the following common feature: every vertex $v\in S$ is checked and if the pruning condition is met for any $v$, $S$ along with any of its extensions cannot be a valid quasi-clique and are thus pruned.

The only exception is Theorem~\ref{prune:deg2} Condition~(i), which prunes $S$'s extensions but not $S$ itself. Of course, if any of the other Type~II pruning condition is met, $S$ is also pruned. Therefore, {\em only when all Type~II pruning conditions except for Theorem~\ref{prune:deg2} Condition~(i) are not met, will we consider $S$ as a candidate for a valid quasi-clique.}

Also note that {\em the computation of bounds $U_S$ and $L_S$ may also trigger Type~II pruning}. For example, in Eq~(\ref{eq:us}), if a valid $t$ cannot be found, then any extension of $S$ can be pruned though $G(S)$ is still a candidate to check. % toGuimu: when implementing, please check G(S) before returning true in algo 1; ok to check t=0 but no cheaper
In contrast, in Eq~(\ref{eq:lsmin}), if a valid $t$ cannot be found (including $t=0$), then $S$ and its extensions are pruned; this also applies to Eq~(\ref{eq:ls}).

\vspace{2mm}

\noindent{\bf (T4) Iterative Nature of Type~I Pruning.} Recall that we have 3 major Type~I pruning rules in Theorems~\ref{prune:deg1}, \ref{prune:ub1} and~\ref{prune:lb1}, which share the following common feature: every vertex $u\in ext(S)$ is checked and if the pruning condition is met for $u$, $u$ is pruned from $ext(S)$.

%Note that removing $u$ from $ext(S)$ may reduce $d_{ext(S)}(u)$ and $d_S(u)$ (and hence $d_{ext(S)}(v)$ of every $v\in\Gamma(u)\cap S$), which will further update $U_S$ and $L_S$. This essentially means that the Type~I pruning is iterative: each pruned $u$ may change degrees and bounds, which affects the various pruning rules (including Type~I ones), which should be checked again and new vertices in $ext(S)$ may be pruned due to Type~I pruning. As this process is repeated, $U_S$ and $L_S$ become tighter until no more vertex can be pruned from $ext(S)$, which consists of 2 cases:

Note that removing a vertex $u_i$ from $ext(S)$ reduces $d_{ext(S)}(v)$ of every $v\in N(u_i)\cap S$, which will further update $U_S$ (see Figure~\ref{ub}), as well as $L_S$ (see Eq~(\ref{eq:ls})). This essentially means that the Type~I pruning is iterative: each pruned $u$ may change degrees and bounds, which affects the various pruning rules (including Type~I ones), which should be checked again and new vertices in $ext(S)$ may be pruned due to Type~I pruning. As this process is repeated, $U_S$ and $L_S$ become tighter until no more vertex can be pruned from $ext(S)$, which consists of 2 cases:

\begin{itemize}
\setlength\itemsep{0pt}
\item C1: $ext(S)$ becomes empty. In this case, we only need to check if $G(S)$ is a valid quasi-clique;
\item C2: $ext(S)$ is not empty but cannot be shrunk further by pruning rules. Then, we need to check $S$ and its extensions.
\end{itemize}

\begin{algorithm}[t]
\caption{Iterative Bound-Based Pruning}\label{algo:iterative}
{\bf Function:} {\em iterative\_bounding}($S$, $ext(S)$, $\gamma$, $\tau_{size}$)\\
{\bf Output:} $true$ iff the case of extending $S$ (excluding $S$ itself) is pruned; $ext(S)$ is passed as a reference, and some elements may be pruned when the function returns
\begin{algorithmic}[1]
\REPEAT
%----------
\STATE Compute $d_S(v)$ and $d_{ext(S)}(v)$ for all $v$ in $S$ and $ext(S)$
\STATE Compute upper bound $U_S$ and lower bound $L_S$ (Type~II pruning may apply)
% ---- following are Type~II pruning caused by U_S and L_S
%\IF{$U_S<L_S$}
%\STATE {\bf return} $true$
%\ENDIF 
%\IF{$\not\exists\ t$ satisfying Eq~(\ref{eq:lsmin}) or then Eq~(\ref{eq:ls})}
%\STATE {\bf return} $true$
%\ENDIF
\IF{$\forall\ v\in S$ that is a critical vertex}
\STATE $I\gets ext(S)\cap N(v)$
\STATE $S\gets S\cup I$
\STATE $ext(S)\gets ext(S) - I$
\STATE Update degree values, $U_S$ and $L_S$  (Type~II pruning may apply) % note that this indent is correcting pkdd08
% ---- following are Type~II pruning caused by U_S and L_S
%\IF{$U_S<L_S$}
%\STATE {\bf return} $true$
%\ENDIF 
\ENDIF
\FOR{{\bf each} vertex $v\in S$}
\STATE Check Type~II pruning conditions: Theorems~\ref{prune:deg2}, \ref{prune:ub2} and~\ref{prune:lb2}
\IF{some condition other than Theorem~\ref{prune:deg2} Condition~(i) holds for $v$}
\STATE {\bf return} $true$
\ENDIF
\ENDFOR
\IF{Theorem~\ref{prune:deg2} Condition~(i) holds for some $v\in S$}
	\IF{$|S|\geq\tau_{size}$ {\bf and} $G(S)$ is a $\gamma$-quasi-clique}
	\STATE Append $S$ to the result file
	\STATE {\bf return} $true$
	\ENDIF
\ENDIF
\FOR{{\bf each} vertex $u\in ext(S)$}
\STATE Check Type~I pruning conditions: Theorems~\ref{prune:deg1}, \ref{prune:ub1} and~\ref{prune:lb1}
\IF{some Type~I pruning condition holds for $u$}
\STATE $ext(S)\gets ext(S)-u$
\ENDIF
\ENDFOR
%----------
\UNTIL{$ext(S)=\emptyset$ {\bf or} no vertex in $ext(S)$ was Type-I-pruned} % nothing to prune
%----------
\IF{$ext(S)=\emptyset$}
\IF{$|S|\geq\tau_{size}$ {\bf and} $G(S)$ is a $\gamma$-quasi-clique}
\STATE Append $S$ to the result file
\ENDIF
\STATE {\bf return} $true$
\ENDIF
\STATE {\bf return} $false$
\end{algorithmic}
\end{algorithm}
\setlength{\textfloatsep}{2mm}

\vspace{2mm}
\noindent{\bf (T5) The Iterative Pruning Subprocedure.} Given a vertex set $S$, and the set of vertices $ext(S)$ to extend $S$ into valid quasi-cliques, Algorithm~\ref{algo:iterative} shows how to apply our pruning rules to (1)~shrink $ext(S)$ and to (2)~determine if $S$ can be further extended to form a valid quasi-clique. In Algorithm~\ref{algo:iterative}, the return value is of a boolean type indicating whether $S$'s extensions (but not $S$ itself) are pruned, and the input $ext(S)$ is passed as a reference and may be shrunk by Type~I pruning when the function returns.

As (T4) indicates, the application of pruning rules is intrinsically iterative since the shrinking of $ext(S)$ may trigger more pruning. This iterative process is described by Lines~1--21, and the loop ends if the condition in Line~21 is met which corresponds to the two cases C1 and C2 described in (T4).

We design function {\em iterative\_bounding}($S$, $ext(S)$, $\gamma$, $\tau_{size}$) to guarantee that it returns $false$ only if  $ext(S)\neq\emptyset$. Therefore, if the loop of Lines~1--21 exits due to $ext(S)$ becoming $\emptyset$, we have to return $true$ (Line~25) as there is no vertex to extend $S$, but we need to first examine if $G(S)$ itself is a valid quasi-clique in Lines~23--24; note that here, $G(S)$ is not pruned by Type~II pruning as otherwise, the loop will directly return $true$ (see Lines~10--12).

Now let us focus on the loop body in Lines~2--20 about one pruning iteration, which can be divided into 3 parts: (1)~Lines~2--8: critical vertex pruning, (2)~Lines 9--16: Type~II pruning, and (3)~Lines~17--20: Type~I pruning. To keep Algorithm~\ref{algo:iterative} short, we omit some details but they are included in our descriptions.

First, consider Part~1. We compute the degrees in Line~2, which are then used to compute $U_S$ and $L_S$ in Line~3. In Line~2, we do not need to compute EE-degrees since they are only used by Type~I pruning; we actually compute it right before Part~3, since if any Type~II pruning applies, the function returns and the computation of EE-degrees is saved. %toGuimu: move them from before Part 2 to before Part 3 !!!
In Line~3, Type~II pruning may apply when computing $U_S$ and $L_S$ (see the paragraphs below Eqs~(\ref{eq:us}) and~(\ref{eq:ls}), respectively), in which case we return $true$ to prune $S$'s extensions. Note that for $U_S$'s case, we still need to examine $G(S)$, and the actions are the same as in Lines~23--25. In Line~3, after we obtain $U_S$ and $L_S$, if $U_S<L_S$ we also directly return $true$ to prune $S$ and its extensions; note that since $L_S\geq 1$, $S$ is not a valid quasi-clique as it needs to add at least $L_S$ vertices to be valid.

Then, Lines~4--7 then apply the critical-vertex pruning of Theorem~\ref{prune:cv}. 
Line~4 first checks the condition of a critical vertex in Definition~\ref{def:cv} which uses $L_S$. Lines~5--7 then performs the movement of $N(v)\cap ext(S)$, which will change the degrees and hence bounds and so they are recomputed in Line 8. Similar to Line~3, Line~8 may trigger type~II pruning so that the function returns $true$. Also similar to Line~3, after we obtain $U_S$ and $L_S$ in Line~8, if $U_S<L_S$ we also directly return $true$ to prune $S$ and its extensions.

In our actual implementation, if $ext(S)$ is found to be empty after running Line~7, we directly exit the loop of Lines~1--21, to skip the execution of Lines 8--21.

In Quick, each iteration only finds one critical vertex and moves its neighbors from $ext(S)$ to $S$. We propose to find all critical vertices in $S$ and move their neighbors from $ext(S)$ to $S$. Such movement will update degrees and bounds in Line~8 which may generate new critical vertices in the updated $S$, therefore, we actually loop Lines~4--8 until there is no more critical vertex in $S$.

Recall that Theorem~\ref{prune:cv} does not prune $S$ itself, and it is possible that the expanded $S$ leads to no valid quasi-cliques, making $G(S)$ a maximal quasi-clique. We therefore actually first check $G(S)$ as in Lines~23--24 before expanding $S$ with $N(v)\cap ext(S)$. The original Quick does not examine $G(S)$ and thus may miss results. While our algorithm may output $S$ while $G(S)$ is not maximal, but just like in Quick, we require a postprocessing phase to remove non-maximal quasi-cliques anyway. %toGuimu: add this check before moving !!!

Next, consider Part~2 on Type~II pruning. Lines~9--12 first check the pruning conditions of Theorems~\ref{prune:deg2}, \ref{prune:ub2} and~\ref{prune:lb2} on every vertex $v\in S$. If any condition other than Theorem~\ref{prune:deg2} Condition~(i) applies, $S$ along with its extensions are pruned and thus Line~12 returns $true$. Otherwise, if Theorem~\ref{prune:deg2} Condition~(i) applies for some $v\in S$, then extensions of $S$ are pruned but $G(S)$ itself is not, and it is examined in Lines~14--16.

Finally, Part~3 on Type~I pruning checks every vertex $u\in ext(S)$ and tries to prune $u$ using a condition of Theorems~\ref{prune:deg1}, \ref{prune:ub1} and~\ref{prune:lb1}, as shown in Lines~17--20. The shrinking of $ext(S)$ may create new pruning opportunities for the next iteration.

To summarize, Quick+ improves Quick for iterative bounding in 3 aspects. (1)~In Quick, each iteration only finds one critical vertex and moves its neighbors from $ext(S)$ to $S$, while we find all critical vertices to move their neighbors to $S$ to improve pruning. (2)~Type-II pruning may apply when computing $U_S$ and $L_S$ (c.f., (P4 \& P5)), and Quick+ handles these boundary cases and returns $true$ to prune $S$'s extensions. (3)~in both critical vertex pruning and degree-based Type-II pruning, $G(S)$ itself should not be pruned which is properly handled by Quick+, but not Quick.

\begin{algorithm}[t]
\caption{Mining Valid Quasi-Cliques Extended from $S$}\label{algo:main}
{\bf Function:} {\em recursive\_mine}($S$, $ext(S)$, $\gamma$, $\tau_{size}$)\\
{\bf Output:} $true$ iff some valid quasi-clique $Q\supset S$ is found
\begin{algorithmic}[1]
\STATE  $\mathcal{T}_{Q\_found}\gets false$
\STATE  Find cover vertex $u\in ext(S)$ with the largest $C_S(u)$
\STATE \{If not found, $C_S(u)\gets\emptyset$\}
\STATE Move vertices of $C_S(u)$ to the tail of the vertex list of $ext(S)$
\FOR{{\bf each} vertex $v$ in the sub-list $\left(ext(S)-C_S(u)\right)$}
\IF{$|S|+|ext(S)|<\tau_{size}$}
\STATE {\bf return} $\mathcal{T}_{Q\_found}$
\ENDIF
\IF{$G(S\cup ext(S))$ is a $\gamma$-quasi-clique}
\STATE Append $S\cup ext(S)$ to the result file
\STATE {\bf return} $true$
\ENDIF
\STATE $S'\gets S\cup v$,\ \ $ext(S)\gets ext(S)-v$
\STATE $ext(S')\gets ext(S)\cap\mathbb{B}(v)$
\IF{$ext(S')=\emptyset$}
	\IF{$|S'|\geq\tau_{size}$ {\bf and} $G(S')$ is a $\gamma$-quasi-clique}
	\STATE  $\mathcal{T}_{Q\_found}\gets true$
	\STATE Append $S'$ to the result file
	\ENDIF
\ELSE
\STATE $\mathcal{T}_{pruned}\gets$ {\em iterative\_bounding}($S'$, $ext(S')$, $\gamma$, $\tau_{size}$) % toGuimu:: no need to pass lb and ub !!! update code
\STATE \{here, $ext(S')$ is Type-I-pruned and $ext(S')\neq\emptyset$\}
\IF{$\mathcal{T}_{pruned}=false$ {\bf and} $|S'|+|ext(S')|\geq\tau_{size}$}
\STATE $\mathcal{T}_{found}\gets$ {\em recursive\_mine}($S'$, $ext(S')$, $\gamma$, $\tau_{size}$)
\STATE $\mathcal{T}_{Q\_found}\gets \mathcal{T}_{Q\_found}$ {\bf or} $\mathcal{T}_{found}$
	\IF{$\mathcal{T}_{found}=false$ {\bf and} $|S'|\geq\tau_{size}$ {\bf and} $G(S')$ is a $\gamma$-quasi-clique}
	\STATE  $\mathcal{T}_{Q\_found}\gets true$
	\STATE Append $S'$ to the result file
	\ENDIF
\ENDIF
\ENDIF
\ENDFOR
\STATE {\bf return} $\mathcal{T}_{Q\_found}$ 
\end{algorithmic}
\end{algorithm}
\setlength{\textfloatsep}{2mm}

\vspace{2mm}

\noindent{\bf (T6) The Recursive Main Algorithm.} Given a vertex set $S$, and the set of vertices $ext(S)$ to extend $S$ into valid quasi-cliques, Algorithm~\ref{algo:main} shows our algorithm for mining valid quasi-cliques extended from $S$ (including $G(S)$ itself). This algorithm is recursive (see Line~21) and starts by calling {\em recursive\_mine}($v$, $\mathbb{B}_{>v}(v)$, $\gamma$, $\tau_{size}$) on every $v\in V$ where $\mathbb{B}_{>v}(v)$ denotes those vertices in $\mathbb{B}(v)$ whose IDs are larger than $v$, as according to Figure~\ref{qc}, we should not consider the other vertices in $\mathbb{B}(v)$ to avoid double counting.

Recall from (P7) that we have a degenerate cover-vertex pruning method that finds the vertex $v_{max}$ with the maximum degree, so that any quasi-clique generated from only $v_{max}$'s neighbors cannot be maximal (as it can be extended with $v_{max}$). To utilize this pruning rule, we need to recode the vertex IDs so that $v_{max}$ has ID 0, while vertices of $N(v_{max})$ have larger IDs than all other vertices, i.e., they are listed at the end in the first level of the set-enumeration tree illustrated in Figure~\ref{set_enum} (as they only extend with vertices in $N(v_{max})$).
If we enable ID recoding, {\em recursive\_mine}($v$, $\mathbb{B}_{>v}(v)$, $\gamma$, $\tau_{size}$) only needs to be called on every $v\in V-N(v_{max})$.

Algorithm~\ref{algo:main} keeps a boolean tag $\mathcal{T}_{Q\_found}$ to return (see Line~26), which indicates whether some valid quasi-clique $Q$ extended from $S$ (but $Q\neq S$) is found. Line~1 initializes $\mathcal{T}_{Q\_found}$ as $false$, but it will be set as $true$ if any valid quasi-clique $Q$ is found.

Algorithm~\ref{algo:main} examines $S$, and it decomposes this problem into the subproblems of examining $S'=S\cup v$ for all $v\in ext(S)$, as described by the loop in Line~5. Before the loop, we first apply cover vertex pruning as described in (P7) of Section~\ref{ssec:prune}: for the selected cover vertex $u\in ext(S)$ (Line~2), we move its cover set $C_S(u)$ to the tail of the vertex list of $ext(S)$ (Line~4), so that the loop in Line~5 ends when $v$ reaches a vertex in $C_S(u)$. This is correct since Line~11 excludes an already examined $v$ from $ext(S)$ and so the loop in Line~5 with $v$ scanning $C_S(u)$ corresponds to the case of extending $S'$ using $ext(S')\subseteq ext(S)\subseteq C_S(u)$ (see Lines~11-12) which should be pruned. If we cannot find a cover vertex (see Line~2), then Line~5 iterates over all vertices of $ext(S)$.

Note that in Line~2, we need to check every $u\in ext(S)$ and keep the current maximum value of $|C_S(u)|$; if for a vertex $u$ we find when evaluating Eq~(\ref{eq:cover}) that $|N_{ext(S)}(u)|$ is already less than the current maximum, $u$ can be skipped without further checking $N(v)$ for $v\in S-N(u)$.

Now let us focus on the loop body in Lines~6--25. Line~6 first checks if $S$ extended with every vertex not yet considered in $ext(S)$ can generate a subgraph larger than $\tau_{size}$ (note that already-consid-%===== note the split
ered vertices $v$ are removed from $ext(S)$ by Line~11 in previous iterations which automatically guarantees the ID-based deduplication illustrated in Fig~\ref{set_enum}); if so, current and future iterations cannot generate a valid quasi-clique and are thus pruned, and Line~7 directly returns $\mathcal{T}_{Q\_found}$ which indicates if a valid quasi-clique is found by previous iterations.

For a vertex $v\in ext(S)$, the current iteration creates $S'=S\cup v$ for examination in Line~11. Before that, Lines~8--10 first checks if $S$ extended with the entire current $ext(S)$ creates a valid quasi-clique; if so, this is a maximal one and is thus output in Line~9, and further examination can be skipped (Line~10). This pruning is called the lookahead technique in~\cite{quick}. Note that $G(S\cup ext(S))$ must satisfy the size threshold requirement as Line~6 is passed, and thus Line~8 does not need to check that condition again.

Now assume that lookahead technique does not prune the search, then Line~11 creates $S'=S\cup v$ (the implementation actually updates $G(S)$ into $G(S')$), and excludes $v$ from $ext(S)$. The latter also has a side effect of excluding $v$ from $ext(S)$ of all subsequent iterations, which matches exactly how the set-enumeration tree illustrated in Figure~\ref{set_enum} avoids generating redundant nodes for $S$.

Then, Line~12 shrinks $ext(S)$ into $ext(S')$ by ruling out vertices more than 2 hops away from $v$ according to (P1) of Section~\ref{ssec:prune}, which is then used to extend $S'$. If $ext(S')=\emptyset$ after shrinking, then $S'$ has nothing to extend, but $G(S')$ itself may still be a candidate for a valid quasi-clique and is thus examined in Lines~14--16. We remark that \cite{quick}'s original Quick algorithm misses this check and thus may miss results.

If $ext(S')\neq\emptyset$, Line~18 then calls {\em iterative\_bounding}($S'$, $ext(S')$, $\gamma$, $\tau_{size}$) (i.e., Algorithm~\ref{algo:iterative}) to apply the pruning rules. Recall that the function either returns $\mathcal{T}_{pruned}=false$ indicating that we need to further extend $S'$ using its shrunk $ext(S')$; or it returns $\mathcal{T}_{pruned}=true$ to indicate that the extensions of $S'$ should be pruned, which will also output $G(S')$ if it is a valid quasi-clique (see Lines~22--25 and 14--16 in Algorithm~\ref{algo:iterative}).

If Line~18 decides that $S'$ can be further extended (i.e., $\mathcal{T}_{pruned}=false$) and extending $S'$ with all vertices in $ext(S')$ still has the hope of generating a subgraph with $\tau_{size}$ vertices or larger (Line~20), we then recursively call our algorithm to examine $S'$ in Line~21, which returns $\mathcal{T}_{found}$ indicating if some valid maximal quasi-cliques $Q\supset S'$ are found (and output). If $\mathcal{T}_{found}=true$, Line~22 will update the return value $\mathcal{T}_{Q\_found}$ as $true$, but $G(S')$ is not maximal.
Otherwise (i.e., $\mathcal{T}_{found}=false$), $G(S')$ is a candidate for a valid maximal quasi-clique and is thus examined in Lines~23--25.

Finally, as in Quick, Quick+ also requires a postprocessing step~to remove non-maximal quasi-cliques from the results of Algorithm~\ref{algo:main}. Also, we only run Quick+ after the input graph is shrunk by the $k$-core pruning of (P2). To summarize, besides Quick's cover vertex pruning, Quick+ also supports a top-level degenerate pruning using $v_{max}$ as mentioned in (P7), and checks if $G(S')$ is a valid quasi-clique when $ext(S')$ becomes empty after the diameter-based pruning of (P1). Quick misses this check and may miss results.

Additionally, we find that the vertex order in $ext(S)$ matters (Algorithm~\ref{algo:main} Line 7) and can significantly impact the running time. To maximize the success probability of the lookahead technique in Lines~8--10 of Algorithm~\ref{algo:main} that effectively prunes the entire $T_S$, we propose to sort the vertices in $ext(S)$ in ascending order of $d_S(v)$ (tie broken by $d_{ext(S)}(v)$) following~\cite{Bayardo98} so that high-degree vertices tend to appear in $ext(S)$ of more set-enumeration tree nodes.

\section{Parallel G-thinker Algorithms}\label{sec:algo}
\noindent{\bf Divide-and-Conquer Algorithm.} 
We next adapt Algorithm~\ref{algo:main} to run on the redesigned G-thinker, where a big task (judged by $|ext(S)|$) is divided into smaller subtasks for concurrent processing. If a task $t=\langle S, ext(S)\rangle$ is spawned from a vertex $v$, we only pull vertices with ID $>v$ into $S$ and $ext(S)$, which avoids redundancy (recall Figure~\ref{set_enum}). 
Whenever we say a task $t$ pulls a vertex $u$ hereafter, we implicitly mean that we only do so when $u>v$ that spawns $t$.

Recall from Theorem~\ref{prune:size} that any vertex with degree less than $k=\lceil\gamma\cdot(\tau_{size}-1)\rceil$ cannot be in a valid quasi-clique. %While Quick~\cite{quick} does not utilize this simple pruning, we find that applying this pruning can speed up mining significantly. 
Therefore, our implementation shrinks the subgraph $g$ of any task $t$ into the $k$-core of $g$ before mining. We adopt the $O(|E|)$-time peeling algorithm~\cite{kcore_peel} for this purpose.

Recall that users write a G-thinker program by implementing two UDFs, and here we spawn a task from each vertex $v$ by pulling vertices within two hops from $v$, to construct $v$'s two-hop ego-network from $\mathbb{B}(v)$. Of course, we only pull vertices with ID $> v$ here and prune vertices with degree $< k$, so that the resulting subgraph to mine is effectively a $k$-core.
Moreover, if we would like to use the initial degenerate cover-vertex pruning described in (P7), we need to recode the vertex IDs. Specifically, we load the ID and degree (or, $|N(v)|$) into memory, find $v_{max}$ and assign it ID 0, and assign vertices in $N(v)$ IDs $(|V|-|N(v)|), \cdots, (|V|-2), (|V|-1)$; for the other vertices, we sort them in ascending order of degree and assign IDs $1, 2, \cdots, (|V|-|N(v)|-1)$ to allow effective look-ahead pruning. We can then use the old-to-new ID mapping table to recode the IDs in the adjacency lists.

\begin{algorithm}[t]
\caption{UDF task\_spawn$(v)$}\label{algo:spawn}
Define $k=\lceil\gamma\cdot(\tau_{size}-1)\rceil$.
\begin{algorithmic}[1]
\IF{$|N(v)|\geq k$}
\STATE Create a task $t$
\STATE $t.iteration \gets 1$
\STATE $t.root \gets v$ \ \ \{spawning vertex\}
\STATE $t.S \gets v$
\FOR{{\bf each} $u\in N(v)$ with $u>v$}
\STATE $t.pull(u)$
\ENDFOR
\STATE {\em add\_task}$(t)$
\ENDIF
\end{algorithmic}
\end{algorithm}

We first consider UDF task\_spawn$(v)$ as given by Algorithm~\ref{algo:spawn}. Specifically, we only spawn a task for a vertex $v$ if its degree $\geq k$ (Lines~1--2). The task is initialized to be at iteration~1 (Line~3, to be used by Line~1 of Algorithm~\ref{algo:compute} later), with spawning vertex $v$ (Line~4, recorded so that future iterations only pull vertices larger than it) and $S=\{v\}$ (Line~5). The task then pulls the adjacency lists of $v$'s neighbors (Lines~6--7) and gets itself added to the system for further processing (Line~8).

\begin{algorithm}[t]
\caption{UDF compute$(t, frontier)$}\label{algo:compute}
Define $k=\lceil\gamma\cdot(\tau_{size}-1)\rceil$
\begin{algorithmic}[1]
\IF{$t.iteration=1$}
\STATE iteration\_1$(t, frontier)$
\ELSIF{$t.iteration=2$}
\STATE iteration\_2$(t, frontier)$
\ELSE
\STATE iteration\_3$(t)$
\ENDIF
\end{algorithmic}
\end{algorithm}

Next, UDF compute$(t, frontier)$ runs 3 iterations as shown in Algorithm~\ref{algo:compute}. The first iteration adds the pulled first-hop neighbors of $v$ into the task's subgraph $t.g$ with proper size-threshold based pruning, and then pulls the second-hop neighbors of $v$. The second iteration adds the pulled second-hop neighbors into $t.g$ with proper size-threshold based pruning, and since $t$ does not need to pull any more vertices, $t$ will not be suspended but rather run the third iteration immediately. The third iteration then mines quasi-cliques from $t.g$ using our recursive algorithm (Algorithm~\ref{algo:main}), but if the task is big, it will create smaller subtasks for concurrent computation. We next present the algorithms of Iterations~1--3, respectively.

\begin{algorithm}[t]
\caption{iteration\_1$(t, frontier)$}\label{algo:it1}
\begin{algorithmic}[1]
\STATE $v\gets t.root$
\STATE $t.\mathbb{N}\gets V(frontier)\cup v$
\STATE $V_1\gets$ vertices in $frontier$ with degree $\geq k$
\STATE $V_2\gets$ vertices in $frontier$ with degree $< k$ 
\STATE Construct subgraph $t.g$ to include vertices $V_1\cup v$
\FOR{{\bf each} vertex $u$ in $t.g$}
\FOR{{\bf each} vertex $w\in N(u)$}
\IF{$w\geq v$ and $w\not\in V_2$}
\STATE Add $w$ to $u$'s adjacency list in $t.g$
\ENDIF
\ENDFOR
\ENDFOR
\STATE $t.g\gets$ $k$-core$(t.g)$
\STATE{\bf{if} $v\not\in V(t.g)$ {\bf then}\ \ \ \ \ \ {\bf return}} $false$
\FOR{{\bf each} vertex $u$ in $t.g$}
\FOR{{\bf each} vertex $w\in N(u)$}
\IF{$w\geq v$ and $w\not\in t.\mathbb{N}$}
\STATE $t.pull(w)$
\ENDIF
\ENDFOR
\ENDFOR
\STATE $t.iteration\gets 2$
\STATE {\bf return} $true$\ \ \ \ \{continue Iteration~2\}
\end{algorithmic}
\end{algorithm}

The algorithm of Iteration~1 is given by Algorithm~\ref{algo:it1}, where $v$ is the task-spawning vertex (Line~1). In Line~2, we collect $v$ and its neighbors already pulled inside $frontier$ into a set $\mathbb{N}$ which records all vertices within 1 hop to $v$, which will be used in Line~14 to filter them when pulling the second-hop neighbors. Then, we divide the pulled vertices into two sets: $V_1$ containing those with degree $\geq k$ (Line~3) and $V_2$ containing those with degree $<k$ (Line~4) which should be pruned.

We then construct the task's subgraph $t.g$ to include vertices $V_1\cup v$ in Line~5, and Lines~6--9 prune the adjacency lists of vertices in $t.g$ by removing a destination $w$ if it is smaller than $v$ or if it is in $V_2$ (i.e., has degree $<k$). Note that the adjacency list of a vertex $u$ in $t.g$ may contain a destination $w$ that is 2 hops away from $v$; since we do not have $N(w)$ yet, we cannot compare the degree of $w$ with $k$ for pruning.

After the adjacency list pruning, a vertex $u$ in $t.g$ may have its adjacency list shorter than $k$, and therefore we run the peeling algorithm over $t.g$ to shrink $t.g$ into its $k$-core (Line~10); here, a destination $w$ that is 2 hops away from $v$ in an adjacency list stays untouched and we only remove vertices in $V_1\cup v$ (though $w$ is counted for degree checking). If $v$ becomes pruned from $t.g$, compute$(t, frontier)$ returns $false$ to terminate $t$ since $t$ is to find quasi-cliques that contain $v$ (Line~11).

Next, Lines~12--15 pull all second-hop vertices (away from $v$) in the adjacency lists of vertices of $t.g$. Note that Line~14 makes sure that a vertex $w$ to pull is not within 1 hop (i.e., $w\not\in\mathbb{N}$) and $w>v$. In the actual implementation, we add all such vertices into a set and then pull them to avoid pulling the same vertex twice when checking $N(v_a)$ and $N(v_b)$ of different $v_a,v_b\in V(t.g)$. Finally, Line~16 sets $t.iteration$ to 2 so that when compute$(t, frontier)$ is called again, it will execute iteration\_2$(t, frontier)$.

\begin{algorithm}[t]
\caption{iteration\_2$(t, frontier)$}\label{algo:it2}
\begin{algorithmic}[1]
\STATE $v\gets t.root$
\STATE $\mathbb{B}\gets V(frontier)\cup t.\mathbb{N}$
\FOR{{\bf each} vertex $u$ in $frontier$}
\IF{$|N(u)|\geq k$}
\STATE Add $u$ into $t.g$
\FOR{{\bf each} vertex $w\in N(u)$}
\IF{$w\geq v$ and $w\in\mathbb{B}$}
\STATE Add $w$ to $u$'s adjacency list in $t.g$
\ENDIF
\ENDFOR
\ENDIF
\ENDFOR
\STATE $t.g\gets$ $k$-core$(t.g)$
\STATE{\bf{if} $v\not\in t.g$ {\bf then}\ \ \ \ \ \ {\bf return}} $false$
\STATE $t.iteration\gets 3$
\STATE $t.S\gets \{v\}$,\ \ \ \ $t.ext(S)\gets V(g)-v$
\STATE {\bf return} $true$\ \ \ \ \{continue Iteration~3\}
\end{algorithmic}
\end{algorithm}

Algorithm~\ref{algo:it2} gives the computation in Iteration~2. Line~2 first collects $\mathbb{B}$ as all vertices within 2 hops from $v$, which is used in Line~7 to filter out adjacency list items of those vertices in $frontier$ that are 3 hops from $v$. Recall that $t.\mathbb{N}$ is collected in Line~2 of Algorithm~\ref{algo:it1} to contain the vertices within 1 hop from $v$, and that we are finding $\gamma$-quasi-cliques with $\gamma\geq 0.5$ and hence the quasi-clique diameter is upper bounded by 2.

Lines~3--8 then add all second-hop vertices in $frontier$ with degree $\geq k$ into $t.g$ (Lines~4--5), but prunes a destination $w$ in an adjacency list if $w<v$ or $w$ is not within 2 hops from $v$ (i.e., $w\not\in\mathbb{B}$). Since adjacency lists may become shorter than $k$ after pruning, Line~9 then shrinks $t.g$ into its $k$-core, and if $v$ is no longer in $t.g$, compute$(t, frontier)$ returns $false$ to terminate the task (Line~10). Finally, Line~11 sets $t.iteration$ to 3 so that when compute$(t, frontier)$ is called again, it will execute iteration\_3$(t)$ which we present next. Since $t$ does not pull any vertex in Iteration~2, G-thinker will schedule $t$ to run Iteration~3 right away.

\begin{algorithm}[t]
\caption{iteration\_3$(t)$}\label{algo:it3}
\begin{algorithmic}[1]
\IF{$|t.ext(S)|\leq\tau_{split}$}
\STATE {\em recursive\_mine}($t.S$, $t.ext(S)$, $\gamma$, $\tau_{size}$)
\ELSE
\STATE  Find cover vertex $u\in t.ext(S)$ with the largest $C_S(u)$
\STATE \{If not found, $C_S(u)\gets\emptyset$\}
\STATE Move vertices of $C_S(u)$ to the tail of vertex list $t.ext(S)$
\FOR{{\bf each} vertex $v$ in the sub-list $\left(t.ext(S)-C_S(u)\right)$}
\STATE{{\bf if} $|t.S|+|t.ext(S)|<\tau_{size}$ {\bf then}\ \ \ \ \ \ \ \ {\bf return} $false$}
\IF{$G(t.S\cup t.ext(S))$ is a $\gamma$-quasi-clique}
\STATE Append $t.S\cup t.ext(S)$ to the result file
\STATE {\bf return} $false$
\ENDIF
\STATE Create a task $t'$
\STATE $t'.S\gets t.S\cup v$,\ \ $t.ext(S)\gets t.ext(S)-v$
\STATE $t'.ext(S)\gets t.ext(S)\cap\mathbb{B}(v)$
\IF{$|t'.S|\geq\tau_{size}$ {\bf and} $G(t'.S)$ is a $\gamma$-quasi-clique}
	\STATE Append $t'.S$ to the result file
\ENDIF
\STATE $\mathcal{T}_{pruned}\gets$ {\em iterative\_bounding}($t'.S$, $t'.ext(S)$, $\gamma$, $\tau_{size}$) % toGuimu:: no need to pass lb and ub !!! update code
\IF{$\mathcal{T}_{pruned}=false$ {\bf and} $|t'.S|+|t'.ext(S)|\geq\tau_{size}$}
\STATE $t'.g\gets$ subgraph of $t.g$ induced by $t'.S\cup t'.ext(S)$
\STATE $t'.iteration\gets$ 3
\STATE {\em add\_task}$(t')$
\ELSE
\STATE Delete $t'$
\ENDIF
\ENDFOR
\ENDIF
\STATE {\bf return} $false$\ \ \ \ \{task is done\}
\end{algorithmic}
\end{algorithm}

Now that $t.g$ contains the $k$-core of the spawning vertex's 2-hop ego-network, Algorithm~\ref{algo:it3} gives the computation in Iteration~3 which mines quasi-cliques from $t.g$. Since the task can be prohibitive when $t.g$ and $ext(S)$ are big, we only directly process the task using Algorithm~\ref{algo:main} when $|ext(S)|$ is small enough (Lines~1--2); otherwise, we divide it into smaller subtasks to be scheduled for further processing (Lines~3--23), though the execution flow is very similar to Algorithm~\ref{algo:main}.

Recall that Algorithm~\ref{algo:main} is recursive where Line~21 extends $S$ with another vertex $v\in ext(S)$ for recursive processing, and here we will instead create a new task $t'$ with $t'.S=t.S\cup v$ (Lines~12--13). However, we still want to apply all our pruning rules to see if $t'$ can be pruned first; if not, we will add $t'$ to the system (Line~21) with $t'.iteration=3$ so that when $t'$ is scheduled for processing, it will directly enter iteration\_3$(t')$. Here, we shrink $t'$'s subgraph to be induced by $t'.S\cup t'.ext(S)$ so that the subtask is on a smaller graph, and since $t'.ext(S)$ shrinks (due to pruning) at each recursion and $t'.g$ also shrinks, the computation cost becomes smaller.

Another difference is with Line~23 of Algorithm~\ref{algo:main}, where we only check if $G(S')$ is a valid quasi-clique when $\mathcal{T}_{found}=false$, i.e., the recursive call in Line~21 verifies that $S'$ fails to be extended to produce a valid quasi-clique. In Algorithm~\ref{algo:it3}, however, the recursive call now becomes an independent task $t'$ in Line 12, and the current task $t$ has no clue of its results. Therefore, %if the pruning rule fails to prune $t'$, 
we check if $G(t'.S)$ is a valid quasi-clique right away in Line~15 in order to not miss it. A subtask may later find a larger quasi-clique containing $t'.S$, rendering $G(t'.S)$ not maximal, and we resort to the postprocessing phase to remove non-maximal quasi-cliques.

Due to cover-vertex pruning, a task $t$ can generate at most $|t.ext(S)\\%===== note the split
-C_S(u)|$ subtasks (see Line~7) where $u$ is the cover vertex found.

\begin{figure}[t]
\centering
\includegraphics[width=\columnwidth]{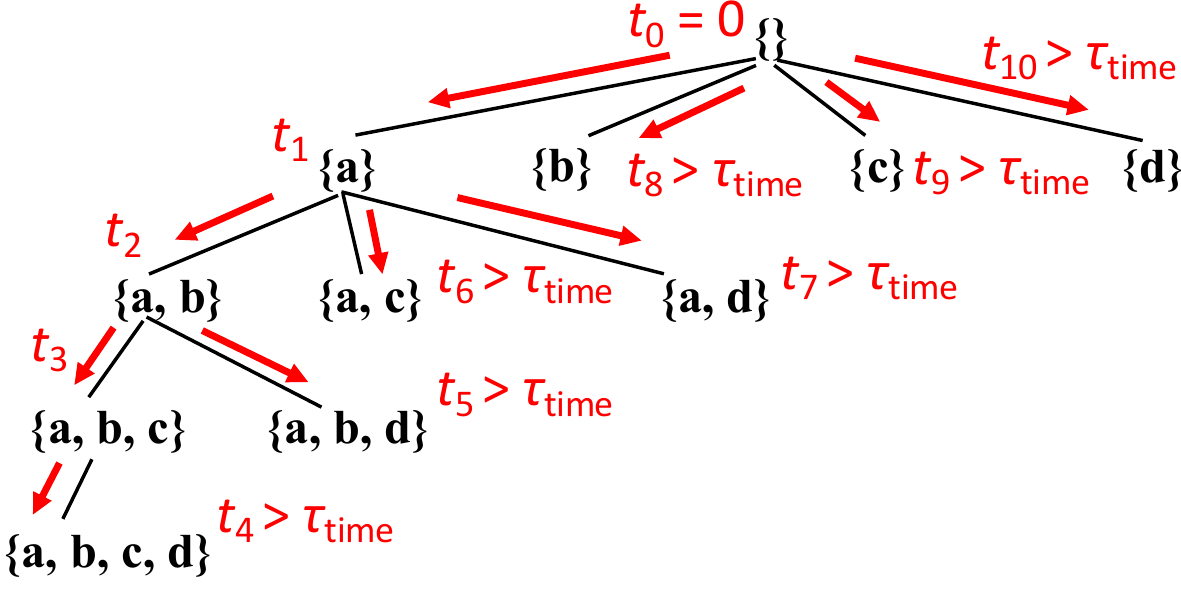}
\caption{Timeout-Based Divide and Conquer}\label{time_delay}
\end{figure}

\vspace{2mm}
\noindent{\bf Timeout-Based Task Decomposition.} So far, we decompose a task $\langle S, ext(S)\rangle$ as long as $|ext(S)|>\tau_{split}$ but due to the large time variance caused by the many pruning rules, some of those tasks might not be worth splitting as they are fast to compute, while others might not be sufficiently decomposed and need an even smaller $\tau_{split}$. We, therefore, improve our UDF {\em compute}($t$, {\em frontier}) further by a timeout strategy where we guarantee that each task spends at least a duration of $\tau_{time}$ on the actual mining of its subgraph by backtracking (which does not materialize subgraphs) before dividing the remaining workloads into subtasks (which needs to materialize their subgraphs). Figure~\ref{time_delay} illustrates how our algorithm works. The algorithm recursively expands the set-enumeration tree in depth-first order, processing 3 tasks until entering $\{a,b,c,d\}$ for which we find the entry time $t_4$ times out; we then wrap $\{a,b,c,d\}$ as a subtask to be added to our system, and backtrack the upper-level nodes to also add them as subtasks (due to timeout). Note that subtasks are at different granularity and not over-decomposed.

\begin{table*}[!t]
\centering
\caption{Graph Datasets}\label{data}
\vspace{2mm}
\includegraphics[width=2.1\columnwidth]{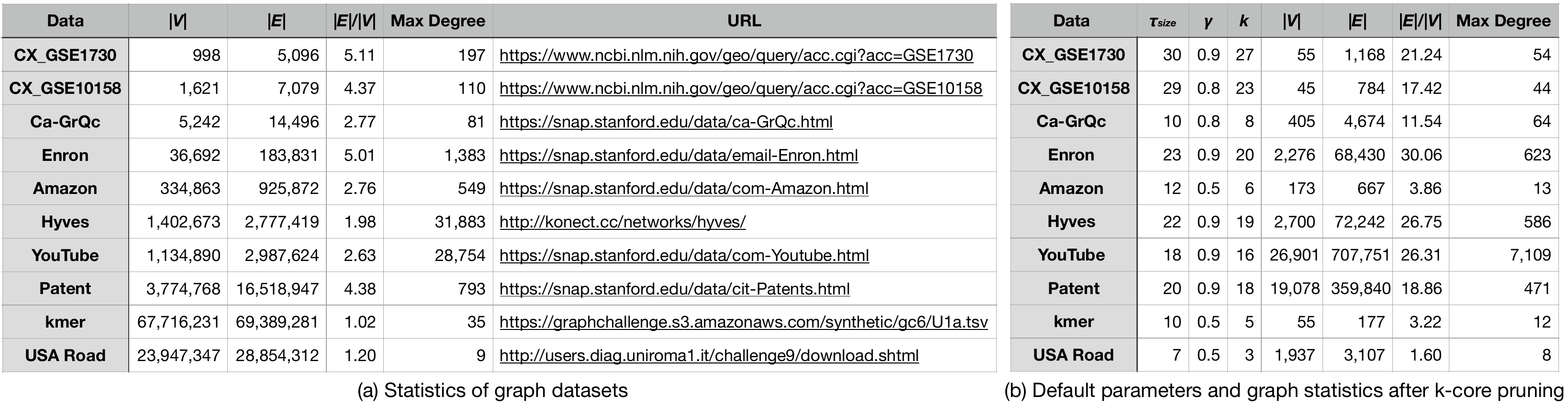}
\end{table*}

\begin{algorithm}[!t]
\caption{iteration\_3$(t)$ with Timeout Strategy}\label{algo:it3_2}
\begin{algorithmic}[1]
\STATE {\em time\_delayed}($t.S$, $t.ext(S)$, {\em initial\_time})
\STATE {\bf return} $false$\ \ \ \ \{task is done\}
\end{algorithmic}
\end{algorithm}

\begin{algorithm}[!t]
\caption{{\em time\_delayed}($S$, $ext(S)$, {\em initial\_time})}\label{algo:tddq}
\begin{algorithmic}[1]
\STATE  $\mathcal{T}_{Q\_found}\gets false$
\STATE  Find cover vertex $u\in ext(S)$ with the largest $C_S(u)$
\STATE \{If not found, $C_S(u)\gets\emptyset$\}
\STATE Move vertices of $C_S(u)$ to the tail of the vertex list of $ext(S)$
\FOR{{\bf each} vertex $v$ in the sub-list $\left(ext(S)-C_S(u)\right)$}
\STATE {\bf if} $|S|+|ext(S)|<\tau_{size}$ {\bf then:\ \ \ \  return} $false$
\IF{$G(S\cup ext(S))$ is a $\gamma$-quasi-clique}
\STATE Append $S\cup ext(S)$ to the result file;\ \ \ \ {\bf return} $false$
\ENDIF
\STATE $S'\gets S\cup v$,\ \ $ext(S)\gets ext(S)-v$
\STATE $ext(S')\gets ext(S)\cap\mathbb{B}(v)$
\IF{$ext(S')=\emptyset$}
	\IF{$|S'|\geq\tau_{size}$ {\bf and} $G(S')$ is a $\gamma$-quasi-clique}
	\STATE  $\mathcal{T}_{Q\_found}\gets true$
	\STATE Append $S'$ to the result file
	\ENDIF
\ELSE
\STATE $\mathcal{T}_{pruned}\gets$ {\em iterative\_bounding}($S'$, $ext(S')$, $\gamma$, $\tau_{size}$) % toGuimu:: no need to pass lb and ub !!! update code
\STATE \{here, $ext(S')$ is Type-I-pruned and $ext(S')\neq\emptyset$\}
\IF{{\em current\_time $-$ initial\_time} $> \tau_{time}$} % ---> note the addition
\IF{$\mathcal{T}_{pruned}=false$ {\bf and} $|S'|+|ext(S')|\geq\tau_{size}$}
\STATE Create a task $t'$;\ \ \ $t'.S\gets S'$
\STATE $t'.ext(S)\gets ext(S')$;\ \ \ $t'.iteration\gets$ 3
\STATE {\em add\_task}$(t')$
\ENDIF
\IF{$|t'.S|\geq\tau_{size}$ {\bf and} $G(t'.S)$ is a $\gamma$-quasi-clique} % ---> note the addition
	\STATE Append $t'.S$ to the result file
\ENDIF
\ELSIF{$\mathcal{T}_{pruned}=false$ {\bf and} $|S'|+|ext(S')|\geq\tau_{size}$}
\STATE $\mathcal{T}_{found}\gets$ {\em time\_delayed}($S'$, $ext(S')$, {\em initial\_time})
\STATE $\mathcal{T}_{Q\_found}\gets \mathcal{T}_{Q\_found}$ {\bf or} $\mathcal{T}_{found}$
%------
\IF{$\mathcal{T}_{found}=false$ {\bf and} $|S'|\geq\tau_{size}$ {\bf and} $G(S')$ is a $\gamma$-quasi-clique}
\STATE  $\mathcal{T}_{Q\_found}\gets true$
\STATE Append $S'$ to the result file
\ENDIF
%------
\ENDIF
\ENDIF
\ENDFOR
\STATE {\bf return} $\mathcal{T}_{Q\_found}$
\end{algorithmic}
\end{algorithm}

With the timeout strategy, the third iteration of our UDF {\em compute}($t$, {\em frontier})  is given by Algorithm~\ref{algo:it3_2}. Line~1 calls our recursive backtracking function {\em time\_delayed}($S$, $ext(S)$, {\em inital\_time}) detailed in Algorithm~\ref{algo:tddq}, where {\em inital\_time} is the time when Iteration~3 begins. Line~2 then returns $false$ to terminate this task.

Algorithm~\ref{algo:tddq} now considers 2 cases. (1)~Lines~18--24: if timeout happens, we wrap $\langle S', ext(S')\rangle$ into a task $t'$ to be added for processing just like in Algorithm~\ref{algo:it3}, and since the current task cannot track whether $t'$ will find a valid quasi-clique that extends $S'$, we have to check if $G(S')$ itself is a valid quasi-clique (Lines~23--24)  in order not to miss it if it is maximal. (2)~Lines~25--30: we perform regular backtracking just like in Algorithm~\ref{algo:main}, where we recursively call {\em time\_delayed}(.) to process $\langle S', ext(S')\rangle$ in Line~26.

\section{Experiments}\label{sec:results}
This section reports our experiments. We have released the code of our redesigned G-thinker and quasi-clique algorithms on GitHub~\cite{code}.

\vspace{1mm}
\noindent {\bf Datasets.} We used 10 real graph datasets as Table~\ref{data}(a) shows: biological networks {\em CX\_GSE1730} and {\em CX\_GSE10158}, arXiv collaboration network {\em Ca-GrQc}, email communication network {\em Enron}, product co-purchasing network {\em Amazon}, social networks {\em Hyves} and {\em YouTube}, patent citation network {\em Patent}, protein $k$-mer graph {\em kmer} and USA road network {\em USA Road}. These graphs are selected to cover different graph type, size and degree characteristics.

\begin{table*}[t]
\centering
\caption{System Comparison for \protect\cite{gthinker}'s Experiments}\label{regular}
\includegraphics[width=2\columnwidth]{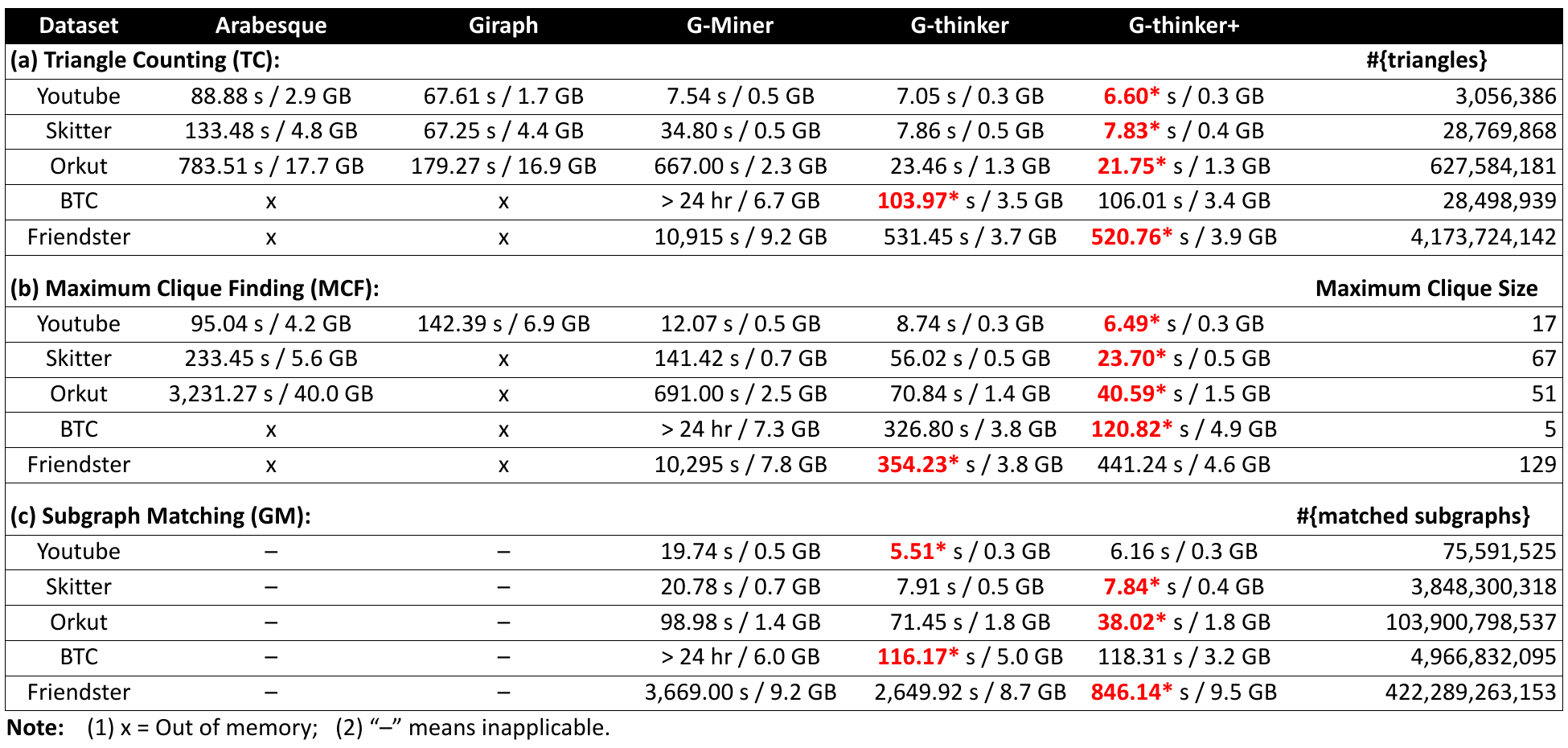}
\end{table*}

\vspace{1mm}
\noindent {\bf Algorithms \& Parameters.} We test our 3 algorithms: (1)~$\mathcal{A}_{base}$: one where a task spawned from a vertex mines its set-enumeration subtree in serial without decomposition; (2)~$\mathcal{A}_{split}$: one that splits tasks by comparing $|ext(S)|$ with size threshold $\tau_{split}$ (c.f.\ Algorithm~\ref{algo:it3}); (3) $\mathcal{A}_{time}$: one that splits tasks based on timeout threshold $\tau_{time}$ (c.f.\ Algorithm~\ref{algo:it3_2}). %Since $\mathcal{A}_{time}$ is superior, we also use it to test the scalability and effect of parameters $(\tau_{split}, \tau_{time})$. 
Note that even $\mathcal{A}_{time}$ and $\mathcal{A}_{base}$ need $\tau_{split}$ which is used by {\em add\_task}($t$) to decide whether a task $t$ is be put to the global queue or a local queue. We have repeated G-thinker paper~\cite{gthinker}'s experiments using our new engine as shown in Table~\ref{regular}, where column ``G-thinker'' refers to the old engine while ``G-thinker+'' refers to our redesigned engine. We observe improvements of our redesigned engine compared with using the old engine in most cases, and in the remaining cases the performance is similar; also, G-thinker is much faster than all prior systems.

%In the serial setting, we compare our recursive algorithm described in Section~\ref{sec:quick} (denoted by $\mathcal{A}_{serial}$) with the state-of-the-art Quick algorithm~\cite{quick} to show how the performance of $\mathcal{A}_{serial}$ improves over Quick. In the distributed setting, we test our 2 algorithms in Section~\ref{sec:algo}: one that splits tasks by comparing $|ext(S)|$ with size threshold $\tau_{split}$ (denoted by $\mathcal{A}_{split}$), the other that splits tasks based on timeout threshold $\tau_{time}$ (denoted by $\mathcal{A}_{time}$). Since $\mathcal{A}_{time}$ is superior, we also use it to test the scalability and effect of parameters $(\tau_{split}, \tau_{time})$. Note that even $\mathcal{A}_{time}$ needs $\tau_{split}$ which is used by {\em add\_task}($t$) to decide whether a task $t$ is be put to the global queue or a local queue.

\begin{table}[!t]
\centering
\caption{Effect of $\gamma$}\label{ratio}
\vspace{2mm}
\includegraphics[width=\columnwidth]{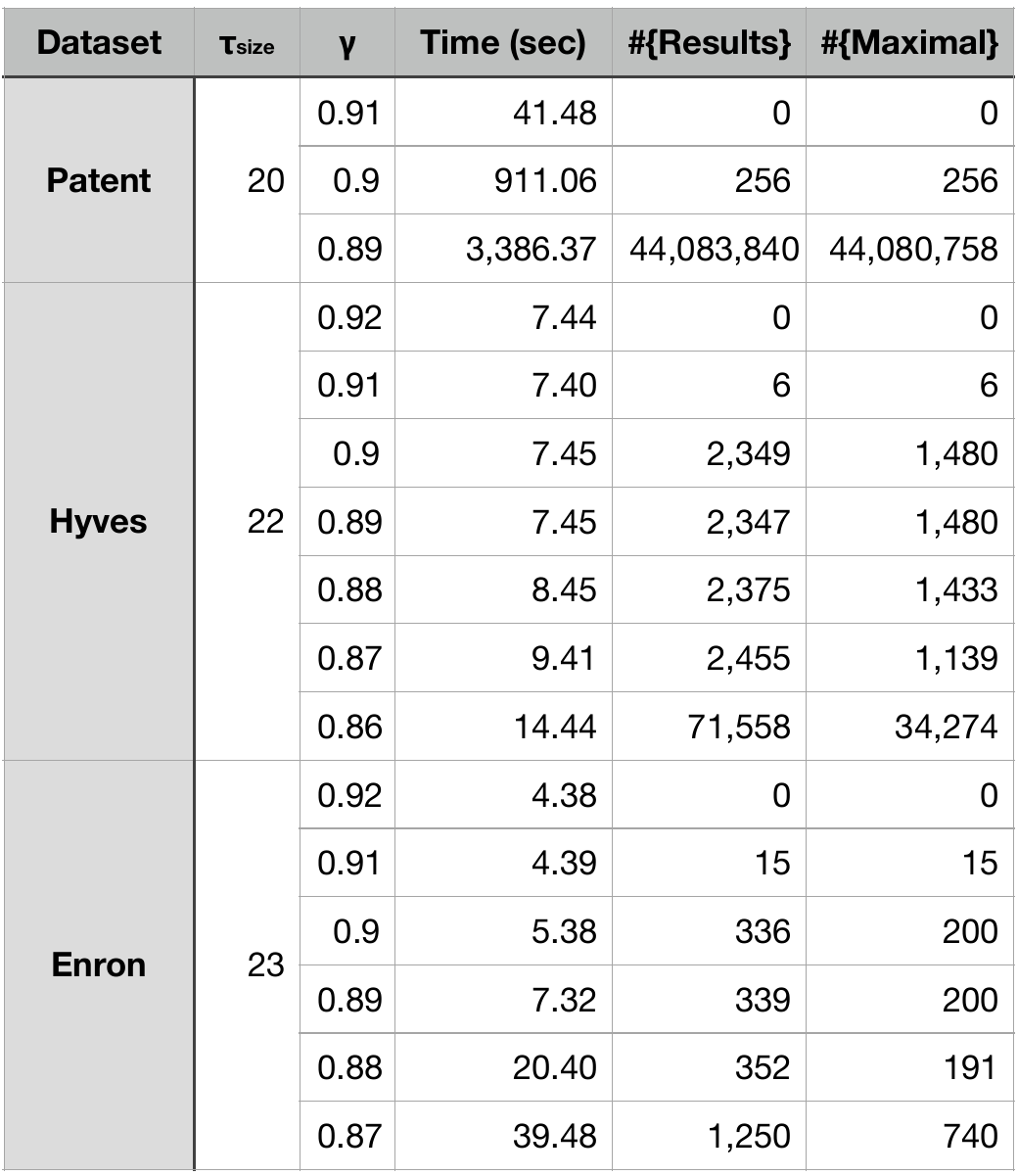}
\end{table}

\begin{table}[!t]
\centering
\caption{Effect of $\tau_{size}$}\label{minsize}
\vspace{2mm}
\includegraphics[width=\columnwidth]{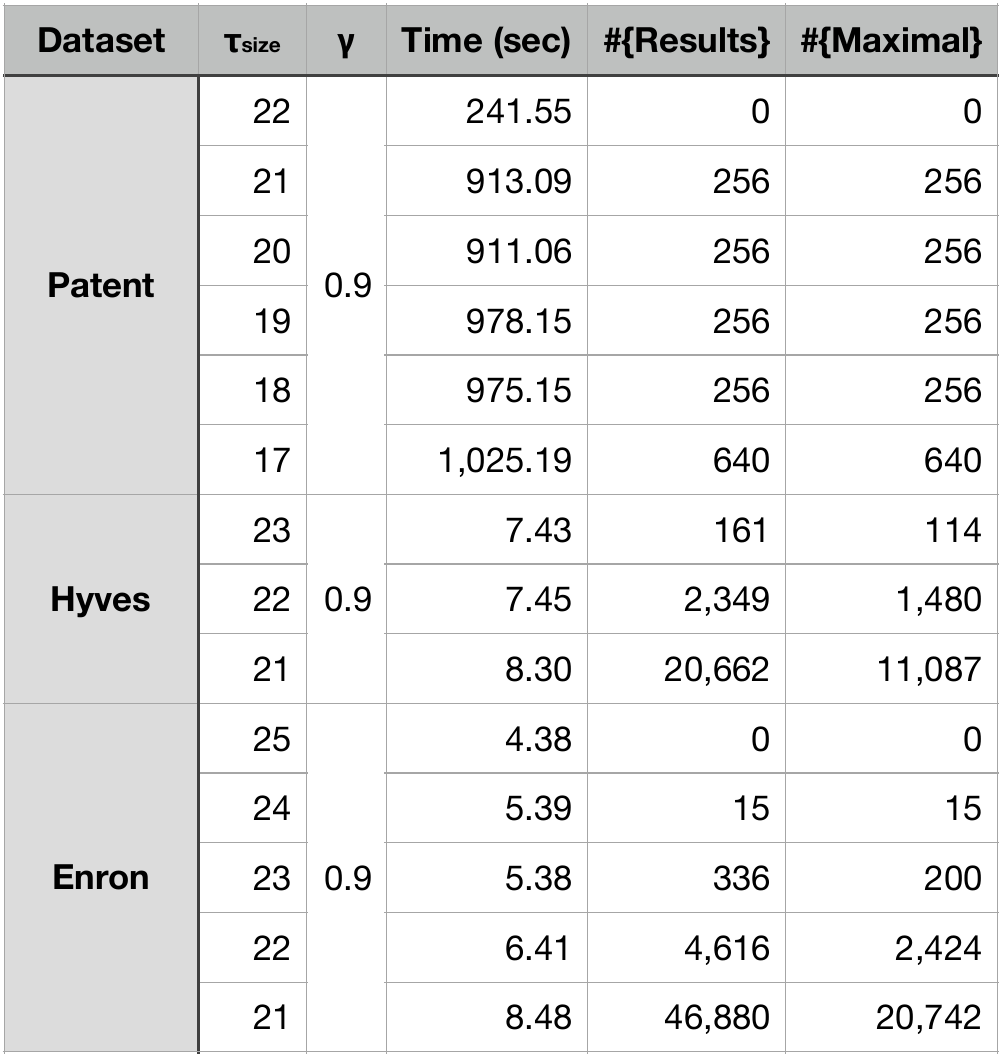}
\end{table}

We remark that $(\tau_{split}, \tau_{time})$ are algorithm parameters for parallelization. We also have the quasi-clique definition parameters $(\gamma, \tau_{size})$ (recall Definition~\ref{def}) at the first place. Interestingly, small value perturbations of $(\gamma, \tau_{size})$ can have significant impact on the result number: if the parameters are too large, there will be 0 results; while if the parameter is too small, there can be millions or even billions of results and run for a long time.
Table~\ref{ratio} (resp.\ Table~\ref{minsize}) shows the number of results (\#\{Results\}) found by $\mathcal{A}_{base}$ and the maximal ones after postprocessing (\#\{Maximal\}) along with the job time spent when we vary $\gamma$ (resp.\ $\tau_{size}$) slightly, where we can see that the result number is quite sensitive to the parameters. For example, when changing $(\gamma, \tau_{size})$ from $(20, 0.9)$ to $(20, 0.89)$ on {\em Patent}, the result number increases from 256 to over 44 million; and when changing $(\gamma, \tau_{size})$ from $(23, 0.9)$ to $(21, 0.9)$ on {\em Hyves}, the result number increases from 114 to 11,087.
Since our goal is to find the pool of largest valid subgraphs for prioritized examination, trials of different parameters are necessary and it is important that each trial should run an efficient algorithm like ours.

We also remark that the post-processing cost of removing non-maximal results is negligible compared with the job running time, by using a prefix tree organization of the result vertex sets. For example, post-processing the 256 results of {\em Patent} when $\gamma=0.9$ takes 0.002 seconds, while post-processing the over 44 million results when $\gamma=0.89$ takes 282.38 seconds.

Table~\ref{data}(b) shows the default values of $(\gamma, \tau_{size})$ for each dataset that we find to return a reasonable number of result subgraphs for human examination. Note that this immediately allows $k$-core pruning of the input graphs where $k=\lceil\gamma\cdot(\tau_{size}-1)\rceil$. We additionally prunes any vertex whose two-hop neighbor set has size $<\tau_{size}$, and statistics of the resulting dense graphs after pruning are shown in Table~\ref{data}(b) where {\em YouTube} and {\em Patent} are still large.

\vspace{1mm}
\noindent {\bf Experimental Setup.} All our experiments were conducted on a cluster of 16 machines each with 64 GB RAM, AMD EPYC 7281 CPU (16 cores and 32 threads) and 22TB disk.
All reported results were averaged over 3 repeated runs. G-thinker requires only a tiny portion of the available disk and RAM space in our experiments.

\begin{table*}[!t]
\centering
\caption{Performance of $\mathcal{A}_{base}$, $\mathcal{A}_{split}$ and $\mathcal{A}_{time}$ on All Datasets}\label{all}
\includegraphics[width=2.1\columnwidth]{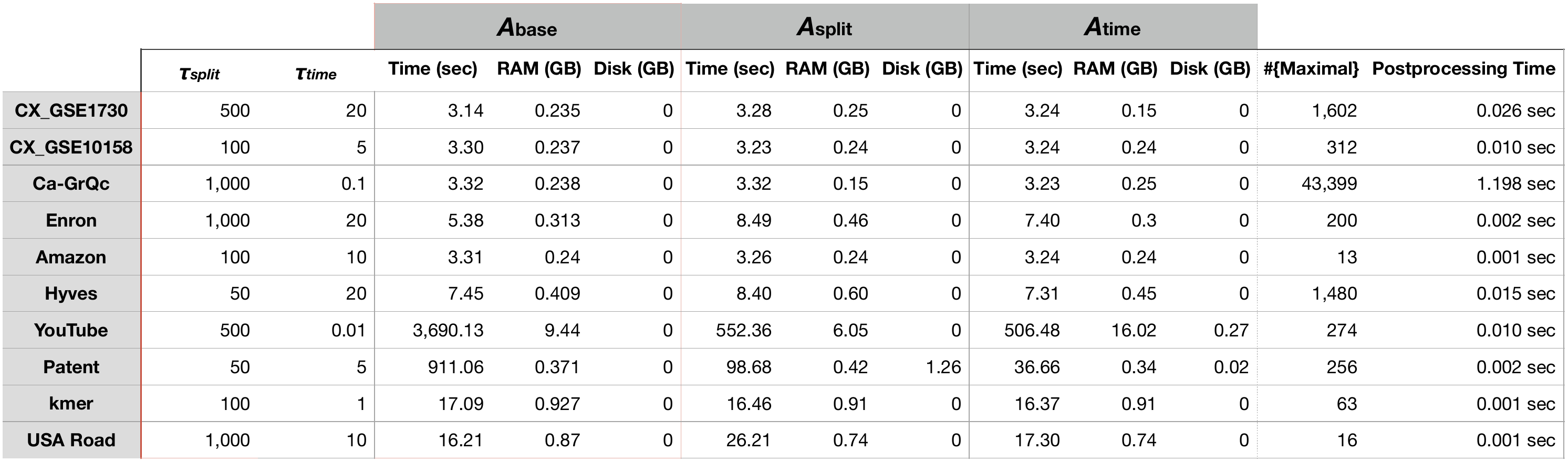}
\end{table*}

\begin{table*}[!t]
\centering
\caption{Effect of $\mathbf{(\tau_{split}, \tau_{time})}$ on All Datasets}\label{table34}
\includegraphics[width=1.8\columnwidth]{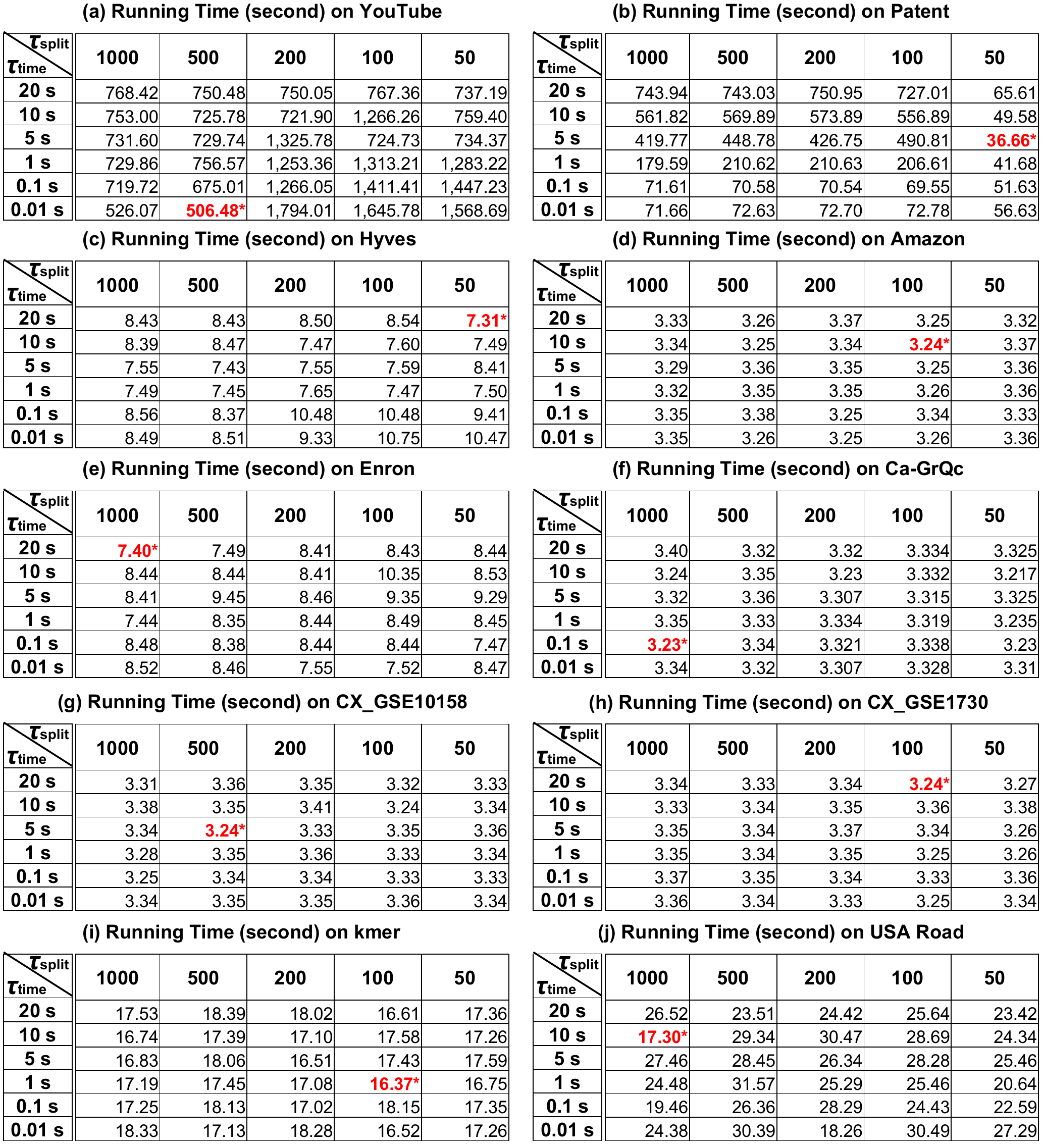}
\end{table*}

\begin{table*}[!t]
\centering
\caption{Performance of~\protect\cite{bigdata18}}\label{bigdata18cmp}
\vspace{2mm}
\includegraphics[width=1.2\columnwidth]{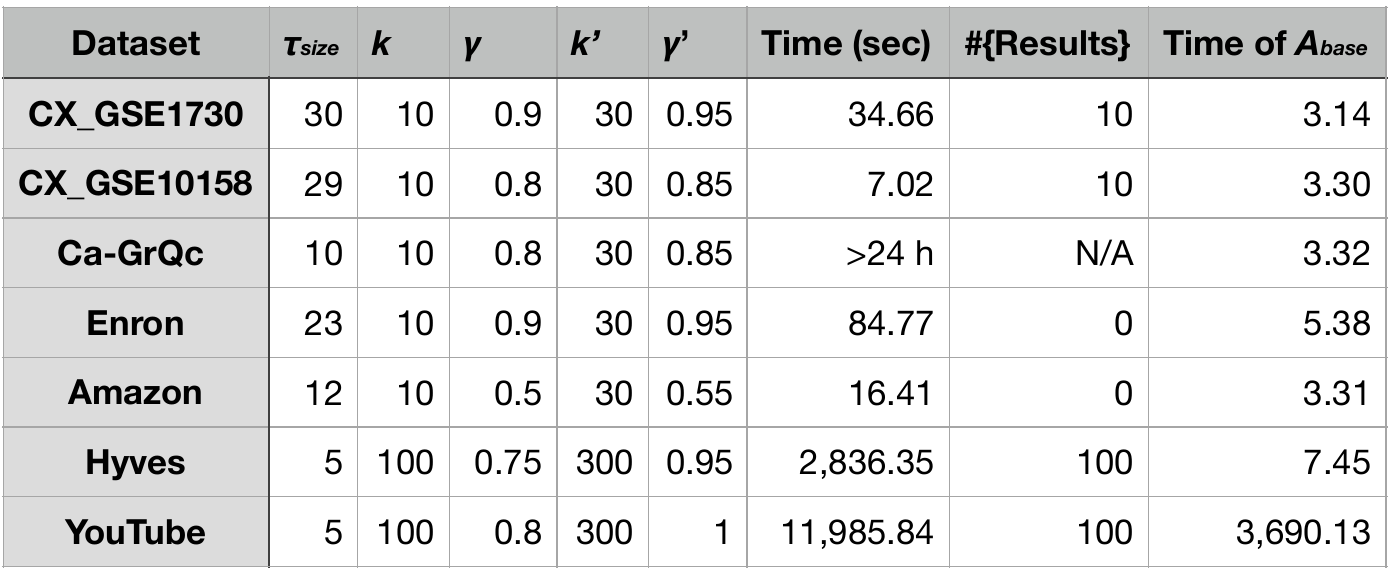}
\end{table*}

\begin{table*}[!t]
\centering
\caption{Scalability of $\mathcal{A}_{time}$}\label{table5}
\vspace{2mm}
\includegraphics[width=1.76\columnwidth]{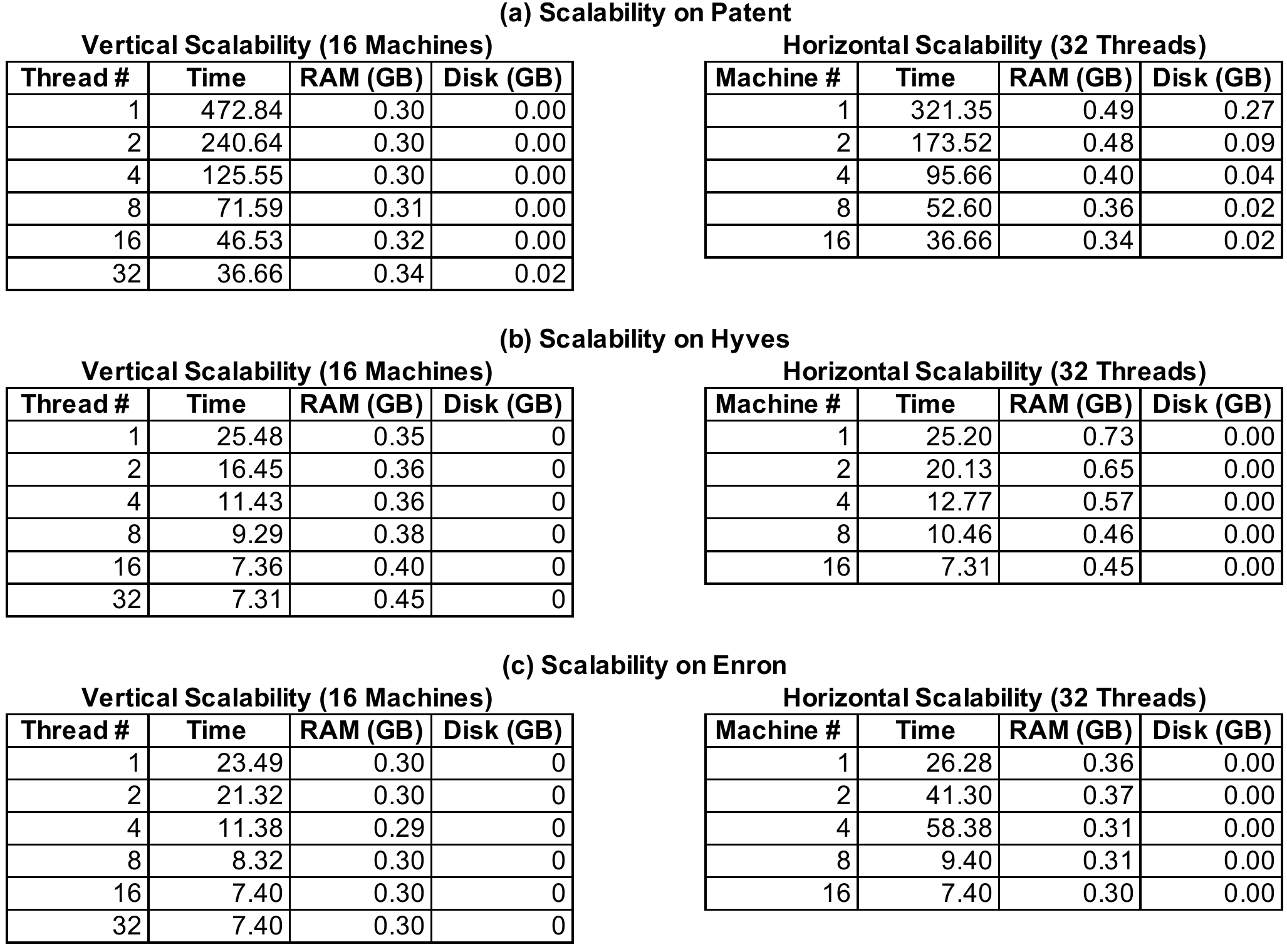}
\end{table*}

\vspace{1mm}
\noindent{\bf Comparison of $\mathcal{A}_{base}$, $\mathcal{A}_{split}$ and $\mathcal{A}_{time}$.} Table~\ref{all} shows the performance of our three G-thinker algorithm variants on all the datasets using the default $(\gamma, \tau_{size})$ values in Table~\ref{data}(b), and $(\tau_{split}, \\
\tau_{time})$ tuned to achieve the best performance. There, we report the job running time, and the peak memory and disk usage on a machine. We can see that for graphs that are time-consuming to mine with $\mathcal{A}_{base}$, i.e., {\em YouTube} and {\em Patent}, $\mathcal{A}_{split}$ significantly speeds it up, which is in turn further accelerated by $\mathcal{A}_{time}$. For example, on {\em Patent}, $\mathcal{A}_{base}$, $\mathcal{A}_{split}$ and $\mathcal{A}_{time}$ takes 911, 98.68 and 36.66 seconds, respectively. This shows the need to task decomposition to handle the straggler problem, and the advantage of our timeout strategy. In fact, if there is no straggler, $\mathcal{A}_{split}$ can be much slower than $\mathcal{A}_{base}$ as on {\em USA Road} due to excessive task decomposition, but $\mathcal{A}_{time}$ does not suffer from this issue.  We also tested other parameters and the results are similar; for example, when mining {\em Patent} with $(\gamma, \tau_{size})=(0.89, 20)$, $\mathcal{A}_{base}$, $\mathcal{A}_{split}$ and $\mathcal{A}_{time}$ take 3,386.37, 194.54, and 126.19 seconds, respectively.

Also, the RAM usage is small, in fact less than 1GB except for on {\em YouTube}. There is also almost no task spilling on disk, with the exception of {\em Patent} where a machine may keep up to 1.28GB spilled tasks, potentially due to a lot of decomposed tasks generated at some point of time. Overall, space is not an issue.

\vspace{1mm}
\noindent{\bf Effect of $\mathbf{(\tau_{split}, \tau_{time})}$.} We have tested the various pairs of values for $(\tau_{split}, \tau_{time})$ on our datasets, and the results are shown in Tables~\ref{table34}(a)-(j). We can see that $(50, 5$ sec$)$ consistently delivers either the best or close to the best performance for $\mathcal{A}_{time}$ in all our datasets. However, other settings may lead to significant increase in time. For example, on {\em Patent}, when fixing $\tau_{split}=1,000$ and varying $\tau_{time}=20, 10, 5, 1, 0.1$ seconds, respectively, the job running time is 743.94, 561.82, 419.77, 179.59, 71.61 seconds, respectively; while if we fix $\tau_{time}=5$ sec and vary $\tau_{split}=1000, 500, 200, 100, 50$ seconds, respectively, the job running time is 419.77, 448.78, 426.75, 	490.81, 36.66 seconds, respectively.

\vspace{1mm}
\noindent{\bf Comparison with~\cite{bigdata18}.} Recall from Section~\ref{sec:related} that \cite{bigdata18} first mines quasi-cliques with $\gamma'>\gamma$, then finds the top-$k'$ largest result subgraphs as ``kernels'' which are then expanded to generate $\gamma$-quasi-cliques and return top-$k$ maximal ones from the results. Thus, a job of~\cite{bigdata18} takes a parameter quadruple $(\gamma', k', \gamma, k)$. We use their code~\cite{bigdata18code} for comparison, and set $k'=3k$ following~\cite{bigdata18}'s  setting.

We observe that they cannot find the exact top-$k$ quasi-cliques. For example, on {\em GSE10158}, with $(\gamma', k', \gamma, k)=(0.9,30,0.8,10)$, the maximum subgraph found has 31 vertices while the true one has 32. If we reduce $\gamma'=0.85$ to include more results, it finds only 5 subgraphs with 32 vertices, but there are actually 6 maximal 0.8-quasi-cliques with 32 vertices. This happens even if we reduce $\gamma'$ to 0.81 (very close to $\gamma$). As another study, on {\em Amazon}, with $(\gamma', k', \gamma, k)=(0.501,300,0.5,100)$, only 9 subgraphs are returned with 6 with 13 vertices, and 3 with 12 vertices. In contrast, there are actually 13 0.5-quasi-cliques with 12 vertices or more.

Their program is also slower than our G-thinker's solution. For example, running their program on {\em YouTube} and {\em Hyves} with exactly the same parameters as in~\cite{bigdata18} (where $\tau_{size}=5$), {\em YouTube} takes 11,985.84 seconds just to get the top-100 results while even our slowest $\mathcal{A}_{base}$ runs for only 3,690.13 seconds to find all the 750 results (247 of which are maximal); {\em Hyves} takes 2,836.35 seconds to get the top-100 results while even our slowest $\mathcal{A}_{base}$ runs for only 7.45 seconds to find all the 2,349 results (1,480 of which are maximal). In fact, even the serial Quick+ takes only 348.49 seconds on  {\em Hyves} to find those results, thanks to our new degenerate cover-vertex pruning technique.

The other datasets we use were not considered in~\cite{bigdata18}. Here, we try different parameters and find that even with smaller parameters ($k'=30, k=10$) to allow faster running time, the program is not faster than our slowest exact programs $\mathcal{A}_{base}$ as reported in Table~\ref{all}. The results are reported in Table~\ref{bigdata18cmp}, where we set $\gamma'=\gamma+0.05$ in most cases since a larger $\gamma'$ leads to zero results in our tests.

We can speed up the approach of~\cite{bigdata18} by parallelization in G-thinker with minor system revision. Specifically, we revise the maximum clique mining program of G-thinker~\cite{gthinker} to find top-$k$ largest cliques (instead of only one biggest clique). We then revise G-thinker so that each machine initially loads a portion of clique ``kernels'' $S$ to construct tasks $t_S=\langle S, ext(S)\rangle$ for mining, which are initially loaded to the global queue. The difference here is that we no longer have a spawning vertex $v$ so we will pull 2-hop neighbors of all vertices in $S$ with $k$-core pruning to construct $ext(S)$, and then mine task subgraph $G(S\cup ext(S))$ with proper task decomposition. Each machine no longer spawns tasks from individual vertices in the local vertex table.

Note, however, that each task $t_S$ can no longer only pull vertices with ID larger than those in $S$, or we will miss maximal results that can be obtained by expanding a ``kernel'' with a vertex with a smaller ID, but~\cite{bigdata18} seems still does so to allow faster mining. So if $k$ is large, we will have a lot of redundant search space exploration by different ``kernel'', even degrading the performance. Of course, we can compromise the result maximality requirement and only pull vertices with ID larger than those in $S$ to eliminate redundancy, but our current implementation is not considering this.

\begin{table*}[!t]
\centering
\caption{Kernel Expansion in G-thinker}\label{kernel}
\vspace{2mm}
\includegraphics[width=2.1\columnwidth]{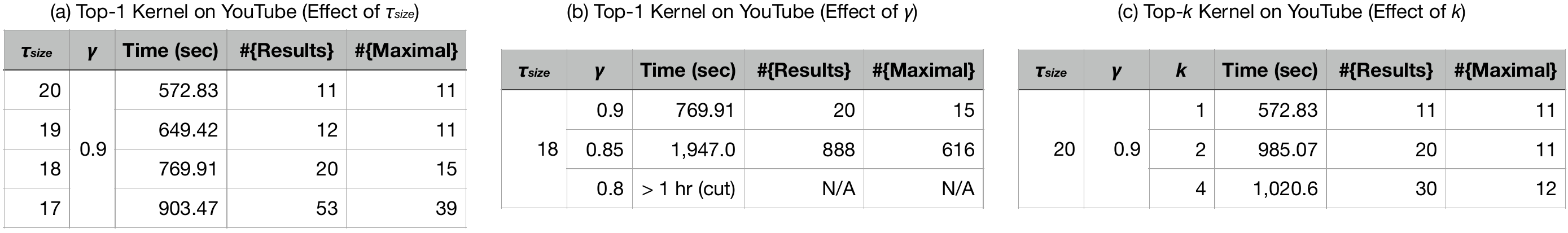}
\end{table*}

Table~\ref{kernel}(a) shows the result when we use top-1 kernel to expand quasi-cliques with different $\tau_{size}$. 
Table~\ref{kernel}(b) shows the result when we use top-1 kernel to expand quasi-cliques with different $\gamma$. 
Table~\ref{kernel}(c) shows the result when we use top-$k$ kernel to expand quasi-cliques with different $k$. 
Here, we do not observe obvious performance improvement compared with our exact solution. 

\vspace{1mm}
\noindent{\bf Scalability.} Table~\ref{table5} shows the scalability results of $\mathcal{A}_{time}$ on {\em Patent}, {\em Hyves} and {\em Enron}. For vertical scalability experiments, we use all 16 machines but change the number of threads on each machine, while for horizontal scalability experiments, we run all 32 threads on each machine but change the number of machines. We can see that $\mathcal{A}_{time}$ scales well along both directions, which verifies that our solution is able to utilize the computing power of all machines in a cluster.

\begin{table}[!t]
\centering
\caption{Mining v.s.\ Subgraph Materialization on {\em Patent}}\label{split}
\includegraphics[width=\columnwidth]{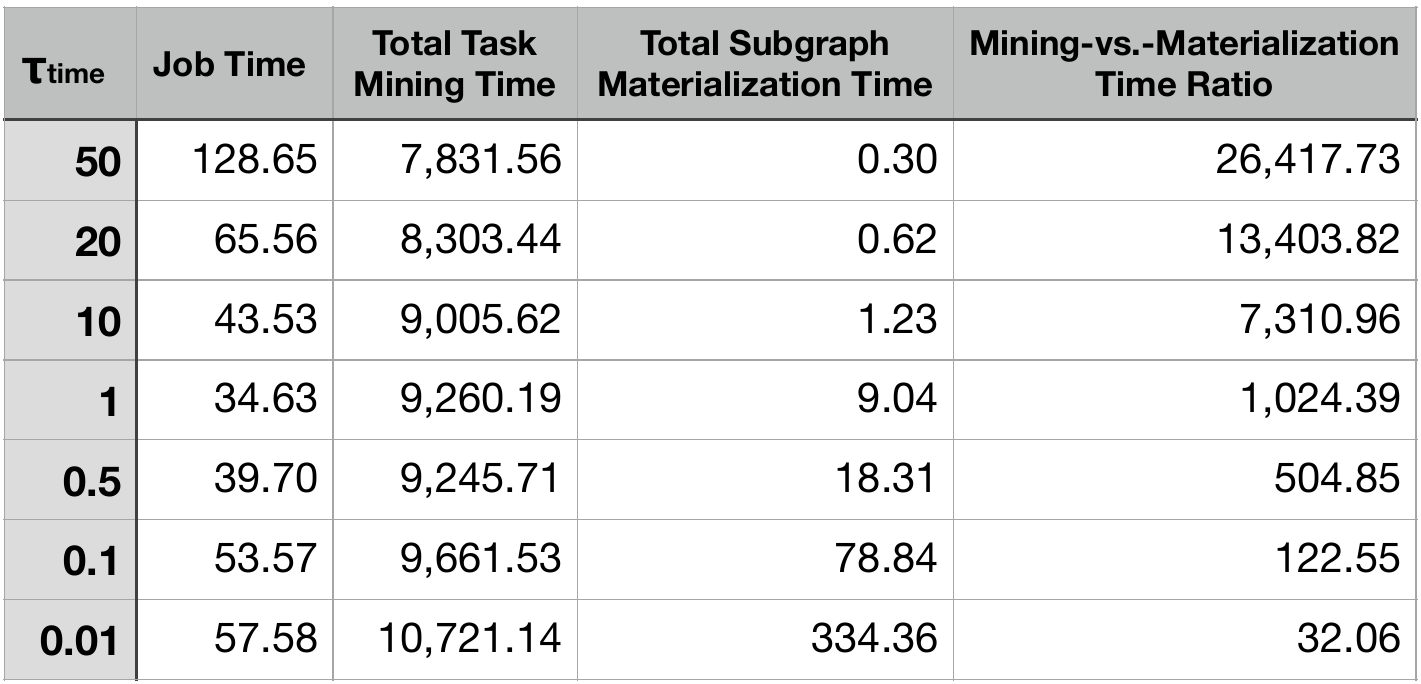}
\end{table}

\begin{table}[!t]
\centering
\caption{Mining v.s.\ Subgraph Materialization on {\em YouTube}}\label{table6_youtube}
\includegraphics[width=\columnwidth]{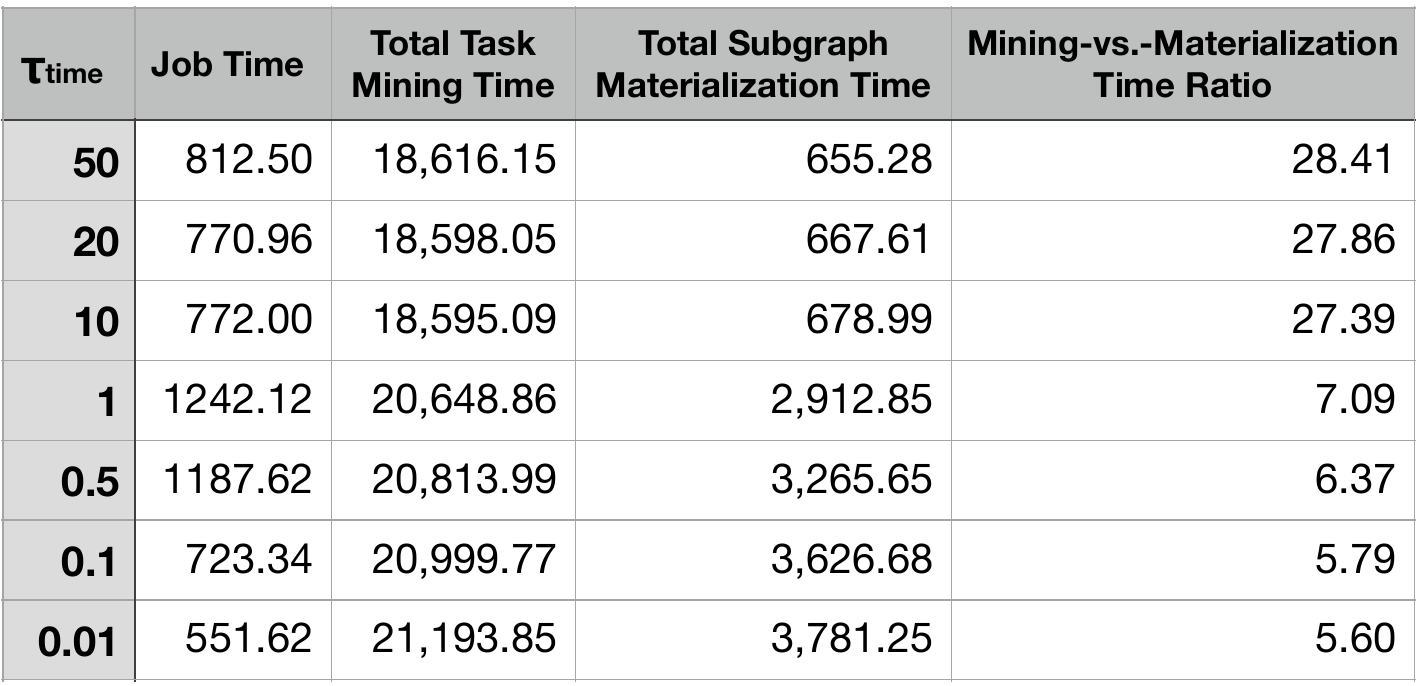}
\end{table}

\begin{table}[!t]
\centering
\caption{Mining v.s.\ Subgraph Materialization on {\em Hyves}}\label{table6_hyves}
\includegraphics[width=\columnwidth]{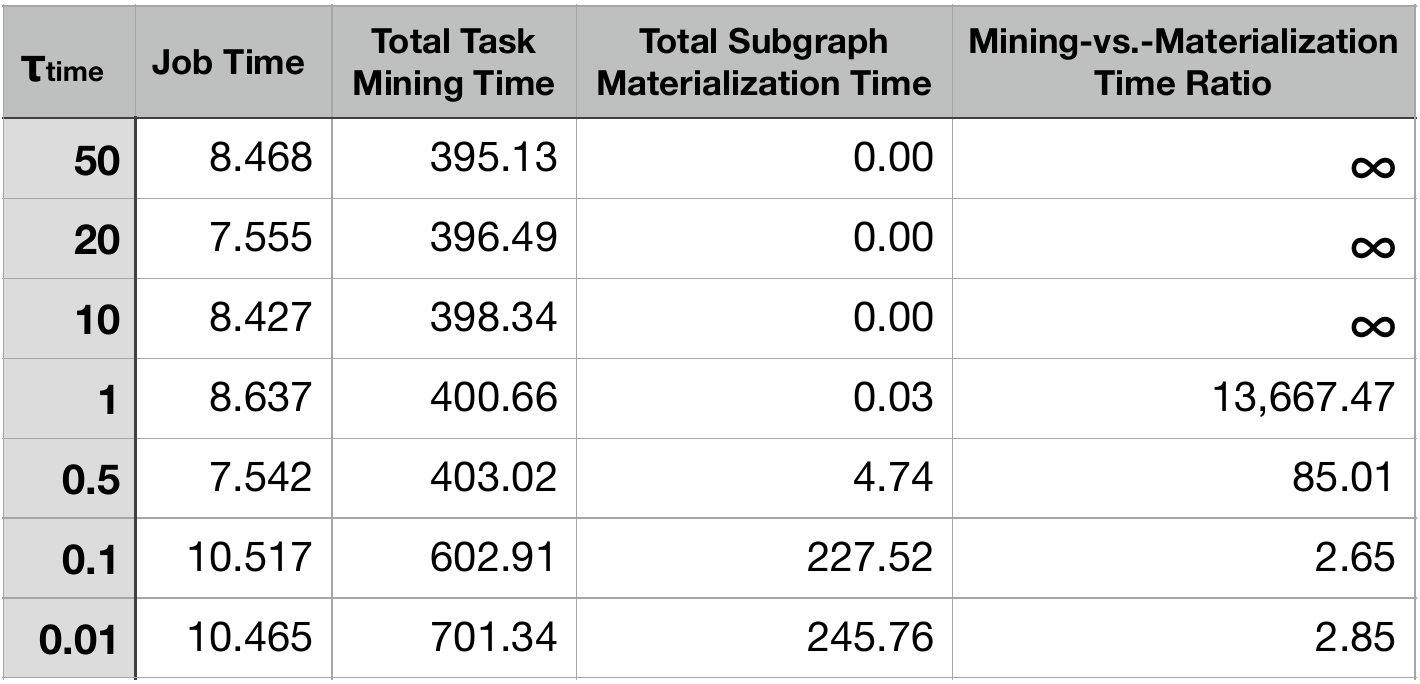}
\end{table}

\vspace{1mm}
\noindent{\bf Cost of Task Decomposition.} Recall from Algorithm~\ref{algo:tddq} that if a timeout happens, we need to generate subtasks with smaller overlapping subgraphs (see Lines~18-22), the subgraph materialization cost of which is not part of the original mining workloads. The smaller $\tau_{time}$ is, the more often task decomposition is triggered and hence more subgraph materialization overheads are generated.

Our tests show that the additional time spent on task materialization is not significant compared with the actual mining workloads. For example, Table~\ref{split} shows the profiling results on {\em Patent}, including the job running time, the sum of mining time spent by all tasks, the sum of subgraph materialization time spent by all tasks, and a ratio of the latter two. We can see that decreasing $\tau_{time}$ does increase the fraction of cumulative time spent on subgraph materialization due to more occurrences of task decompositions, but even with $\tau_{time}=0.01$ sec, the materialization overhead is still only 1/32 of that for mining, so only a small cost is paid for better load balancing. Tables~\ref{table6_youtube} and~\ref{table6_hyves} show the profiling results on {\em YouTube} and {\em Hyves} where we observe similar results and hence conclusion.

\begin{table}[!t]
\centering
\caption{Quick+ v.s.\ Quick}\label{quick}
\includegraphics[width=0.64\columnwidth]{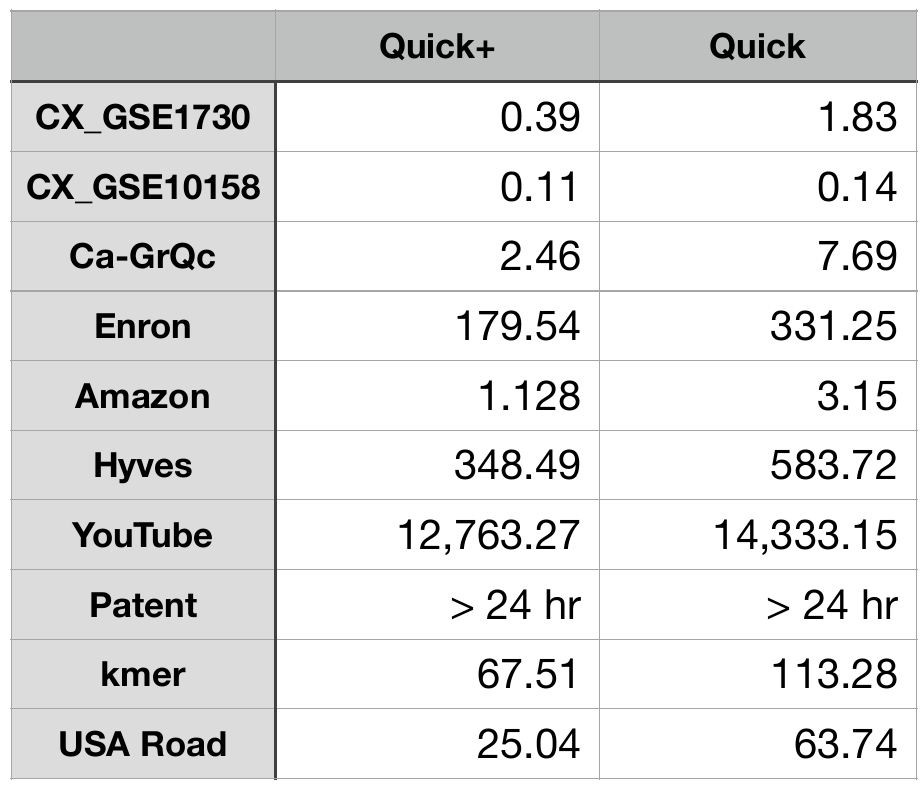}
\end{table}

\vspace{1mm}
\noindent{\bf Quick+ v.s.\ Quick.} We have compared our Quick+ with the original Quick algorithm on all the datasets in the single-threaded setting, the results of which are reported in Table~\ref{quick} where we can observe that Quick+ improves over Quick for up to over 4$\times$ w.r.t.\ running time.

Also, Quick did miss results although rare. For example, on {\em CX\_GSE1730} (resp.\ {\em Ca-GrQc}), Quick finds 1,601 of the 1,602 valid quasi-cliques (resp.\ 43,398 of the 43,499 valid quasi-cliques), i.e., misses 1 result.

\begin{table}[!t]
\centering
\caption{Cost of Different Pruning Phases}\label{phases}
\includegraphics[width=\columnwidth]{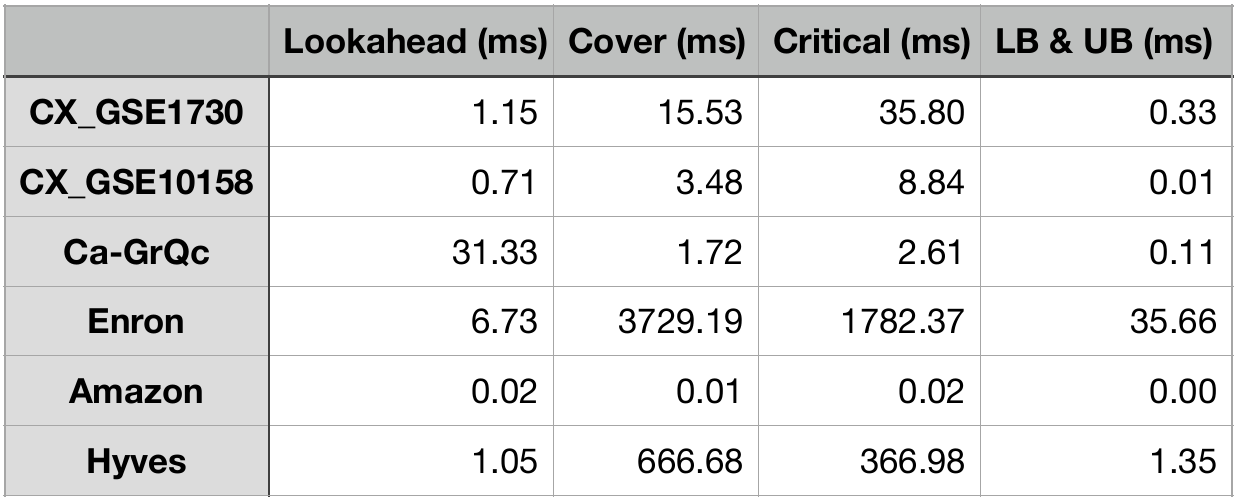}
\end{table}

In terms of how the costs of different phases of Quick+ distribute, we consider 4 important phases related to pruning rules: (1)~the check by lookahead pruning, (2)~the check by cover-vertex pruning, (3)~the check by critical-vertex pruning, and (4)~the check by lower- and upper-bound pruning. The results are shown in Table~\ref{phases} for 6 graphs, where we can see that cover-vertex pruning and critical-vertex pruning consumes a lot of the time, while the other two prunings are very fast. However, our test shows that it is still well worth to conduct cover-vertex pruning and critical-vertex pruning as otherwise, the increased search space adds significantly more time to the overall mining than the cost needed by the pruning rule checking.

\section{Conclusion}\label{sec:conclude}
This paper proposed an algorithm-system codesign solution to fully utilize CPU cores in a cluster for mining maximal quasi-cliques. We provided effective load-balancing techniques such as timeout-based task decomposition and  big task prioritization.

\vspace{2mm}
\noindent{\bf Acknowledgment.} This work was supported by NSF OAC-1755464, NSF DGE-1723250 and the NSERC of Canada. Guimu Guo acknowledges financial support from the Alabama Graduate Research Scholars Program (GRSP) funded through the Alabama Commission for Higher Education and administered by the Alabama EPSCoR.

%This paper proposed an algorithm-system codesign solution to fully utilize CPU cores of all machines in a cluster for mining maximal quasi-cliques. We are able to handle the million-node graph of {\em Hyves} in 51 seconds, and that of {\em YouTube} in 2.59 hours where serial mining would otherwise take 40 days. In fact, the most expensive mining task spawned from a vertex in {\em YouTube} would take over 100 hours to mine in serial. We provided a lot of effective techniques such as time-delayed task decomposition, and prioritized big task processing in our reforced G-thinker, besides effective pruning.

% space-saving !!!!!!!!!! move to CORR and then
% - save the proofs of pruning rules, summative way of rule presentation
% - introduce iter 1 and 2 briefly only

\bibliographystyle{abbrv}
\bibliography{ref_qc}

\end{document}